\documentclass{report}

\newcommand{\nn}{\newline \newline \noindent}
\newcommand{\ti}{\textit}
\newcommand{\tb}{\textbf}
\newcommand{\te}{\text}
\newcommand{\bse}{\begin{subequations}}
\newcommand{\ese}{\end{subequations}}
\newcommand{\bal}{\begin{aligned}}
\newcommand{\eal}{\end{aligned}}
\newcommand{\eg}{\emph{e.g.} }
\newcommand{\ie}{\emph{i.e.} }
\newcommand{\et}{\emph{et al.} }
\newcommand{\BG}{Barbieri-Giudice }
\newcommand{\AC}{Anderson-Casta\~{n}o }
\newcommand{\AM}{Athron-Miller }
\newcommand{\lcc}{$\Lambda_{\text{CC}}$}

\usepackage{amsmath, amssymb, color, graphicx, hyperref, fancyhdr, physics, slashed, mathtools, titlesec}
\usepackage[shortlabels]{enumitem}
\usepackage{epigraph}
\usepackage[utf8]{inputenc}
\usepackage[titletoc,toc]{appendix}
\usepackage{setspace}
\usepackage{tocloft}

\usepackage[paperwidth=23cm,paperheight=28cm]{geometry}
\graphicspath{{./Figures/}}

\titlespacing*{\subsubsection}{0.2pt}{0.8\baselineskip}{0pt}
\begin{document}	
	\begin{titlepage}	
		\begin{center}
			\textbf{	Candidate: Casper Dani\"{e}l Dijkstra. }\\
			\textbf{	Student Number: s2026104.}
		\end{center}
		
		\begin{center}
			\large{Master's thesis for Philosophy of a Specific Discipline.} 
			
			\large{Specialization: Natural Sciences. }
		\end{center}
		
		\begin{center}
			\huge{Naturalness as a reasonable scientific principle in fundamental physics}
		\end{center}
		
		\begin{figure}[h!]
			\centering
			\includegraphics[scale=0.2]{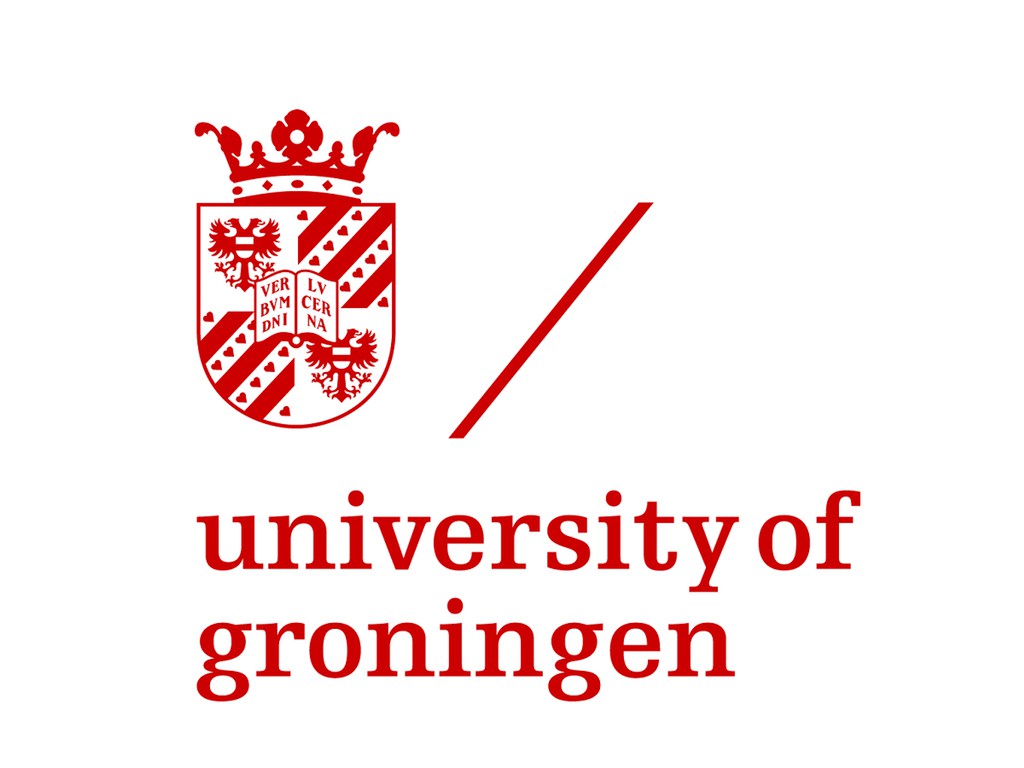}
		\end{figure}
		
		\vspace{5cm}
		
		\begin{minipage}{.9\linewidth}
			\begin{flushleft}                 \begin{tabbing}          
				\textbf{Supervisor:} \ \ \ \ \ \ \ \ \ \= \textbf{Prof. Dr. Simon Friederich.} \\
				\textbf{Second examiner:} \> \textbf{Prof. Dr. Leah Henderson} \\
				\textbf{Third examiner: }
				\> \textbf{Prof. Dr. Frank Hindriks} \\
				\textbf{Academic year:} \> \textbf{2017/2018.} 
			\end{tabbing}
			\end{flushleft} 
		\end{minipage}
		
	\end{titlepage}

\newpage 
	
\begin{center}
\large{ \textbf{	Abstract}}
\end{center}
{ \small	
Underdetermination by data hinders experimental physicists to test legions of fundamental theories in high energy physics, since these theories predict modifications of well-entrenched theories in the deep UV and these high energies cannot be probed in the near future. 
Several heuristics haven proven successful in order to assess models non-emperically and ``naturalness" has become an oft-used heuristic since the late 1970s. The utility of naturalness is becoming progressively more contested, for several reasons. The fact that many notions of the principle have been put forward in the literature (some of which are discordant) has obscured the physical content of the principle, causing confusion as to what naturalness actually imposes. Additionally, naturalness has been criticized to be a ``sociological instrument"  (Grinbaum 2007, p.18), an ill-defined dogma (\eg by Hossenfelder (2018a, p.15)) and have an ``aesthetic character" which is fundamentally different from other scientific principles
(\eg by Donoghue (2008, p.1)). 
\nn
I aim to clarify the physical content and significance of naturalness. Physicists' earliest understanding of naturalness, as an autonomy of scales (AoS) requirement 
provides the most cogent definition of naturalness 
and I will assert that
i) this provides a uniform notion which undergirds a myriad prominent naturalness conditions, ii) this is a reasonable criterion to impose on EFTs and iii) the successes and violations of naturalness are best understood when adhering to this notion of naturalness.
I argue that this principle is neither an aesthetic nor a sociologically-influenced principle. 
I contend that naturalness may only be plausibly argued to be an aesthetic/sociological principle when formal measures of naturalness and their use in physics communities are conflated with the central dogma of naturalness - the former may indeed be argued to be sociologically-influenced and somewhat arbitrary - but these formal measures of naturalness are in fact less successful than AoS naturalness. 
I put forward arguments as to why AoS naturalness is well-defined and why it was reasonable for physicists to endorse this naturalness principle on both theoretical and empirical grounds.
\nn 
Two dire violations of naturalness - the Higgs mass and the cosmological constant -
have already been discovered several decades, this fueled physicists' surmise that these problems can be rendered natural in more fundamental laws of nature.
``If one has to summarise in one word what drove the efforts in physics beyond the Standard
Model of the last several decades, the answer is naturalness'' as was correctly pointed out by Giudice (2017, p.3).
The surmise that these several parameters \emph{may actually be unnatural} has recently gained credibility due to compelling \emph{anthropic arguments} which entail that natural cosmological constant and Higgs vacuum expectation values would have led to life-hostile universes in which observers could not have emerged (Weinberg 1996, Donoghue 2007). 

Up to date, no compelling reasons have appeared as to why the laws of nature should \emph{generically} decoupling into quasi-autonomous physical domains. Chaotic phenomena provide a clear exception to this rule within classical physics and we should take into account the possibility of violations of this dogma in quantum physics as well.
A decoupling of scales in the quantum realm is often claimed to be entailed by the Decoupling Theorem (\eg by Cao and  Schweber (1993)), yet I will show that this theorem is too weak to underwrite quasi-autonomous physical domains in quantum field theories because one should additionally impose that parameters be natural. 
Violations of naturalness would then have ontological import - unnatural parameters would not be accurately described by effective field theories but rather by field theories exhibiting some kind of UV/IR interplay. }

\newpage

\begin{spacing}{0.8}
	\tableofcontents
	\addtocontents{toc}{\protect\thispagestyle{empty}}
\end{spacing}
	\thispagestyle{empty}	
	
\newpage

\begin{center}
	\large{\textbf{Notation and conventions}}
\end{center}
\addcontentsline{toc}{chapter}{Notation and conventions}
\setcounter{page}{1}
\pagenumbering{roman}
\subsubsection*{Natural units}
Natural units simplify particle physics considerably in relativistic quantum mechanics, since quantum mechanics introduces factors of $\hbar$ and special relativity introduces factors of $c$ which obfuscate equations. We can bypass this by using natural units where $\hbar = h/2\pi =  1$ (turns Joule into inverse seconds) and $c = 1$ (turns meters into seconds). This makes all quantities have dimensions of energy (or mass, using $E = mc^2$) to some power. \newline \newline
\noindent
Quantities with positive mass dimension, (e.g. momentum $p$ and $\partial_\mu$) can be
thought of as energies and quantities with negative mass dimension (e.g. position $x$ and time $t$) can
be thought of as lengths. Denoting the dimensionality of $`\cdots'$ by $[\cdots]$, some examples are:
\begin{subequations}
	\begin{align}
	\left[ \partial_{\mu} \right] = [p_\mu] = [k_\mu] = [m] &= M, \label{M1} \\
	[\text{velocity}] = [x] / [t] &= M^0, \\
	[dx] = [x] = [t] &= M^{-1}, \\
	\left[d^4x \right] &= M^{-4}.
	\end{align}
\end{subequations}
The fact that the action $\left[ S \right] = \int \left[d^4x \ \mathcal{L} \right] = 0$ is a dimensionless quantity implies that the Lagrangian density has dimensionality energy to the power four, i.e.
$\left[ \mathcal{L} \right] = M^4$.\footnote{
	This only holds true for  field theories in \emph{four dimensions} but the dimensional analysis can easily be generalized.
	One obtains $[\mathcal{L}] = M^{D}$
	in $D$ spacetime dimensions
	(\eg string theory is formulated in $D>4$ dimensions).} 
\nn
It is worth mentioning the exact expressions of several important constants in natural units. The Einstein equation in general relativity (GR) is given by
\begin{equation}\label{key}
\mathcal{R}_{\mu \nu} - \frac{1}{2}\mathcal{R}g_{\mu \nu} = \kappa \tau_{\mu \nu},
\end{equation}
where $\tau_{\mu \nu}$ is the energy-momentum tensor  and $\kappa$ is the constant which guarantees that Newtonian mechanics is recovered for low energies. Its expression is well-known in SI units, namely: $\kappa = 8 \pi G/c^4$ and this thus reduces to $8\pi G$ in natural units.
It is often desirable to display the Planck mass (which sets the scale where gravitational quantum effects become important) instead of $\kappa$ in gravitational equations, let us therefore mathematically relate these quantities. The Planck mass is defined as
\bse
\begin{equation}\label{key}
m_{pl} := \sqrt{\frac{\hbar c}{G}} \simeq 1.2 \cdot 10^{19} \text{ GeV}
\end{equation}
and since $\hbar = c = 1$ in natural units we can relate $\kappa$ and $m_p$ ($\kappa = 8\pi m_{pl}^{-2}$), which can be written more conveniently after introducing the \textit{reduced Planck mass} which already contains the requisite factor of $8\pi$:
\begin{equation}\label{eq:M_P}
M_P := \sqrt{\frac{\hbar c}{8 \pi G}} \simeq 2.4 \cdot 10^{19} \text{ GeV}.
\end{equation}
\ese
\subsubsection*{Dimensional analysis} 
Free massive spin-0 particles are described by the Klein-Gordon equation
\begin{equation}\label{key}
\mathcal{L}_{\text{free}} = \frac{1}{2} \left[ \partial_\mu \phi  \partial^\mu \phi - m^2 \phi^2 \right]
\end{equation}
so the dimensionality of the boson field can immediately be deduced using Eq. \ref{M1}, namely: $[\phi] = M$. Free spin 1/2 particles are described by the Dirac equation
\begin{equation}\label{key}
\mathcal{L}_{\text{free}} = \bar{\psi} (i \slashed{\partial} - m) \psi,
\end{equation}
where the slashed notation $\slashed{\cdots}$ should be read as $\slashed{\cdots} = \gamma^{i} (\cdots)$. The fact that \emph{one partial derivative} and \emph{one mass term occurs} in this Lagrangian density implies that $[\psi] = M^{3/2}$ and this can be generalized to other types of bosonic fields. 
The dimensionality of coupling parameters can be straightforwardly obtained in a similar fashion. In scalar $\lambda \phi^4$ theory for instance, the quartic couplic is dimensionless whereas $\lambda = M$ for cubic interactions (defined by $\lambda \phi^3$).

\subsubsection*{Indices and the Einstein summation convention}
Latin indices $i,j,k$  are used for spatial components only and thus take on values 1,2,3 in a (3+1)-space-time and values $1,2,\cdots,D-1$ in a $D$-dimensional space-time. Greek indices $\mu, \nu, \rho$ \emph{etc.} run over both spatial components and the temporal component.
Moreover, the Einstein summation convention, which states that repeated indices are summed over, is used in order to achieve notational brevity. It implies that
\bse
\begin{equation}\label{key}
\sum_{\mu=0}^{4} x_{\mu}x^{\mu} \equiv x_{\mu}x^{\mu}  \ (=x^2).
\end{equation}
and the Einstein summation convention likewise allows for the following short-hand notation 
\begin{equation}\label{key}
\mathbf{x} \cdot \mathbf{y} = x^{i}v_{i}.
\end{equation}
\ese
\newpage
\subsubsection*{Acronyms} 
The following acronyms are used repeatedly throughout this thesis. Although these abbreviations are introduced in the text as well, the confused reader may always consult this page to clarify the meaning of an acronym.
\begin{table}[h!]
	\begin{tabular}{|l | l|}
		\hline
		\textbf{Acronym}	& \textbf{Meaning} \\	
		\hline
		AC	& Anderson-Casta\~{n}o \\
		AoS	&	Autonomy of Scales \\
		AP &	Anthropic Principle \\
		BG	&	Barbieri-Giudice	\\
		BSM &	Beyond the Standard Model \\
		CC	&	Cosmological Constant	\\
		DT	&	Decoupling Theorem	\\
		EDM	&	Electric Dipole Moment \\
		EFT	&	Effective Field Theory \\
		GR	&	General Relativity		\\
		GSW	&	Glashow-Salam-Weinberg \\
		h.c. & hermitian conjugate	\\
		HEP	&   High Energy Physics \\
		LHC	&	Large Hadron Collider	\\
		LHS	&	Left Hand Side	\\
		LNH	&	Large Number Hypothesis	\\
		MSSM&	Minimal Supersymmetric Standard Model \\
		NMSSM&	Next-to-Minimal Supersymmetric Standard Model \\
		QCD	&	Quantum ChromoDynamics \\
		QFT	&	Quantum Field Theory \\
		RG(F)	&	Renormalization Group (Flow) \\
		RHS	&	Right Hand Side		\\
		SUSY &	SUperSYmmetry			\\
		vev		& vacuum expectation value \\
		WIMP	&	Weakly Interacting Massive Particle \\
		\hline
	\end{tabular}	
\end{table}

\setcounter{page}{1}
\pagenumbering{arabic}
\chapter{Introduction}	
I will first elucidate in \S \ref{sectionguidingprinciples} that underdetermination by data hinders experimental verification of theories in high energy physics (HEP). Several heuristics have proven successful in the history of particle physics
for non-emperical assessments of hypotheses, these guiding principles will be introduced in \S \ref{subsec:guides}. Naturalness is an often used guide in HEP, yet its scientific character and utility is considerably more contested than any of the other guides (Donoghue 2007). 
The utility of this guiding principle will be evaluated in the ensuing chapters of this thesis.
I will already
give a rough characterization of what naturalness aims to achieve in \S \ref{subsec:natural} and subsequently introduce various criticisms of the principle which have been put forward in the literature. I will then introduce the outline of the thesis in \S \ref{sec:outline} and signpost where the relevant questions will be answered in the ensuing chapters. The central question of this thesis is whether the models of HEP have to be natural.
\section{Guiding principles for fundamental theoretical physics
\label{sectionguidingprinciples}
}
 Fundamental theoretical physics 
suffers, more than ever, from underdetermination by data.
Underdetermination by data poses a grand problem for the high-energy frontier (HEP) of fundamental physics, most notably within the fields of particle physics and quantum gravity. The well-entrenched Standard Model and general relativity are typically modified at high energies\footnote{When employing natural units, the high-energy regime is equivalent to the small distance regime. The high-energy regime is referred to as the \emph{ultraviolet (UV) regime} and the low-energy regime is called the \emph{infrared (IR) regime} in the physics literature.} (Donogue 2008) which cannot be probed using currently available experiments, hindering physicists from experimentally verifying a myriad HEP theories which been around in the physics literature for a while (Giudice 2013).
I am not claiming that 
the unfeasability of
connecting theory to experiment is a novel problem, quite the contrary.\footnote{Underdetermination of data has recurringly played a big role throughout the history of physics. The mutually incompatible Big Bang model and Steady State model (Bondi and Gold 1948) hypotheses could initially not be tested because astronomy was still in its infancy.
	One should keep in mind that this conundrum was solved rather quickly in 1965, when the serendipitous discovery of the Cosmic Microwave Background (Penzias and Wilson 1965) ineluctably confirmed the Big Bang model.}
What I do claim is that the problem of underdetermination has become 
more severe in contemporary physics, 
because experiments can no longer put any HEP theory to the test. 
\nn
 The Standard Model has met numerous experimental confirmations and is arguably the greatest achievement in the history of physics (Arkani-Hamed 2012). Despite its emperical success, the model fails to provide a complete catalog of all building blocks of the universe, for instance because it does not describe dark matter, dark energy and the graviton (Feng 2013).\footnote{The problems of the SM will be discussed extensively in \S \ref{sectioneft}.} In order to ameliorate the paradigmatic model, physicists customarily modify it at high energy scales which cannot be probed by particle colliders yet.
Several theoretical frameworks predict new physics around the TeV scale (for instance low-energy SUSY) and these can be tested in near-future runs of the LHC. However, many other theories (high-energy SUSY (Arkani-Hamed \et 2005), composite Higgs models (Kaplan \et 1984), loop quantum gravity (Biswas \et 2012)) predict the emergence of novel phenomena at energies far beyond the TeV scale, posing an urgent problem for the falsification of these theories.\footnote{Underdetermination of data plays a pivotal role within other (high-energy) disciplines of theoretical physics as well including quantum gravity, which aims to provide a quantum description of gravitation but typically modifies the well-entrenched theory of general relativity in the deep ultraviolet (\ie at very high energies; these gravitational energy scales have not been probed yet).}
Nowadays,  whole legions of incommensurable rival theories give plausible explanations of identical phenomena. Both string theory and loop quantum gravity for instance aim to construct a viable unification of quantum mechanics and gravity.
\nn
This raises the philosophical question whether there are any good guiding principles in order to asses the viability of models in a non-empirical fashion. 
It was correctly pointed out by Hossenfelder (2018b) that
``[t]esting all possible hypotheses is simply infeasible, most of the scientific enterprise today - from academic degrees to peer review to guidelines for scientific conduct - is dedicated to identifying good hypotheses to begin with.''
Much of the scientific enterprise today is dedicated to selecting
hypotheses worth testing, where physicists have learned, all through their education, to identify viable hypotheses then assess these by means of successfully proven heuristics (guiding principles) and so-acquired experience
(Hossenfelder 2018a, p.1). These heuristics 
have become progressively more indispensable for our assessment of fundamental theories, since HEP hypotheses cannot be experimentally verified. 
A handful of successfully proven heuristics 
for the evaluation of hypotheses can be distilled from the history of particle physics.\footnote{See Holton (1973, \S5-10) for a discussion of these guides in the context of GR.}
The following guiding principles
	were of course not handed down to scientists on stone tablets, they were
	arrived at by much trial and error. I will argue that 
symmetries, unification, renormalizability and unitarity are fruitful guides, whereas naturalness 
(although it is a well-motivated principle)
sometimes gives poor counsel.
\subsection{Symmetries, unification, renormalizability and unitarity \label{subsec:guides} }
 \subsubsection*{Symmetries}
Symmetries play an important role in quantum field theories and dictate (through Noether's theorem) which quantities are conserved (energy, angular momentum, baryon number, \emph{et cetera})
(Noether 1918). 
Symmetries have provided a powerful engine which fuelled the formulation of the Standard Model and, afterwards, guided the
grand majority of attempts to go beyond this paradigmatic model.
The underlying symmetries of the standard model are characterized by the $SU(3) \otimes SU(2) \otimes U(1)$ Lie groups and many predictions of the standard model originate in these symmetries, including
the fact that protons and neutrons interact in the same
ways (the neutrons are related by an internal symmetry - isospin - of the nuclear force)\footnote{Of course, the proton and neutron do not interact the same way in electromagnetic interactions, because the proton is electrically charged whereas the neutron is not. Besides that, the neutrons interact identically.} and the photon is massless due to gauge symmetry (Nelson 1985).\footnote{Other examples are the custodial symmetry of the Higgs model in the limit of vanishing hypercharge
and quark mass difference, flavour symmetry in the limit of vanishing Yukawa couplings and chiral symmetry in the pion Lagrangian. }

Another example is the $SU(2) \otimes U(1)$ symmetry underlying the Glashow-Weinberg-Salam theory of electroweak interactions. The four Lie generators of this symmetry group entail the existence of 
four gauge bosons;
the massless photon, and the massive $W$, $Z$ and Higgs bosons. The existence of this Higgs boson was initially disputed by several scientists\footnote{Among others by Iliopoulos 1979, why the existence of the Higgs boson was disputed will be discussed later in this thesis since we firstly need to understand the naturalness criterion and its relation to the degree of fine-tuning in theoretical models.}, but faith in the explanatory power of symmetries was soon to be restored due to previous successes of symmetries in field theories. 
Consensus regarding the existence of the Higgs boson had been established long before this experimental verification at CERN's Large Hadron Collider (LHC) in 2012 (Atlas Collaboration 2012).
\nn
 Nelson (1985) argues that ``[s]ymmetry soon became routinely accepted as a valid principle for reducing problems of numerical naturalness to questions of structure'' and symmetries have indeed played the role of indispensable guiding principles in quantum field theories.\footnote{
 Symmetries may no longer be useful to describe physics at the smallest distances.
 The possibility that \emph{global symmetries} are absent in quantum gravity has been pursued by Kallosh \emph{et al.} (1995) and more recently by Banks and Seiberg (2011). The argumentation is that
 	global symmetries
 	can only be accidental and approximate and therefore be emergent properties
 	in the IR-regime of the most fundamental laws of nature. An easy-to-grasp example can be given for rotational symmetry - 
 	particles (where one may also conceive of ``macroscopic'' objects including Earth as particles, albeit not fundamental particles)
 	can be modelled as spherically symmetric point particles within EFTs which describe phenomena at large distances compared to the size of the particles - referred to in the literature as the ``spherical-cow approximation.'' A rotational $SO(3)$ symmetry emerges at large distances, whereas the particles in fact turn out to be \emph{spherically asymmetric} at smaller distances (Giudice 2017). Increasing the resolution may therefore reduce the symmetry of the system and even break every possible symmetry exhibited by low-energy effective Lagrangians. Despite the tremendous successes of symmetries in
 	the IR, they may be inconsequential for the truly fundamental theory in the UV (Witten 2017).} 
Symmetries are indispensable for assessments of Beyond the Standard Model (BSM) models; they can
either render BSM models ``viable'' (or at least ``potentially true'') or rebut these models straight away. 
Certain models (based on, say, $SU(6)$ symmetry) do not contain the SM and other symmetry groups may predict a Noether's current whose conservation has been ruled out  experimentally.\footnote{I will discuss the indispensiable role of symmetries more extensively in section \ref{sectiontechnicalnaturalness} and its relation to  \emph{naturalness} will be established in that section.}
\subsubsection*{Unification}
Throughout the history of physics, scientists have pursued \emph{unification} (Nelson 1985). What I mean by unification is that classes of many complicated things should be reducible to fewer, simpler things; this can be thought of as an incarnation of Ockham's razor which favors simplicity over complexity. 
The history of particle physics provides an
illuminating illustration of this desirable optimization of predictive power of theories. 
Newtonian mechanics has been superseded by general relativity, which recovers Newtonian mechanics in its non-relativistic static limit.\footnote{This does not entail that general relativity is correct and Newtonian mechanics is false. Laplace would find this hard to believe; theories were either wrong or right according to him (Merz 1904, p. 350). 
Newton's laws are perfectly correct within its range of validity (the non-relativistic limit) ries, as
sociating with each a position in a sequence. Newton's
law is perfectly correct within its range of validity, but has to be replaced by a more fundamental theory under relativistic circumstances. ``We would no more discard it than we would hydrodynamics, even though we know that fluids are not really
continuous'' as was succinctly argued by Nelson (1985, pp. 61-62).}

Many examples of unification can be provided in the context of particle physics. Molecules were divided into smaller numbers of atoms and subsequently atoms were describes in terms of their microconstituents. This astounding success subsequently led to the idea
that protons and neutrons too must have smaller constituents (initially called partons, later quarks) long before their existence was experimentally verified (Nelson 1985, p.61). Moreover, both electromagnetic and weak interactions can be described in a single framework called the \emph{electroweak theory} (Schwartz 2014).
\nn
Among other things, 
scientists aim to include quantum mechanical gravitational effects and solve the ``electroweak hierarchy problem'' (this will be discussed in chapter \ref{chapterviolations}).
Extensions of the Standard Model should recover the $SU(3) \times SU(2) \times U(1)$ symmetry groups for low energies in order to be consistent with the well-entrenched Standard Model. This poses constraints on theories beyond the Standard Model, where the simplest extension of the Standard Model would be a \emph{grand unified theory} (GUT) with an underlying $SU(5)$ symmetry (Dimopoulos \et 1981). Other examples of appropriate symmetry groups 
which
``contain the Standard Model'' 
are $SU(10)$ and $E_6$, all of which contain the 
$SU(3) \otimes SU(2) \otimes U(1)$
symmetry group of the SM as a subgroup.\footnote{Likewise, viable theories of quantum gravity should be invariant under \emph{diffeomorphism symmetries} in order to consistent with GR (see Wald (1984) for an extensive discussion of diffeomorphism invariance). }
New physics emerges at higher energies (for instance due to supersymmetric particles (Chan \et 1988), stringy excitations (Atick and Witten 1988), warped extra dimension (Randral and Sundum 1999)), while the Standard Model is recovered in the IR.

At the high-energy frontier of particle physics, physicists investigate
how the Standard Model can be \emph{extended}.\footnote{I will discuss in chapter \ref{chaptermotivation} why the Standard Model cannot provide the most fundamental description of particle physics.} 
The dream of a ``Theory
of Everything" (TOE) - a consistent microtheory including quantum gravity, and therefore
capable of accurately describing physics at  the Planck scale, and also
yielding the well-entrenched Standard Model at low energies - is the motivating force for the
study of superstrings and their descendants (M-theory, p-branes, \emph{etc}.). 
TOEs consist of two important ingedrients:
supersymmetry (a symmetry between bosonic
and fermionic particles) and duality (a symmetry connecting the weak and strong
coupling sectors of the theory).

 \subsubsection*{Renormalizability and unitarity}
An important mathematical property which is customarily required from quantum field theories is \textit{renormalizability}. All field theories of the Standard Model are renormalizable, meaning that unphysical infinities can be cured order by order by using regularization schemes (see Williams (2015, p.3) and Zee (2010) for a playful introduction to renormalization). Nonrenormalizable theories require infinitely many counterterms and these renormalized parameters must be taken from experiment. These theories are consequently thought to be unphysical; one would need to perform infinitely many experiments before being able to calulate anything!

The quantum formulation of general relativity is however not perturbatively renormalizable (Burgess 2003).
This has fueled scientists' belief that a \emph{more fundamental theory of gravity} is required to describe gravitational phenomena at high energies
(Donoghue 1994, \S3).\footnote{
	The nonrenormalizability of Einstein's theoretical framework is reflected by singularities which emerge on both the classical and the quantum level (Biswas \emph{et al.} 2012). On the classical level, one encounters astrophysical spacetime singularities (within black holes) and the notorious cosmological singularity known as the Big Bang.
	Scientists interpret these as indications that the theory is pushed beyond its range of applicability, 
	since both kind of singularies emerge in the UV-regime of GR. These singularities have prompted worldwide research into quantum gravity, which would have to provide a viable description of gravity in the UV-regime of general relativity (Biswas \et 2013).}  
Ultimately, physicists aim to construct an appropriate theory of quantum gravity because gravitational interactions are no longer negligible in particle physics at sufficiently high energies (around the Planck scale). All forces should therefore be expressible within a unified field theory - and this theory should be renormalizable.
\nn
Another important theoretical principle which quantum field theories should respect is \textit{unitarity} (discussed elaborately in Peskin and Schroeder 1995). This mathematical property guarantees that the sum of quantum mechanical probabilities is conserved. I will discuss one important application of this principle.\footnote{Another important application may be given for quantum gravity. Physical ghost-like degrees of freedom render theories non-unitary, their occurrence in field theories should therefore be circumvented. Several theories of modified gravity have been ruled out due to their non-unitarity, for instance Stelle's Fourth Order Gravity (Stelle 1977) and local theories of gravity which involves modifications of the Ricci $\mathcal{R}_{\mu \nu}$ and Riemann 
	$\mathcal{R}_{\mu \nu \rho \sigma}$
	tensors (Biswas \et2012).\footnote{Effective field theories, which are introduced in section \ref{sectioneft}, are however exempt of the rule of unitarity when the ghost is introduced at energies beyond the domain of applicability of the effective field theory. 
		Stelle's Fourth Order Gravity may therefore be a viable theory of gravity at energies beyond general relativity but have to be replaced by a unitary theory of gravity at even higher energies.}
	Modifications involving Ricci scalars (so-called $f(\mathcal{R})$ theories) do not introduce ghost-like DOFs, however, these do not ameliorate the UV-behavior of the theory either with respect to GR ($f(\mathcal{R})$ theories thus remain nonrenormalizable). This trade-off between nonrenormalizability and unitarity only appears in \emph{local} theories of gravity and is circumvented when \emph{non-local} theories of gravity are considered. The reasonable requirement that the fundamental quantum formulation of gravity
	is both renormalizable and unitary implies that it is necessarily \emph{non-local} (Conroy 2017).}
\begin{itemize}
	\item {}
Fermi introduced his theory of nuclear beta decay in 1933/34; this theory of the weak interactions is described by
\begin{equation}\label{eq:fermi}
\mathcal{L}_{\textbf{Fermi}} = \frac{G_F}{\sqrt 2} J_{\mu} J^{\mu}
\end{equation}
(where $G_F = 1.17\times 10^{-5}$GeV$^{-2}$ is the Fermi constant\footnote{The Fermi scale is therefore given by $M_F = G_F^{-1/2} = 293$ GeV.} and the currents $J_{\mu}$ are bilinears in the fermions).
Fermi theory has been ravishingly successful: 
all experimental results for weak interactions were accurately described by a Lagrangian of this
form for slightly more than forty years (Dine 2007, \S4.0.1). 

The Lagrangian however exhibits violations of unitarity just below the TeV scale. From a purely mathematical point of view, 
Fermi theory
cannot be conceived of as a fundamental field theory for weak interactions, but instead should be considered an \emph{accurate low-energy field theory} (\ie an effective field theory - effective field theories will be discussed extensively in \S \ref{sectioneft}). Indeed, renormalizability is retained because deviations from the Lagrangian in equation \ref{eq:fermi} were detected when the energy of bosons approached the mass of the $Z$-bosons ($E \simeq 91.91$ GeV) and the subsequent Glashow-Salam-Weinberg (GWS) theory of electroweak interactions could account for these deviations while retaining renormalizability as well.\footnote{The entire Standard Model is renormalizable and could, based on mathematical consistency, be a fundamental theory of particle physics. Physicists have put forward other compelling reasons as to why this is not the case, as will be discussed in \S \ref{sectioneft}.}
\end{itemize}
Unfortunately, a plethora of incommensurable
models pass aforementioned selection criteria with flying colors. These models would therefore be equally viable. 
If the guides that were introduced in this subsection would constitute an exhaustive list of guiding principles, only future experiments would allow physicists to distinguish good from bad theories. 
An additional guiding principle called ``naturalness'' has been put forward in the physical literature in the 1970s to further assess 
 loop quantum gravity, supersymmetry, little Higgs models and other BSM models (Giudice 2007). Naturalness purportedly helps physicists to recognize the good theories
 from a large landscape of otherwise equally plausible models. Because  naturalness is said to sometimes give poor counsel (Donoghue 2007), its physical motivation is often obscured in the literature (Williams 2015) and the concept is hard to quantify in measures of naturalness (as I will argue in \S \ref{sectiontechnicalnaturalness}-\ref{sec:bayesian}),
 it is more contested than aforementioned guides. I will provide arguments in my thesis supporting my claims
 that naturalness is a reasonable scientific criterion and
  that problems of naturalness do constitute genuine problems that scientists should aim to solve.
\subsection{Naturalness \label{subsec:natural} }
To come to a prelimary notion of naturalness, let me first elaborate on how it came about.
The term ``natural" first enters HEP while the SM is rising. To my knowledge it was first discussed by Weinberg (1972) who aimed to find
a ``natural explanation of the approximate symmetries in nature". This was soon ensued by other embeddings of naturalness in the physics literature\footnote{Other examples of early discussions and applications of naturalness are
	\begin{itemize}[nolistsep]
		\item {} “A natural mechanism for mass hierarchy” (Georgi and Glashow,
		1972).
		\item {} “Calculability and naturalness in gauge theory”
		(Georgi and Pais, 1974).
\end{itemize}}
and most physicists agree in hindsight that naturalness has played a grand role in
theory choice within HEP in the last decades\footnote{See Arvanitaki \et (2014) and Dine (2015) for similar claims}:
\begin{center}
If one has to summarise in one word what drove the efforts in physics beyond the Standard
Model of the last several decades, the answer is naturalness. (Giudice 2017, p.3)
\end{center}
The principle was introduced in the literature as a plausible
assumption (Castellani 2018, p.2) and has been strongly advocated by large physics communities (\eg Nelson (1985), Giudice (2008), Wells (2015), Williams (2015)).
Nelson advocates the utility of this guiding principle and
contends that
``naturalness seems to be one of the best-kept secrets of
physicists from the public, a secret weapon for evaluating and motivating theories of the world on its deepest
levels'' (Nelson 1985, p.61).\footnote{Others take an opposing view. Grinbaum describes naturalness as a ``last resort'' for theory choice; when no other compelling scientific arguments can be put forward as to which model (of potentially incommensurable models) will be in better accordance with future data then scientists rely on naturalness.}

Although naturalness has indeed become an important guide in HEP, the fruitfulness of the principle is becoming progressively more contested (Dine 2015). 
The guiding principle has been critically reviewed by several physicists and philosophers of quantum field theory, some of whom claim that the criterion is a ``sociologial heuristic'' (Grinbaum 2007), a ``philosophical principle'' (Shifman 2012) or merely a theorists' prejudice 
(Hossenfelder 2018a). These criticisms may give the impression that the principle can be ignored if one does not like it (Shifman 2012 actually claims that) and it is of fundamental importance to evaluate the scientific character of naturalness. Several physicists have argued that naturalness \emph{may} no longer be a guiding principle in future research (Williams (2015), Giudice (2018)) and other that it \emph{should} no longer be a guide (Hossenfelder (2017), (2018a)). It is therefore of grand importance to evaluate the utility of naturalness and evaluate whether it actually constitutes a fruitful guiding principle.
 
In doing so, I think it is important to realize that many formulations of the principle coexist in the scientific literature\footnote{I will show in section \ref{sec:bayesian} that landscape naturalness is based on a disparate notion of naturalness}. For this reason, one should always carefully examine \emph{which notion} of naturalness is criticized. Often ``naturalness'' is criticized 
and it is left to the reader's subjectivity which notion of naturalness has been adhered to
(see Hossenfelder (2018a) for a good example). Different definitions of naturalness seemingly differ in physical content - among other things it is defined in terms of i) symmetry considerations ('t Hooft 1979), ii) infinitesimal variation of parameters (Barbieri and Giudice 1988), iii) a prohibition of correlations between widely separated physical scales (Susskind 1979), \emph{etc.}. 

We should first study whether these notions
are disparate or whether they, upon closer examination, are \emph{superficially discordant} (they would then formalize the same idea). Perhaps some notions may be said to be ``aesthetic principles'' or a ``theorists' prejudice'' while other notions are exempt from these criticisms and constitute fruitful guidance.
\nn
This brief discussion of naturalness already gives rise to several important questions:
\begin{itemize}[nolistsep]
	\item {} How should naturalness be defined?
	\item {} Was naturalness successful or not in guiding research?
	\item {} Is it an aesthetic or scientific principle?
	\item {} Should it be redefined or abandoned?
\end{itemize}
The aim of my thesis is to answer these questions and clarify the (often misunderstood) physical significance of naturalness. In order to so, I will first explain what the physical motivation for physicists' early understanding of naturalness has been and assert that this notion of naturalness is a well-motivated requirement to impose on EFTs. I will now briefly introduce the \emph{central tenet} of this notion to clarify why effective field theories have to be introduced 
in \S \ref{sectioneft}
before naturalness can be discussed in depth.

The pivotal assumption underlying the \emph{earliest notion of naturalness} is the that no special
correlations occur between phenomena occurring at vastly different physical scales (Giudice 2013).
Naturalness imposes that low-energy physics is not sensitively dependent on high-energy physics (Williams 2015, p.2).
This autonomy of scales naturalness condition is \emph{deeply
rooted in the logic of effective field theories} and naturalness problems only emerge in the context of EFTs. 
\section{Outline of the thesis \label{sec:outline}}
My thesis is organized in the following way:
\begin{itemize}
	\item {\textit{Chapter \ref{chaptermotivation}: Motivating and defining naturalness.} 
		
	I will
	first motivate EFTs and describe in great detail how these field theories work in \S \ref{sectioneft}. 
I will then introduce the earliest definition of naturalness - as an autonomy of scales (AoS) requirement - and argue in this chapter that AoS provides the most cogent definition of naturalness in \S \ref{sectionnaturalprohibitionscales} (more arguments will be provided in chapter \ref{chapterviabledefinition}).
AoS naturalness ensures that \emph{widely separated scales decouple} and that this is an important feature that EFTs should exhibit in order to yield meaningful predictions.
	I argue that naturalness is not deeply rooted in EFTs (as is argued, among others, by Giudice 2008) but is instead a necessary additional condition which guarantees the explanatory power of EFTs. This is guaranteed because quantum corrections cannot be arbitrarily large in natural theories. 
\nn		 
Many inequivalent formulations of naturalness have been put forward in the literature (among others, 't Hooft's technical naturalness (\S \ref{sectiontechnicalnaturalness}) and the \BG measure (\S \ref{sectiongiudicebarbieri})). I will examine the compatibilities and disagreements between these oft-used definitions in these sections and assert these definitions are too restrictive to capture the central dogma of AoS naturalness.
More recently, a reminiscent - but in fact mutually incompatible - notion of naturalness has become fashionable in the mid-90s, where naturalness has acquired a statistical character. The most prominent definition - the \AC measure - will be discussed in \S \ref{sec:anderson}. I will assert that naturalness is better understood as an AoS requirement because the philosophy behind the latter requirement can be defended well, whereas the philosophy behind the former cannot. In this chapter, I will thus put forward arguments as to why AoS naturalness is a cogent and well-motivated notion of naturalness, while other definitions are ineluctably prone to criticisms which would undermine the validity of these notions of naturalness.
	}

	\item {\textit{Chapter \ref{chapterviabledefinition}: Defending autonomy of scales naturalness}.
		
	In this chapter I refute two often expressed criticisms of naturalness, namely that it is an \emph{aesthetic principle} and \emph{that the principle is ill-defined}.
	These criticisms could undermine the validity of the principle and I will argue that these criticisms are not sound when AoS naturalness is concerned. 
	
	I will elucidate in \S \ref{sec:aesthetics} that arguments of aesthetics give poor counsel in physics and will critically evaluate Dirac's Large Number Hypothesis and enunciate that aesthetic arguments have led Dirac's description of the universe in sterile byways. I will then distill the reasons why there is no logical connection between aesthetics and the laws of nature in \S \ref{subsec:implicationnaturalness}. 
	Naturalness would be utterly useless if it is indeed an aesthetic criterion, I will therefore assess whether it actually an aesthetically-motivated principle. I put forward arguments as to why naturalness is not an aesthetic principle in \S \ref{sec:naturalnessnotaesthetic}.
	\nn
	I will then disentangle naturalness from its alleged ill-defined character. The AoS notion of naturalness, when conjoined with the renormalization group equations and physical intuition, allows us to recognize both natural and unnatural parameters. The remaining important question is whether \emph{the laws of nature are necessarily natural} - this will be discussed in chapter \ref{chapterviolations}.
}

	\item {\textit{Chapter \ref{chapterviolations}: Assessing the utility of naturalness}
		
	In order to assess the utility of naturalness, I first introduce several successes of the naturalness criterion (some of which have only been realized in hindsight) in \S \ref{sectionsuccessesnaturalness}, ensued by severe violations of naturalness in the context of both cosmology and particle physics in \S \ref{sec:violations}. The biggest violations are caused by the 125 GeV Higgs mass and the unnaturally small cosmological constant, both of 
	exhibit a strong sensitivity on high-energy physics. Whether the Higgs mass in the electroweak theory does or does not respect naturalness depends entirely on whether new physics is discovered beyond the TeV-scale. I will introduce several popular resolutions of this naturalness problem in the context of Beyond the Standard Model (BSM) models in \S \ref{sec:solvehierarchy}  and assert that the prospects of any of these natural theories
	providing a viable resolution to this naturalness problem 
	seems dim in light of current experimental data. \emph{Could certain parameters perhaps be unnatural?}

The unnaturally low values of both the Higgs mass and the cosmological constant can be understood by imposing selection criteria in the context of a multiverse. For instance, if the cosmological constant would have its natural value in this universe then galaxies could not have formed and life could not have emerged. This explanation is an \emph{anthropic argument} - since observers are located in this universe the cosmological constant could impossibly have been a lot larger than its actual value (Donoghue 2007, p.7). Solutions involving selection criteria will be extensively discussed in \S \ref{subsec:selection}. 
Naturalness is usually thought to be entailed by the \emph{Decoupling Theorem}, but I will elucidate why generic effective field theories do not satisfy the assumptions underlying this theorem.
I will argue that this is not the case, and we should take into account the possibility that the most fundamental laws are not describable by EFTs. In \S \ref{subsecdecnotsat} I will show that, \emph{even field theories which do meet the conditions of the Decoupling Theorem do not necessarily exhibit a decoupling of widely separated scales} - this is only guaranteed whenever these field theories are \emph{natural}.
Finally, I will put forward arguments as to why unnatural parameters in quantum field theories are not describable by effective field theories but could be described by theories exhibiting an \emph{UV/IR interplay} in \S \ref{subsec:UV/IR}.
}

	\item {\textit{Chapter \ref{chapterconclusions}: Conclusions} are briefly summarized in the last chapter.
	}
\end{itemize}

\chapter{Motivating and defining naturalness \label{chaptermotivation}}
Because naturalness problems arise in the context of effective field theories (EFTs) (Williams 2015, p.1), my first goal is to elucidate how these field theories work in \S \ref{sectioneft}. I will clearly state i) on which assumptions their ability to accurately describe the universe hinges and 
ii)
explicate that these assumptions are intimately related to scientists' earliest understanding of naturalness 
as an ``autonomy of physical scales" (AoS) requirement. 
This notion was expressed clearly in the earliest discussions of the concept, among others by Ovrut and Schnitzer (1980) and Susskind (1979). This notion of naturalness
will be introduced in \S \ref{sectionnaturalprohibitionscales}, where
the corresponding restrictions entailed by this notion of naturalness on EFTs (naturalness prohibits large quantum corrections) are discussed as well. I will argue that naturalness is a reasonable criterion to impose on EFTs, because it delivers the decoupling of energy scales on which the fruitfulness of EFTs hinges.

Ensuing definitions of naturalness have been proposed in order to formalize this notion of naturalness.
The most prominent notions of naturalness
(absolute and technical naturalness) will be discussed in \S \ref{sectiontechnicalnaturalness}, where I will assert that these definitions obscure the physical content of naturalness and merely provide \emph{sufficient mathematical criteria} for 
a
prohibition of widely separated correlations (which is entailed by the AoS notion of naturalness). These notions thus constitute impoverished formulations of the earliest formulation of a shared unease about widely separated correlation functions.
There is a close relationship between unnatural parameters and \emph{fine-tuned parameters} (\S \ref{subsecfinetuning}) and several scientists have, as a result, misidentified naturalness as ``a prohibition of fine-tuned parameters'' (\eg Dine (2015)).
I will discuss why this identification of naturalness with (prohibitions of) fine-tuning ignores essential features of the naturalness dogma in \S \ref{subsecfinetuning} and argue that the \BG measure consequently cannot capture the central dogma of AoS naturalness in \S \ref{sectiongiudicebarbieri}. 
\nn
A discordant notion of naturalness - called \emph{landscape naturalness} - has become fasionable in the physics literature from the 1990s onward. This is a Bayesian notion of naturalness 
employing a multiverse approach along with 
suitably chosen
selection criteria. 
When adhering to landscape naturalness (the most prominent one - the \AC measure - will be evaluated in \S \ref{sec:anderson}) the corresponding notion of landscape does not formalize our initial notion of naturalness. Unnatural parameters are now thought of as \emph{unlikely in the landscape}.
Naturalness, when conceived of in this way, aims to explicate a disparate sense of unease 
rather than a strong sensitivity of IR parameters on UV physics.
For instance, the value of the cosmological constant is very unnatural in the autonomy of scales notion because the parameter is quartically divergent in the UV regime (Arvanitaki \et 2017), while it is considered ``natural'' in landscape naturalness conjoined with an anthropic selection criterion (Susskind 2004).
\nn
The AoS and landscape notions of naturalness ``can and will come apart'' as was recently put forward by Williams (2018). It is therefore imperative to avoid equivocation in discussions of naturalness and clearly state which notion of naturalness one adheres to. As I already discussed, the AoS notion is what I claim to be the most cogent definition of naturalness. This notion will therefore be adhered to in all ensuing chapters. 
\nn
Briefly summarized, the goals of this chapter are threefold.
I will i) discuss how EFTs work, ii)
argue why the autonomy of scales notion introduced in \S \ref{sectionnaturalprohibitionscales} provides the most accurate and transparent understanding of naturalness and iii) put forward arguments as to why qualitative measures of naturalness fail to capture the central tenets of AoS naturalness.
\section{Effective field theories
	\label{sectioneft}
}
The Standard Model (SM) provides the well-entrenched theoretical framework  in which all (hitherto) experimentally verified elementary particles and corresponding interactions are accurately described. 
With a handful of free parameters\footnote{The Yukawa coefficients fix the masses of the six quarks 
	$(u, d, c, s, t, b)$	
	and three lepton flavors
	$(e, \mu, \tau)$, the Higgs mass $m_{\text{H}}$ and vacuum expectation value (vev) $\nu$ (which multiplies
	the Yukawa coefficients to determine the fermion masses), three angles and one phase of
	the CKM (Cabibbo-Kobayashi-Maskawa) matrix (which mixes quark weak- and stronginteraction
	eigenstates), a phase for the QCD vacuum, and three
	coupling constants $g_1, g_2, g_3$ of the gauge group underlying the Standard Model: $SU(3) \times SU(2) \times U(1)$. The 
	experimentally verified existence of neutrino oscillations implies that neutrinos have a very small but nonzero mass, implying that there are seven more parameters (three masses and
 four CKM matrix elements).}, the SM has enabled scientists to accurately predict the properties of matter
from very small distances (down to about $10^{-18}$ m)\footnote{Although very small from a human perspective, it is still
	roughly 15 orders of magnitude smaller than $M_P$ where we know that
	(at the very latest) the effects of quantum gravity may no longer be ignored. The
	nature of the quantum dynamics of elementary processes must undergo a dramatic,
	and as yet completely mysterious, alteration. This is one of many incentives for LHC to probe progressively smaller distances.} to the conditions of the early universe one second after the Big Bang (Giudice 2008, p.4).
It constitutes the overarching theoretical framework of three quantum field theories (QFTs) -
quantum electrodynamics (QED), the theory of weak interactions\footnote{
\label{footnote:GWS}	
	QED and the weak interactions are customarily described in the unified framework called Glashow-Weinberg-Salam (GWS) theory of electroweak interactions.} and the theory of strong interactions (called quantum chromodynamics or QCD) 
-
and yields a ravishingly successful description of non-gravitational phenomena in the quantum realm. Extraordinarily accurate agreements between theoretical predictions and experimental data 
(often on the the part-per-billion level) 
have firmly established the accuracy of the theoretical framework
(Arkani-Hamed 2012, p.54). No deviations from SM predictions have been detected up to energies scales of $E \simeq 100$ GeV (Giudice 2017).
\nn
My praising discussion of the SM may give the reader the impression that no anomalies are to be solved in particle physics.
One may consequently wonder why many scientists are actually exploring the high-energy frontier of particle physics, why would we expect deviations from SM predictions at higher energies?
This would be an unduly optimistic view - HEP physicists are practically certain that the SM ceases to be applicable at sufficiently high energies 
for a couple of disparate reasons.

The SM suffers from several epistemological and conceptual problems.
Among other things, the framework fails to account for basic phenomena of the universe including dark matter and dark energy (Dine 2007, \S4.5) and  the origin of non-vanishing neutrino masses (which are responsible for neutrino oscillations) remains mysterious in the SM (Duncan	2012, p.564). 
	Additionally, the myriad of coupling
constants (18 in total) is a puzzling feature of the SM - it seems unlikely that a purportedly fundamental theory contains (i) so many undetermined parameters and (ii) parameters whose values are ``all over the map'' (Dine 2007, pp. 73, 107).\footnote{
Many physicists think that we can learn something fundamental about these undetermined parameters - that they become calculable in a more fundamental theoretical framework.
Another frequently mentioned problem of the Standard Model is the unnaturally low Higgs mass. Since naturalness is a contrived guiding principle and  different definitions of naturalness coexist in the literature, a discussion of this problem will be deferred to a later chapter (\ref{sechierarchy}).}
Despite its impeccable empirical success, 
we are led to conclude that
the SM fails at
providing a complete catalog of the building blocks of our universe (Feng 2004, p.2). In fact, we can even show mathematically that the SM is an incomplete theory.
\subsection{Why the Standard Model consists of effective field theories} 
QFTs possessing abelian gauge symmetries (for instance QED in the SM and $\phi^4$ models) run into problems when they are extrapolated to arbitarily high energy scales (Williams 2015, pp. 3,4). The interactions in QED become progressively stronger at smaller distances (as is reflected in its renormalization group equations)\footnote{This claim hinges on the assumption that the spacetime dimension is greater than 3 (Williams 2015, p.3); we are however not interested in field theories in less than 4 dimensions in this thesis so can safely ignore this caveat.} and  become \emph{infinitely strong for sufficiently high (but finite) energies}. This unphysical so-called \emph{Landau pole} implies that this QFT breaks down on purely mathematical grounds (see Aizenman (1982) 
for a discussion in a perturbative context
and Montvay \& Munster (1997) for a non-perturbative context) and ought to be replaced by a more fundamental theory if one wants to describe QED phenomena at energies beyond the Landau pole. 

The situation is fundamentally different for QFTs which possess non-abelian symmetries (such as QCD). While QED becomes progressively stronger coupled at smaller distances (higher energies), QCD becomes progressively weaker coupled. This property of QCD (and non-abelian theories in general) is called \emph{asymptotic freedom} (Williams 2015, p.3) and entails that non-abelian theories do not contain any Landau poles. This class of field theories remains mathematically consistent up to arbitrarily high energies, guaranteeing renormalizability of non-abelian theories (Williams 2015, p.4). 
Compelling reasons as to why both abelian and non-abelian QFT of the SM should nonetheless be conceived of as \emph{effective rather than fundamental} descriptions of elementary particles have culminated in the 1970s (Burgess 2004, p.2, Dine 2007, \S 4.2).
A serious encumbrance to treating both non-abelian \emph{and abelian} theories of the SM as 
 complete theories
revolves around the absence of gravitational fields in these theories.

It is well-known that each matter field (at least those occurring in the SM) couples to the gravitational field\footnote{Even the massless photon is affected by gravity - this causes gravitational lensing.}, and yet 
gravitational phenomena are 
described in a disjoint theoretical framework in contemporary physics known as general relativity (GR) or Einstein gravity.
Since GR does not allow for a straightforward quantization (the quantum field theoretic description is not perturbatively renormalizable) we lack a viable quantum description of Einstein gravity.\footnote{The nonrenormalizability of Einstein gravity has given physicists a strong incentive to examine modifications of Einstein gravity in the UV regime; the central research topic of quantum gravity. Einstein gravity would then be an EFT of a more fundamental (and renormalizable) theory of gravity.} 
This actually is an important reason why gravity cannot successfully be incorporated into the SM\footnote{It should also be kept in mind that the SM is formulated around a \emph{flat background} while gravitational phenomena should allow to be embedded into curved backgrounds as well.} (Biswas \et 2013, p.1). 
The absence of gravitational interactions in the SM is however unproblematic for most practical purposes. Gravity is by far the weakest force so gravitational interactions can safely be ignored for energies far below the (reduced) Planck scale $E \ll M_{P}$ (Duncan 2012, p.65).\footnote{I recall from page	\pageref{eq:M_P} that the Planck mass is defined as $M_P = \sqrt{\hbar c/ 8\pi G_{\text{N}}} = 2.4 \times 10^{18}$ GeV.}

The situation is fundamentally different
near the Planck scale. Quantum mechanical gravitational effects can no longer be neglected and should consequently be included in our theories of particle physics. This implies that a unified theory of the SM and gravitation is required at high energies. 
Stated differently,
each QFT which purportedly describes phenomena up to arbitrarily high energies should not be trusted above, at the highest, the Planck scale (Williams 2015, p.3).
\nn
These conceptual reasons
suffice to convey to the reader why the SM should not be conceived of as consisting of ``fundamental'' or ``complete'' field theories, but rather of \emph{effective field theories} (EFTs). An EFT provides an accurate, self-contained low-energy approximation to the more fundamental theory (which is usually unknown)\footnote{Sometimes the more fundamental theory \emph{is} known. For instance, we know that the GSW model of electroweak interactions is the more fundamental theory to Fermi theory, but in the context of the SM we do not currently know what its UV completion is.} up to an energy UV cutoff scale $(\Lambda)$, beyond which the EFT becomes inapplicable.\footnote{Unless otherwise noted, the exact value of $\Lambda$ is not of central importance. What really matters is that high energy DOFs are integrated out of the theory and only leave a small imprint on the bare quantities (Williams 2015, pp.6-7).} An EFT is a truncation of the more fundamental theory in which the high-energy DOFs are omitted. Let us consider the Wilsonian action of Fermi's theory of weak interactions (which was discussed in \S \ref{sectionguidingprinciples}) as an example. This Lagrangian can be derived from the Glashow-Salam-Weinberg (GSW) theory (see footnote \ref{footnote:GWS}). Fermi theory constitutes an accurate self-contained low-energy field theory up to $\Lambda \simeq 100$ GeV in which the $W$ and $Z$ bosons contribute no degrees of freedom
(see Dine 2007, \S4.0.1 for a proof). 

Although the EFTs of the SM  have to be superseded by more fundamental field theories in the UV, we should keep in mind that all of these field theories can have different cutoff scales. These $\Lambda$s can be anywhere in the huge desert between the TeV scale and the Planck mass\footnote{These cutoffs should however be as low as possible if naturalness is a good guiding principle, I will come back to this soon.} - lower and higher scales are excluded because the sub-TeV regime has already been probed by LHC and quantum gravity becomes important at the Planck mass.
\nn
Now that I have explained why EFTs are useful (and ubiquitous) in particle physics, my next goal is to elucidate how these field theories actually work.  
\subsection{The assumed decoupling of vastly separated energy scales}
When calculating properties of any EFT using perturbation
theory (assuming the coupling constants are way smaller than one), quantum fluctuations\footnote{Quantum fluctuations are intimately related to the quantum mechanical Heisenberg uncertainty relation: we need
higher energies to probe short distances.} give corrections to all the
parameters of the theory - for instance to the charge of an electron or to the Higgs mass. Perturbation theory describes the various quantities of a theory in a power series in their coupling constants. The calculation
involves summing over the effects of \emph{all virtual states} that are possible in the theory, including those at high energy (Donoghue 2007, p.4). 
The quantum correction refers to the terms in the series that depend on the coupling constants. The ``bare'' value
is the term independent of the coupling constants. The physical measured
value (denoted $q_{\text{phys}}$) is the sum of the bare value and the quantum corrections and can thus generically be decomposed into a \textit{bare term} and additional \textit{quantum fluctuations} or \emph{quantum corrections}:
\begin{equation}\label{eq:naturalness}
\underset{\text{(observable
		term)}}{q_{\text{phys}}} = \underset{\text{(bare term)}}{q_{0}} + \underset{\substack{\text{(quantum} \\ \text{fluctuations)}}}{\Delta q},
\end{equation}
The quantum fluctuations induce a correction to the bare term because the particle is no longer \emph{free} (\emph{i.e.} does not participate in interactions). 
The vacuum at
small distances produces particle/anti-particle
pairs so the vacuum may be thought of
as being filled with quantum fluctuations, where
``virtual particles'' (and virtual anti-particles) pop
in and out of existence on progressively faster timescales at progressively shorter
distances. Likewise, an electron is surrounded by virtual electrons and positrons which contribute to the charge of the electron (Arkani-Hamed 2012, p.54). 
\nn
New degrees of freedom (DOFs) and particle physics phenomenology may emerge at the energy scale where the EFT loses its predictive power.
A relevant question to ask at this point is \textbf{how this extraordinarily accurate agreement between predictions and data from particle colliders can be obtained, given that the high-energy (UV) physics has been neglected in the EFT}.
Future experiments may reveal that the gauge bosons from the SM are actually ``accompanied'' by supersymmetric particles as predicted by supersymmetry (SUSY). Wouldn't the myriad of ``exotic particles" contribute to the scattering amplitudes and cross sections of the established elementary particles? Wouldn't virtual supersymmetric particles contribute to loop corrections of SM particles and induce a mass shift of these particles? I will present arguments to support the claim that, since vastly different energy scales (typically) decouple\footnote{The exceptions are scalar field theories, these will be discussed in \S \ref{subsec:unnaturaltheory}.}, EFTs can yield excellent predictions as self-contained theories. 
\nn
Let us first examine \emph{free field theories} - quantum corrections will be included in the discussion later.
Masses of supersymmetric particles 
(and other exotic particles)
are indubitably higher than hitherto probed energies, otherwise 
these particles would have been discovered already.\footnote{A notable exceptions are weakly interacting massive particles (WIMPs)
and other weakly interacting particles whose masses may be low. These are candidates for dark matter. Since these interacting very weakly with experimentally observed matter they do not rigorously modify the QFTs of the SM.} 	
These high-energy degrees of freedom (DOFs) are not excited in (and can thus be integrated out of) the low-energy field theory. The philosophy behind this is that \emph{the high-energy degrees of freedom are distinct from the dominant ones at lower energies} (Williams 2015, p.4). 

An instructive example can be given for QCD. Strong interactions are mediated by gluons, which act on the ``color'' quantum number of quarks:
\begin{equation}\label{lagqcd}
\mathcal{L} = 
\underbrace{
	\bar{\psi}_i^{(f)} \left[ i \gamma_{\mu} D_{ij}^{\mu} - m \delta_{ij} \right] \psi_j^{(f)}
}
_{\text{quarks}}
- 
\underbrace{
	\frac{1}{4} F_{\mu \nu}^{a} 
	F^{\mu \nu}_{a}
}_{\text{gluons}}.
\end{equation} 
While this field theory (QCD) is formulated in terms of
quarks\footnote{Quarks cannot be observed as single particles in nature due to \emph{color confinement}. The only colorless composite particles which have been observed are $q\bar{q}$ (mesons) and $qqq$, $\bar{q}\bar{q}\bar{q}$ (hadrons).}  and gluons\footnote{The definitions of the covariant derivative $D_{ij}^{\mu}$ and the QCD field strength tensors $F_{\mu \nu}^{a}, F^{\mu \nu}_{a}$ and the flavors $(f)$ are unimportant here
	- what is important here is that the Lagrangian is not moulded in terms of composite particles DOFs.}
the low-energy EFT for strong interactions is not defined in terms of microconstituents of mesons and hadrons.
Instead, it is defined in terms of \emph{hadrons and mesons} and yields accurate results in the IR regime of equation \eqref{lagqcd}. This is completely analogous to how Fermi theory (equation \ref{eq:fermi}) presents an accurate low-energy description of the GSW model.

We can therefore conclude that successive energy scales are insulated from one another in these field theories, giving rise to quasi-autonomous energy regimes.\footnote{I call these regime ``quasi-autonomous'' rather than ``autonomous" because IR physics always retains a small sensitivity on UV physics through the renormalization group equations.} This \emph{decoupling of energy scales}
(or \emph{idea of insulation}) is incredibly important in particle physics, because 
the utility of
EFTs hinges on this assumption. 
A decoupling of energy scales facilitates an understanding of one ``energy shell'' without understanding the physics of the deeper (UV) shells (Nelson 1985, p.62).
\nn
Decouplings of widely separated energy scales are ubiquitous and well-known 
in classical physics.
Since IR phenomena typically remain robust under small variations of their UV constituents (chaotic phenomena are the exception, macroscopic trajectories are sensitively dependent on on small distance perturbations), a decoupling of widely separated energies is fairly common in theories of classical physics. Two examples suffice for the purposes of this thesis:
\begin{itemize}[nolistsep]
	\item {} The trajectory of a football is not sensitively dependent on the details of high-energy atomic physics.
	\item {} The orbit of the Earth around the sun is independent of where human beings are localized on Earth.
\end{itemize}
Violations of such decouplings of scales would have great ontological import.\footnote{In this case one would have to construct field theories which exhibit an UV/IR interplay. This kind of field theories will be discussed in \S \ref{subsec:UV/IR}.}
Were it necessary for
deriving the trajectory of the Moon's orbit to solve the equation of motion of each
individual electron and quark in the lunar interior, how could Newton have obtained
his gravity equation? This UV/IR interplay is usually absent in classical physics and its absence allows us understand each energy shell individually.

Let us now return to decoupling of scales in quantum field theories. We should, and can, do better than pointing at historical precedent and investigate whether the idea of insulation is generically valid in QFTs. An important aspect that should be taken into account is that Heisenberg's uncertainty principle is operative in the quantum realm. The uncertainty principle implies that \emph{high-energy DOFs contribute to the quantum corrections of IR parameters} (Donoghue 2007). 
The quantum fluctuation term $\Delta q$ sums over all virtual states up to the cutoff scale $\Lambda$. 
The bare term $q_0$ on the other hand sums up effects from \emph{energies beyond the cutoff scale} $\Lambda$, where \emph{hiterto unknown effects} take place.\footnote{This entails that the bare term is dependent on the cutoff scale - it changes when employing another value of $\Lambda$.}
This quantity can be regarded ``a black box'' since it can impossibly be inferred where the separate contributions to $q_0$ actually come from although the collective effect can be computed (Friederich 2017).

 Whenever 
the high energy DOFs provide small quantum corrections,
the ``UV physics decouples'' and
one can construct suitable low-energy EFTs which are largely independent of high-energy physics. Although this decoupling of scales is expected to hold true for our laws of nature, it is merely an \emph{assumption} which is largely empirically justified.
The great empirical success of the Standard Model underpins this central dogma of effective field theories. Indeed, the ``low-energy'' physics which is currently probed at particle colliders has to be largely independent of whatever new physics awaits us at higher energy scales (Williams 2015, p.26). 
\nn
Many physicists have surmised that this decoupling of energies should be exhibited by EFTs as well. This has turned into an important criterion - this is the \emph{naturalness criterion} - according to which the collective quantum corrections should be, at most, of the same order of magnitude as the bare parameter. 
Naturalness is based upon the surmise that EFTs should not depend sensitively on the specific details of the more fundamental theory which lives in the UV. 
This guiding principle has proven quite successful in the SM, where only 3 out of its 28 parameters are unnatural (Donoghue 2007).\footnote{The Higgs vacuum expectation value $\nu$, the strong charge-partity violating angle $\theta$ and the cosmological constant \lcc. These violations 
and their potential implications will be discussed extensively in this thesis.}
Besides empirical success of the SM, the most powerful justification of naturalness is provided by the Decoupling Theorem which implies that low-energy phenomena are described \emph{entirely by low-energy DOFs}. 
 \subsubsection{The Decoupling Theorem
\label{subsecdecoupling} 
}
According to many physicists (\eg Cao and Schwinger 1993, Castellani 2002, Bain 2013)
the principle of insulation receives theoretical support from the Decoupling Theorem (DT). This theorem has been derived by Appelquist and Carazzone (1975) and implies that widely separated physical scales are largely independent of one another, in other words, QFTs can be decomposed into ``quasi-autonomous domains.'' Since I will soon assert that naturalness hinges entirely on the DT, I will now extensively discuss the assumptions, features and implications of this DT, since these are of fundamental importance for our ensuing discussion of naturalness.
\nn
Hartmann describes the essential features of the Decoupling Theorem succinctly:
\begin{center}
	For two coupled systems with different energy scales $m_1$ and $m_2$ ($m_2 > m_1$) and described by a renormalizable theory, there is always a renormalization condition according to which the effects of the physics at scale $m_2$ can be effectively included in the theory with the smaller scale $m_1$ by changing the parameters of the corresponding theory. (Hartmann 2001, p.283)
\end{center}
What Appelquist and Carazzone have proven is that perturbatively renormalizable theories $\mathcal{T}$, which contain a set of field $\chi$ (allowed to be a single field) and fields with significantly lower masses (this set is called $\xi$), can be decoupled. What this means is that low-energy processes (where $E \ll m_\chi$) can be obtained from an EFT whose Lagrangian only contains the fields $\xi$, in other words, the heavier fields have been integrated out of this field theory (Williams 2015, p.8). The heavier fields do nonetheless leave an imprint on the low-energy Wilsonian action - the couplings of the fields $\xi$ are modified with respect to the respective coupling in the more fundamental theory $\mathcal{T}$
and the Green functions (correlation functions) are changed slightly, both of them typically logarithmically. These 
changes incorporate all contributions from high-energy physics, since the
heavy fields do
not contribute in any other way
to the correlators of the light fields $\xi$ in the EFT. Only the low-energy degrees of freedom are thus important for low-energy physics in field theories according to the DT. This is analogous to the familiar decoupling of physical scales occurring in classical physics.
\nn
Cao and Schweber argue that
\begin{center}
	with the decoupling theorem and the concept of EFT emerges a hierarchical picture of nature offered by QFT, one that explains why the description at one level is so stable and is not disturbed by whatever happens at higher energies, and thus justifies the use of such descriptions (Cao \& Schweber 1993, p.64)
\end{center}
As a consequence, nature can effectively be understood as being layered in quasi-autonomous domains, each of which contains their own set of physical laws. The Decoupling Theorem indeed licenses an ontology of QFT
which is characterized by such quasi-autonomous domains, \emph{provided that the conditions which are required for the proof of the DT hold true in generic EFTs}. For the time being, we will assume that the DT licenses a decoupling of widely separated energy scales and the conditions underlying the DT will be critically reviewed later (in \S \ref{sec:withoutnaturalness}).  I will elucidate in the next subsection (\S \ref{sectionnaturalprohibitionscales}) that this supposed decoupling of scales is nothing but the AoS naturalness principle.
\subsubsection{A pragmatic motivation for effective field theories}
This decoupling of vastly different energy scales
introduces a strong \emph{pragmatic reason} for treating QFTs as EFTs: low-energy physics is more suitably described in an EFT rather than in a more fundamental UV theory.
In principle one could describe low-energy phenomena using the (more) fundamental UV theory, but one would have to pay the price of less tractable mathematics, because DOFs for vastly different energy scales get mixed up (Batterman 2011).\footnote{Batterman discusses several difficulties of modelling systems over many length scales and a philosophical discussion thereof.}. 
Let us consider the QCD example:
the behavior of hadrons and mesons can be studied utilizing the UV Lagrangian (as introduced in equation \eqref{lagqcd}), however the involved degrees of freedom are distinct from those dominating in the IR regime (Williams 2015, p.4).
Since the low-energy effective theory has already been moulded in the pertinent DOFs (hadrons, mesons) this would yield a more informative qualitative description of the phenomena in question.\footnote{See Burgess (2003, \S2) for a philosophical discussion of the utility of EFTs in the context of gravity.}
Williams has discussed another insightful example:  
scientists think that it should be possible to study (low-energy) ocean wave propagation using the SM, notwithstanding, only deranged scientists would use this theoretical framework for that purpose - that would be a incredibly daunting task (Williams 2015, p.4).\footnote{Rational scientists may however be interested in this question for \emph{academic} reasons (``can we use the SM to describe phenomena in the deep IR?''). What I am claiming here is that nobody would use this theoretical framework for \emph{pragmatic} reasons in this context.} 
Both examples revolve around the idea that the UV degrees of freedom can be integrated out of a high energy theory to obtain a self-contained EFT for low-energetic phenomena, \ie a decoupling of widely separated energy scales.\footnote{Both examples highlight that the more fundamental theory may actually be \emph{a less informative description of the low-energetic phenomena which one aims to describe}. All degrees of freedom are retained in the fundamental field theory, obfuscating what the dominant contributions are, while high-energy degrees of freedom are typically irrelevant for low-energetic phenomena. 
	Since ``irrelevant'' degrees of freedom only provide minute corrections (and these are not ignored in the UV theory) the mathematics becomes significantly more tedious.}

\section{Naturalness as autonomy of scales
	\label{sectionnaturalprohibitionscales}
}
I have discussed in \S \ref{subsecdecoupling} that the DT allows us to integrate out the heavy fields from the more fundamental theory and therefore obtain our low-energy EFT whose coupling constants and Green functions typically receive logarithmic modifications (Wells 2015). An exception to this rule are scalar fields, which receive quadratic quantum corrections in $\Lambda$ and thus \emph{exhibit a strong sensitivity on high-energy physics}.
Two toy models (a scalar EFT and a fermion EFT) will be discussed in this section,
since these field theories already give the reader a rough characterization of what an ``unnatural theory'' amounts to.  I will subsequently state succinctly what is meant by naturalness as a prohibition of correlations among widely separated energies or equivalently  as an \emph{autonomy of scales requirement}. 
\subsection{How to recognize natural and unnatural theories?}
A Lagrangian describing both a scalar $\varphi$ and a fermion $\Psi$ (and all corresponding interactions that are allowed by Lorentz invariance and by the symmetries of the Lagrangian) can be constructed straightforwardly. We obtain
\begin{equation}\label{fullaction}
\mathcal{L}_{\varphi \Psi} = \frac{1}{2} \left[\partial_\mu \varphi \partial^{\mu} \varphi - m^2 \varphi^2 \right] - \frac{\lambda}{4!} \varphi^4 + i\bar{\Psi}\gamma^{\nu} \partial_{\nu} \Psi - M \bar{\Psi} \Psi + g \varphi \bar{\Psi}\Psi,
\end{equation}
where $m$ and $M$ are bare masses of the scalar and fermion fields, respectively, and $g$ denotes the Yukawa coupling constant. Let us now construct low-energy field theories for both the fermion and scalar and elucidate why the former is ``natural'' while the latter is ``unnatural''. 
\subsubsection{A natural theory: $\mathbf{m \gg M}$}
Let us assume that the massive scalar is much heavier than the fermion $(m \gg M)$, which allows us to construct an EFT for energies $E<M$. The fermion field will be integrated out of the theory in this low-energy regime by calculating all effects of the scalar field in the full theory (described by equation \eqref{fullaction}) and subsequenly integrating the scalar field out of the theory while absorbing the effects of the heavy scalar field in the fermion coupling terms. We can now write
\begin{equation}
\mathcal{L}_{\varphi} =  i\bar{\Psi}\gamma^{\nu} \partial_{\nu} \Psi - M^* \bar{\Psi} \Psi + \frac{g*}{2m^2} \varphi \bar{\Psi}\Psi,
\end{equation}
where $M^*$ is the bare mass of the fermion in the EFT
and $g*$ is the modified Yukawa coupling.\footnote{I remind you that these values change because the cutoff has changed. These terms do now take into account all effects occurring beyond $\Lambda$, including the effects of the heavy scalar field.} When including the leading contribution
to the vacuum fluctuations (the one-loop perturbative correction to $\Delta M$) we obtain the following effective fermion mass (Williams 2015, p.9)
\begin{equation}\label{key}
M^* = 
M
\left[
1 +  \frac{g}{16\pi^2} \ln \left( \frac{\Lambda}{M}
\right)
\right].
\end{equation}
We conclude that $\Delta M =  \frac{g M}{16\pi^2} \ln \left( \frac{\Lambda}{M}
\right)$ is proportional to the fermion's bare mass term (where $M \leq \Lambda$, by assumption). Since the bare fermion mass received quantum contributions of the same order of magnitude, the physical mass is not beyond the domain of applicability of the EFT.
The physical mass is not sensitively dependent on high-energy physics so we conclude that naturalness is respected by this EFT. 
\nn 
If we would have obtained huge quantum corrections (say, $\mathcal{O}(\Delta m) \simeq 10^{4}
\times \mathcal{O}(m_{\text{bare}})$) then the physical mass of the particle would most likely lie beyond the domain of applicability of the EFT - the effective field theoretic approach would then become inaccurate. We will now see that scalars suffer from this problem.
Scalars receive huge quantum corrections (because scalars are sensitively dependent on UV physics) and this is intimately related to the unnaturalness of scalars.
\subsubsection{An unnatural theory: $\mathbf{m \ll M}$
\label{subsec:unnaturaltheory}
}
Let us now construct the EFT for a light scalar particle satisfying $m \ll M$. We integrate out the UV fermion fields $\Psi$ in order to construct an EFT and the scalar mass $m$ and Yukawa coupling $g$ will be modified.\footnote{The following substitutions have been used; $m\rightarrow m^*$ and $g \rightarrow g'^*$.} The resulting Lagrangian is given by
\begin{equation}\label{fullaction2}
\mathcal{L}_{\Phi} = \frac{1}{2} \left[\partial_\mu \varphi \partial^{\mu} \varphi - m^2 \varphi^2 \right] - \frac{\lambda}{4!} \varphi^4 + g \varphi \bar{\Psi}\Psi.
\end{equation}
Williams (2015, p.10) has shown that
the effective scalar mass is now given by  
\begin{equation}\label{eq:unnaturalm*}
(m^*)^2 = m^2  + \frac{g}{16\pi^2} \left[ \Lambda^2 + M^2 + m^2 \ln \left(
\frac{\Lambda}{M}
\right)
+
\mathcal{O} \left(
\frac{M^4}{\Lambda^4}
\right)
\right].
\end{equation}
Let us note that this expression exhibits a \emph{quadratic sensitivity on the UV cutoff $\Lambda$}, in other words, the scalar particle receives an unduly large perturbative mass correction.\footnote{Moreover, it receives a large $M^2$ correction (keep in mind that $M > \Lambda$). The logarithmic correction in equation \eqref{eq:unnaturalm*} is not problematic because this introduces a relatively small mass shift.} We started with a small bare scalar mass $m$ (small with respect to $\Lambda$) and yet perturbative corrections drive the physical mass far beyond $\Lambda$. This is problematic because the EFT loses its predictive power at energies $E > \Lambda$.
 We aimed to integrate out all heavy fields 
with masses $M > \Lambda$
of the EFT, however, at the end of the day we retain fields 
in our field theory whose masses greatly exceed the cutoff scale $(M \sim \Lambda^2)$. The low-energetic field theory for scalar remains sensitively dependent on high energy physics because of its	 quantum corrections.
An EFT for scalars should thus not be supposed to produce accurate predictions whenever scalars couple to heavier fermions.
\subsection*{Implication: energy scales do not necessarily decouple}
Several physicists have taken it for granted that EFTs imply an autonomy of scales. Among others, Giudice contends that naturalness (when understood as the ``autonomy of scales'' notion which I advocate in my thesis)
\begin{center}
is deeply rooted in our
description of the physical world in terms of effective theories (Giudice 2013).
\end{center} 
One could, naively, argue that this statement is correct because the high-energy DOFs have been integrated out of the theory. There is however one caveat, pointed out by Williams (2018), with profound implications for naturalness: nothing in the EFT machinery puts any bounds on radiative
corrections to the bare quantities. In other words, no ingredients of EFTs prohibit quantum corrections many orders of magnitude larger than the UV cutoff scale. 
The only problematic cases are those cases where relevant operators are involved (Williams 2015) but these operators do occur in our laws of nature, among others describing the scalar Higgs mass.
We may thus conclude that
\begin{center}
[n]aturalness requires a more stringent autonomy of scales than we are
strictly licensed to expect [in generic EFTs]
(Williams 2018, p.24).
\end{center}
The SM contains one elementary scalar particle - the Higgs with mass $m \approx 125$ GeV. The SM also contains a single particle with even higher mass - the heaviest fermion in the SM is the top quark with $M \approx 172$ GeV. 
Since $m \ll M$, equations \eqref{fullaction2} and \eqref{eq:unnaturalm*} imply that the Higgs mass is sensitively dependent on UV physics. This so-called \emph{hierarchy problem of the Standard Model} 
had already been acknowledged soon after the introduction of naturalness into fundamental physics due to contributions by Wilson (1971), Gildener (1976), 't Hooft (1979) and Weinberg (1979).\footnote{The unnaturalness of the Higgs has led physicists to introduce ``Higgsless theories'' of which technicolor is a well-known example (see Weinberg 1976 and Susskind 1979).}
 Let us recall that this theory is considered unnatural because correlations between widely separated energy are not prohibited: the IR Higgs mass sensitively depends on what happens in the UV regime!\footnote{A plethora of physicists have surmised
that fundamental scalar particles cannot exist due to their quadratic sensitivity on the cutoff scale $\Lambda$. 
The Higgs mechanism was considered to be a \emph{convenient mathematical parametrization} and was not thought to entail the existence of a fundamental scalar particle. Iliopoulis' remarks at the Einstein Symposium of 1979 succinctly capture their collective sense of unease: ``Several people believe, and I share this view, that the Higgs scheme is a convenient parametrization of our ignorance concerning the dynmics of spontaneous symmetry breaking, and elementary scalar particles do not exist'' (Iliopoulos 1979, p. 89).
Other prominent physicists who discarded the possibility of a fundamental Higgs
 include Wilson (1971) and Callaway (1988). Universes without a Higgs  would be fundamentally different from ours, see Quigg (2007, \S 5).}
\nn
An important lesson can be learned from the previous examples. Although the decoupling of scales is deeply rooted in the \emph{logic of effective field theories}, this decoupling of scales is in fact \emph{not entailed} by such effective theories. That is related to the magnitude of quantum corrections which may be many orders of magnitude larger than the bare terms. Since the physical mass is quadratically dependent on the cutoff scales\footnote{It is often mentioned in the literature that scalar masses ``diverge quadratically'', but this is misleading. Since the problem occurs in EFTs (which have a finite cutoff scale) it is not justified to take the $\Lambda \rightarrow \infty$ limit. The scalar mass hence does not become infinitely large, \ie there is not truly a divergence.} the IR scalar mass is very sensitively dependent on UV physics. I will soon discuss that all so-called \emph{relevant} operators in QFTs retain a strong sensitivity on UV physics.
\subsection{Naturalness implies a prohibition of correlations among widely separated scales}
Physicists' early understanding of naturalness amounted to an ``autonomy of scales'' requirement: low-energy physics
should not depend ``sensitively'' on high-energy physics. 
The sensitivity of IR physics on UV physics is described by the renormalization group equations (RGEs), in which the sensitivity of parameters on the cutoff $\Lambda$ becomes apparent.
That the scalar EFT exhibits a strong sensitivity on the cutoff scale can 
easily be deduced from equation \eqref{eq:unnaturalm*}. 
The AoS notion of naturalness now dictates that the vacuum fluctuations should not be extraordinarily sensitive on the exact value of the cutoff $\Lambda$, in other words, they should not be sensitively dependent on UV physics.
This requirement is tantamount to a \emph{prohibition of correlations among widely separated energy scales}: this understanding of naturalness has been advocated in the physics literature for a long time (first introduced in the late 1970s), see Susskind (1979), Ovrut and Schnitzer (1980) and 't Hooft (1979). This understanding has been maintained by several physicists (\eg Georgi(1993), Giudice (2008), Burgess (2013)) but most physicists nowadays adhere to different definitions of naturalness (see \eg Arvanitaki \et (2014) and Athron and Miller (2007)).\footnote{I emphasize once more that I will put forward arguments as to why this early notion of naturalness is actually superior to other definitions which have been put forward in the literature.}
\nn
Several physicists have 
(implicitly or explicitly)
endorsed this notion of naturalness and aimed to \emph{quantify} this notion in order to enable scientists to assign an actual number to the \emph{degree of (un)naturalness} of models.
A myriad technical conditions, 
which supposedly capture the central dogmas of naturalness, have been put forward in the literature. This early understanding of naturalness
undergirds 
many of these naturalness measures,
although the more recent
landscape naturalness has completely shifted the meaning of naturalness. 
I will discuss several attempts to quantify naturalness and evaluate whether (i) these definitions are capable of capturing the essential dogma of naturalness and (ii) whether the assigned degree of naturalness would be meaningful. I will conclude, in similar vein as Craig (2014) and Williams (2016), that most of these superficially discordant formulations of naturalness are attempts to formalize the central dogma of naturalness which has been introduced in this section. I will argue that naturalness is best understood as an autonomy of scales (henceforth called AoS) requirement which was introduced in this section, beccause i) it provides a uniform notion which undergirds the plethora of naturalness conditions which will be introduced in the ensuing subsections (\S \ref{sectiontechnicalnaturalness}-\ref{sec:bayesian}), 
ii) 
it allows us to introduce compelling arguments why naturalness is a reasonable criterion to impose on EFTs (chapter \ref{chapterviabledefinition}),
and iii) the successes and violations of naturalness are best understood when adhering to this notion of naturalness (this will be discussed in \S \ref{sectionsuccessesnaturalness},\ref{sec:violations}). 
\nn
I will first discuss i) and assert that the following definitions obfuscate the central tenet of naturalness and are moreover  overambitious in attempting to formalize naturalness (\ie assigning an actual number to the degree of naturalness). Often, paradigmatically natural parameters are given a high number of unnaturalness. Absolute and technical naturalness are the exceptions - when parameters are absolutely or technically naturalness they guarantee AoS naturalness, but these criteria are only \emph{sufficient} to imply AoS naturalness.


\section{Absolute naturalness and technical naturalness
\label{sectiontechnicalnaturalness}
}
\subsection{Absolute naturalness}
What are the \emph{natural sizes of parameters} in quantum field theories? Dirac 
was among the first to 
(purportedly)
answer this question: Given an operator $\xi$ of the form
$\mathcal{L} \sim c_0 \xi$
in a theory with a fundamental scale $\Lambda$, then
according to Dirac
the natural size of the dimensionless\footnote{A dimensionless number is one that has no	units of measurement associated with it, so that its value is the same in any system
	of measurement.} coefficient $c_0$ in natural units
is\footnote{An implicit assumption is that we are working in $D = 4$ dimensions. 	
The general formula for $c_0$ in arbitrary $D$ dimensions is
(using dimensional analysis) given by
 $c_0 = \mathcal{O}(1)\times \Lambda^{4-D}$. (Craig 2017, p.2)}
$$\boxed{c_0 = \mathcal{O}(1)} .$$
This notion of naturalness is referred to
as 
\emph{absolute naturalness}
or
\emph{Dirac naturalness}
 in the literature (Craig 2017, p.2). 
Many physicists are puzzled by incredibly small or large dimensionless ratios, for instance Zee (2010, p.419) argues that physicists ``naturally expect that dimensionless ratios of parameters in our theories should be of order unity.. say anywhere from $10^{-2}$ or  $10^{-3}$ to $10^{2}$ or $10^{3}$.''
Two dimensionless numbers which are much closer together than their absolute values are also considered ``Dirac unnatural'', because the difference between these numbers would be a small number. Naturalness problems in physics often originate in such small
differences, as will later be exemplified 
by the small neutral kaon mass difference 
(naturalness problem)
in \S\ref{sectionsuccessesnaturalness}.
\nn
The precise range of values implied by 
$\mathcal{O}(1)$ remains a subjective matter in the literature, where one finds claims such as
```order unity' is
interpreted liberally between friends'', 
``factors of $\pi$ have been omitted'' (Zee 2010) and ``[i]n practice one factor of 5 or 10 is still not totally unacceptable'' ('t Hooft 1979, p.141).
I will allow dimensionless $\mathcal{O}(1)$ parameters, and their inverses of course, to be as low as $10^{-3}$. This subjectivity should not be taken lightly: depending on one's tolerance the inverse
fine-structure constant $\alpha^{-1} \simeq 137$ or even the
electron to proton mass ratio $m_e/M_P \simeq 2 \times 10^{-3}$ is deemed either natural or unnatural. Moreover, 
the Yukawa
couplings of the SM range from $10^{-6}$
for the electron to $\sim 1$ for the top quark - 
it is unclear
for many of these parameters whether they cry out for explanation due to the fuzzy definition of $\mathcal{O}(1)$ parameters.
\nn
Leaving this problem of vagueness aside, we may wonder why dimensionless parameters of order unity are special and why these would be related to naturalness. 
Several physicists have argued that there are no fundamental reasons to believe that $\mathcal{O}(1)$ parameters are special, other than being in great accordance with experimental data. Wells and Hossenfelder have taken this position, but disagree as to whether the concept bears fruit.
On the one hand, Wells has argued that
\begin{center}
The principle of [absolute] Naturalness cannot be derived from
first principles, and its invocation in science is more of a
product of intuition against the likelihood of large numbers
conspiring together to give small numbers than it is
on rigorous deduction. Nevertheless, the concept bears
fruit and is satisfied with respect to our QED example
here and other examples to be found in the literature.
(Wells 2013, p.7)
\end{center}
Wells believes that absolute naturalness is a fruitful concept despite its shaky foundation.\footnote{
	Wells strongly advocates the utility of absolute naturalness in Wells (2013). What Wells investigated was whether absolute naturalness as a
	guide to model building would have made particle physics advance 
	\emph{if researchers had firmly
		devoted themselves to the principle in the 1950s} (since absolute naturalness had not been introduced in the literature before the 1970s this is obviously an \emph{a posteriori} analysis). 
	Wells argues that it would have led toward more fundamental
	theories 
	(rather than leading
	theories astray) 
	and provides an illuminating example in the context of QED. I do not endorse his view that physicists should have firmly devoted themselves to absolute naturalness (many parameters in nature are Dirac unnatural - as will be discussed in the next subsection), however, what Wells' a posteriori analysis does successfully show is that the introduction of chirality into QED was in fact necessary from a naturalness point of view.}
On the other hand, Hossenfelder is an outspoken critic of naturalness (in the broadest sense). She contends that
scientists' belief that $c_n \neq \mathcal{O}(1)$ parameters are unnatural is dogmatic, because it
\begin{center}
is usually rationalized by claiming that numbers which are
very large or very small are unlikely. (Hossenfelder 2018a, p.3) 
\end{center}
and she argues that scientists have never come up with compelling reasons for this ``arbitrary" $\mathcal{O}(1)$ parameter desideratum.
I will now discuss that $\mathcal{O}(1)$ 
often imply small quantum corrections and
are therefore intimately related to the decoupling of widely separated energy scales entailed by AoS naturalness. However, \emph{there are indeed no compelling reasons as to why $\mathcal{O}(1)$ parameters generically imply small quantum corrections}. A myriad violations of absolute naturalness have however been discovered, while these parameters are AoS naturalness.
We will soon conclude that absolute naturalness is too restrictive a formalization of the AoS notion of naturalness\footnote{Absolute naturalness was soon replaced by 't Hooft's
\emph{technical naturalness} - a more permissive notion of naturalness. I will introduce technical naturalness in \S \ref{sucsec:technat}
 and discuss why this definition ameliorates absolute naturalness.} but first I will elucidate that $\mathcal{O}(1)$ may imply AoS natural parameters. My position is neither Wellsian not Hossenfelderian - although absolute naturalness can indeed not be proven by rigorous deduction the concept of absolute naturalness is sometimes useful (this is stronger than Hossenfelder's claim but also weaker than Wells' claim). As a consequence it is not a particularly useful concept, while AoS naturalness in fact is a useful concept. The physical motivation for $\mathcal{O}(1)$ parameters is way less transparent 
and less deeply rooted in the logic of EFTs
than that of a prohibition of widely separated energy scales. 
\subsubsection{Why $\mathbf{\mathcal{O}(1)}$ parameters may imply a prohibition of widely separated correlations}
The idea behind absolute naturalness is that
dimensionless couplings $\mathcal{O}$  whose magnitudes satisfy either $\mathcal{O} \gg 1$ or $\mathcal{O} \ll 1$ drag an operator away from
the energy scale where it naturally ``lives.''\footnote{Since every small
	number can be converted into a large number by taking its inverse, these two cases do not have to be distinguished (Hossenfelder 2018, p.2).}
This notion of naturalness can philosophically be backed up by the claim that \emph{dimensionful parameters determine the size of contributions of parameters of the theory}.
For any given physical process, one expects that one would be able to estimate (at tree-level) the
contribution of an operator in an EFT more or less entirely on the physical
scales which are involved in the problem (Williams 2015, p.6).
The ony scales which are involved in the problem are the energy $E$ (which permeates the internal propagators) and UV cutoff $\Lambda$ (beyond which the EFT loses its predictive power).
\nn
Generic EFTs for scalar fields are defined by the following Wilsonian effective action 
\bse
\begin{equation}\label{eq:S_W1}
\bal
S_W &= \int d^4x \ \mathcal{L}_\Lambda(\varphi_\Lambda, \partial_\mu \varphi_\Lambda, \cdots) \\
&= \int d^4x \left[
\frac{1}{2} \left(
\partial_\mu \varphi_\Lambda
\right)^2 + \sum_{n \geq 0} \left[ a_n \mathcal{O}_n + a_n' \mathcal{O}_n'
+ a_n'' \mathcal{O}_n'' + \cdots \right]
\right], \\
\eal
\end{equation}
and generally contain infinitely many terms, both renormalizable and nonrenormalizable.
Operators with more primes in equation \eqref{eq:S_W1} become ``relevant'' at progressively higher energy scales because these correspond to higher derivative terms. We can write equation \eqref{eq:S_W1} more succinctly:  
\begin{equation}
\bal
S_W = \int d^4x \left[
\frac{1}{2} \left(
\partial_\mu \varphi_\Lambda
\right)^2 + \sum_{n \geq 0} \left[ a_n \varphi_\Lambda^{2+n} + a_n' \left(\partial_\mu \varphi_\Lambda \right)^2 \varphi_\Lambda^n
+ a_n'' \left(\partial_\mu \varphi_\Lambda \right)^4 \varphi_\Lambda^{n-2} + \cdots \right]
\right],
\eal
\end{equation}
where ``$\cdots$'' denotes higher derivative terms. The spacetime dependence $(x)$ of the fields has been omitted and ($n = 0,2,4,\cdots$), moreover the fields $\varphi_\Lambda$ are defined as (Williams 2015, p.6)\footnote{The spacetime and frequency-momentum parameters $x$ and $k$ are displayed in cyan in equation \eqref{eq:cyan} to emphasize that these parameters are henceforth omitted.}
\begin{equation}\label{eq:cyan}
\varphi_\Lambda^{n}
{ \color{cyan}
(x)} = \int_{\abs{k}<\Lambda} {d^4k \over (2\pi)^4} \tilde{\varphi}_\Lambda^{n} {\color{cyan} (k)} e^{-ik \cdot x}
\end{equation}
\ese
and exclude field modes of momenta $\abs{k} \geq \Lambda$.
The $a_0(\varphi_\Lambda)^{2}$ and $a_2 (\varphi_\Lambda)^{4}$ are the familiar mass term and dimensionless quartic coupling $(\lambda)$, the infinitude of additional interactions in equation \eqref{eq:S_W1} are described by operators with mass dimension greater than 4 (these are phenomenologically speaking irrelevant in the infrared regime). 
\subsubsection*{Tree level contributions}
Dimensional arguments 
allow us to predict the
contribution of a particular operator at
the energy scale $E$ by counting the dimension of operators. Keeping in mind that $[\mathcal{L}] = E^4$ and that $a_n$ and $a_n'$ also have mass dimensions, we deduce that $a_n$ has mass dimension $(n-2)$ while $a_n'$ has dimension $(-n)$.
It is convenient to rescale the couplings $(a_n, a_n')$ in terms of dimensionless
couplings by extracting the appropriate powers of the cutoff scale:
\begin{equation}\label{eq:couplingsa_n}
\bal
a_n &= c_n \Lambda^{2-n}, \ \ \
a_n'&= c_n \Lambda^{-n}
\eal
\end{equation}
We expect the following contributions for the lowest dimensional operators  in EFTs   
(Duncan 2012, \S16.3):
\begin{equation}\label{eq:couplingsa_n2}
\bal
 \mathcal{O}_n &= \varphi_\Lambda^{2+n} &&\sim  E^{n-2} \\
  \mathcal{O}_n' &=  \left(
  \partial_\nu \varphi_\Lambda
  \right)^2
  \varphi_\Lambda^{n}
  &&\sim
E^{n}\\
\eal
,
\end{equation} 
and straightforwardly obtain (by combining equations \eqref{eq:couplingsa_n} and \eqref{eq:couplingsa_n2})
\begin{equation}\label{eq:O_n}
\bal
a_n \mathcal{O}_n &= c_n \left( {E \over \Lambda} \right)^{n-2} \ \ \ \\
a_n' \mathcal{O}_n' &= c_n' \left( {E \over \Lambda} \right)^{n}
\eal
.
\end{equation}
\begin{figure}
	\centering
	\includegraphics[scale=0.4]{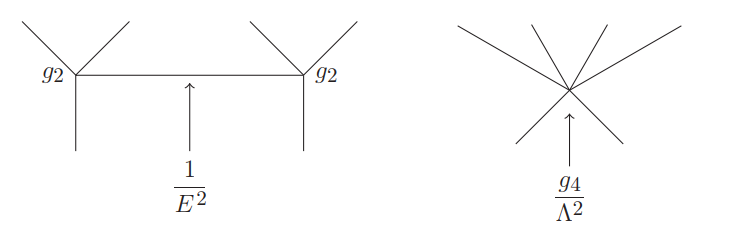}
	\caption{\textit{Two tree graph contributions to a scalar scattering amplitude, this diagram is used in my argumentation in which I convey why dimensionless parameters are expected to be of order unity. The black lines denote scalar fields, and interactions take place at the $c_2 (=g_2)$ and $c_4 (=g_4)$ vertices.}}
	\label{fig:scattering}
\end{figure}
Interestingly, the \emph{scaling behavior} of different operators (relevant, marginal and irrelevant)\footnote{
Relevant, marginal and irrelevant operators have mass dimension $M<4, M=4, M>4$, respectively.	
	The reader should guard against attaching the colloquial meaning of
	terms such as ``irrelevant'' to the physics generated by the corresponding operators:
	the dimension six four-fermion operator of Fermi weak interaction theory is irrelevant (although these processes become dominant in the deep UV), but the associated
	vast phenomenology of beta-decay and radioactivity is, of course, not irrelevant.} can now be deduced. 
Both a relevant $(g_2)$ and marginal $(g_4)$ operator are displayed in  the \emph{tree diagrams} of Fig. \ref{fig:scattering}, where two
contributions to the 2-4 (two incoming and four outgoing particles) scattering amplitude are shown.
\begin{itemize}[nolistsep]
	\item {}
The first graph arises from two quartic
coupling terms $\mathcal{O}_2 = c_2\varphi^4$ (and is 	of order $c_2^2/E^2$, where the incoming and outgoing
momenta are of order $E$). 
\item{} The second graph (which arises from the higher-dimension
term $\mathcal{O}_4 = {c_4 \over 
\Lambda^2} \varphi^6$) is very small, of order $c_4 E^2/\Lambda^2$ \emph{relative to the contribution from the first diagram}. 
\end{itemize}
Experiments have shown that both $c_2$ and $c_4$ are of order unity (Duncan 2012, \S 16.3).
\nn
Absolute naturalness imposes the restriction that \emph{contributions of operators to physical processes are determined, more or less, entirely by the physical scales which are involved in the problem}, namely: $E$ and $\Lambda$ (Williams 2015, p.17).
The contributions of marginal and irrelevant operators are suppressed (rather than enhanced) by positive powers of $\Lambda$ - these contributions thus cannot exhibit a strong sensitivity on high-energy physics! Relevant operators, on the other hand, are enhanced by positive powers of $\Lambda$.
These dimensional arguments should work at arbitrary energy scales below the cutoff ($E<\Lambda$) and imply that $c_n \approx 1$ (Williams 2015, p.17). 
Dimensionless parameters of order unity are compatible with the scaling behavior of different kinds of operators (marginal and irrelevant).
These dimensional arguments are ``ubiquitous, and almost always successful, in effective field theories'' as was already put forward by Williams (2015, pp.17-18)
and Duncan (2012, \S16.3).
\subsubsection*{Contributions from loop diagrams}
\begin{figure}[h!]
	\centering
\includegraphics[scale=0.3]{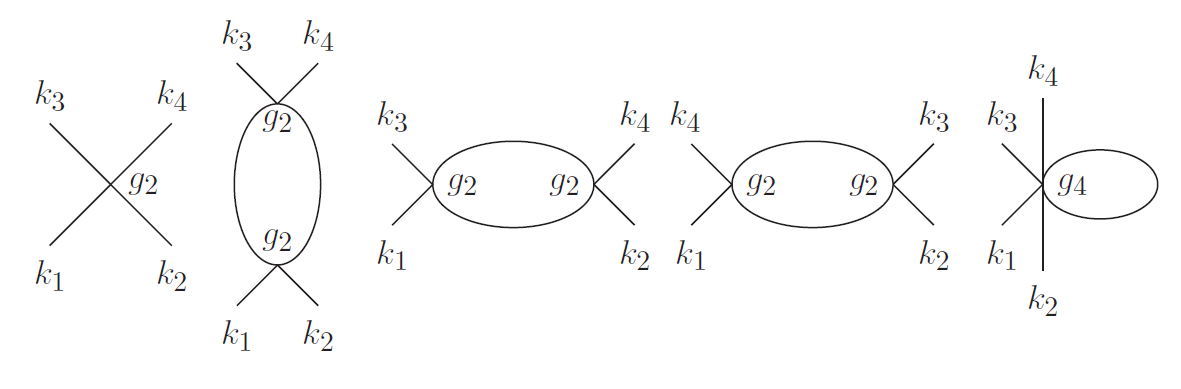}
\caption{\ti{Some tree and leading one-loop contributions to the 2-2 scalar scattering amplitude. The parameters $k_i (i \in {1,2,3,4})$ denote the fourmomenta of pareticles and $g_j (j\in\{2,4\})$ are coupling constants.} This figure has been taken from Duncan (2012), Fig. 16.2.}
\label{fig:scat2}
\end{figure}

We should note that loop effects (quantum corrections) have not been taken into account yet, let us now go beyond the tree level.
When loop effects are included the situation becomes more complicated, and much more interesting. Following Duncan (2012, \S16.3), I take 2-2 scattering as a test case and show that \emph{order unity dimensionless parameters imply quantum corrections produce order unity modifications}. In other words, $\mathcal{O}(1)$ parameters imply that quantum corrections remain small and that the physical parameters remain natural. 
\nn
The leading 2-2 scattering processes are displayed in figure \ref{fig:scat2}.
\noindent
The
lowest-order tree
graph (at the left) corresponds to the dimensionless quartic coupling $c_2\varphi^4$, the three one-loop graphs arise at second
order in $c_2$ and the one-loop graph (at the right) comes from the first-order contribution of the
dimension 6 irrelevant operator $c_4
 \varphi^6/ \Lambda^2$.\footnote{I will drop the ``$\Lambda$'' subscript on
	the fields here with the reminder that it simply instructs us to cut off the momenta
	on all internal propagators at $\abs{k} = \Lambda$.}

The last graph contains the following one-loop cutoff integral
\begin{equation}\label{eq:Theta}
\int \theta(\Lambda^2 - k^2) {1 \over k^2 + m^2}  {d^4k \over (2\pi)^4}
= {1 \over 16\pi^2 } \left[
 \Lambda^2 - m^2 \ln \left(
\frac{\Lambda^2}{m^2}
\right)
\right]
+ \mathcal{O} \left(
m^2 \over \Lambda^2
\right)
\end{equation}
and cancels
the inverse factor of $\Lambda^2$ (in the coupling $c_2
/\lambda^2$). The $\varphi^6$ operator is no longer irrelevant at low energies, the result is now momentum-independent constant contribution to the amplitude, of exactly the same
form as marginal operators at tree level.
\nn
The truncated four-point function arising
from the four graphs involving loop diagrams and the irrelevant $\varphi^6$ operators is given by (Duncan 2012, p. 577)
\begin{equation}\label{eq:Gamma4}
\Gamma^{(4)}(k_1,k_2,k_3,k_4) = g_2 - {3 \over 4\pi^2} g_2^2 \left[
\mathcal{I}(s,m^2, \Lambda^2) + 
\mathcal{I}(t,m^2, \Lambda^2) +
\mathcal{I}(u,m^2, \Lambda^2)
\right]
+ {15 \over 16 \pi^2 } g_4 + \mathcal{O} \left(
{m^2, k_i^2 \over \Lambda^2}
\right),
\end{equation}
where the Mandelstam variable $s,t,u$ and the integral $\mathcal{I}(p^2,m^2, \Lambda^2)$ are given in this footnote.\footnote{
	The integral is given by
$$ \mathcal{I}(p^2,m^2, \Lambda^2) = \int_0^1 
\left[
\ln \left(
\Lambda^2 \over x(1-x)p^2 +m^2 - 1
\right) 
- 1
\right]
dx$$
and the Mandelstam variables are
$$s = (k_1 + k_2)^2, \ \ \ t= (k_1 - k_3)^2, \ \ \ u = (k_1 - k_4)^2.$$
}
The important feature of equation \eqref{eq:Gamma4} is that it entails that, \emph{assuming
that the dimensionless couplings $g_2$, $g_4$ are of order unity}, the momentum modes
of the field between $E$ and $\Lambda$, when integrated out in the path integral, produce order
unity modifications in the effective four-point
coupling strength at the low-energy scale (Duncan 2012, \S 16.3).
The reason is simple - explicit
inverse powers of the cutoff in the coupling factors can be cancelled by positive
powers of the cutoff 
(this also happened in equation \eqref{eq:Theta})
arising from loop-integrals containing vertices corresponding to
these higher-dimension operators.
Although it cannot be proved that $\mathcal{O}(1)$ generically imply small quantum corrections, they typically do (Williams 2015, p.17). 
\begin{center}
Essentially the only instances in which they [dimensional arguments facilitated by $\mathcal{O}(1)$ parameters] break down are when relevant operators are involved. (Williams 2015, p.18)
\end{center}
The ``filtering down'' effect from higher- to lower-dimension operators seems
fairly innocuous for the marginal couplings such as $g_2$, but it implies much more
dramatic consequences for each coefficient of a relevant operator.
The contributions of relevant operators 
$\mathcal{O}_0 = \varphi^2$
are proportional to the ratio $\left(
{\Lambda \over E}
\right)$ to some positive power (Williams 2015, p.6).
The scalar Higgs mass has mass dimension two and is thus
expected to be of order $\Lambda^2$.\footnote{The mass operator grows quadratically as we lower
	the energy. This is hardly surprising if we consider the mass expansion of the free (Euclidean) propagator
	$${1 \over 
	k^2 + m^2}
	= {1 \over 
	k^2}
	- {m^2 \over 
	k^4} +
	{m^4 \over 
	k^6} - \cdots, $$
	where we see that increasing powers of the mass correspond to larger and larger
	contributions in the IR region $k \ll m$ (Duncan 2012, \S16.3).
This implies that its contribution grows for processes occurring at lower and lower energies.} 
This is problematic, because the EFT retains its predictive power only up to energy scales of $\Lambda$. We are led to conclude that a \emph{prima facie} naturalness problem emerges for relevant operators. This is intimitely related to the unnatural of the Higgs mass whose physical mass, despite large quantum corrections, is incredibly low.

Its dimensionless coupling constant should therefore be significantly lower than order unity. The Higgs mass is determined by a relevant operator (of mass dimension 2, namely $\varphi_\Lambda^2$) and the corresponding coupling constant is expected to be $a_0 = c_0 \Lambda^2$ on purely dimensional grounds. The physical mass of the scalar Higgs field is however unsettingly low compared to $\Lambda^2$, in fact one would require a tiny dimensionless coupling constant of $c_0 = m_\varphi^2/\Lambda^2 = 10^{-34}$ 
at the Planck scale, assuming that the SM remains valid all the way up to the Planck scale. The reason behind this tiny coupling constant ultimately resides in the quadratic sensitivity of the Higgs on UV physics. 
\nn
 I will now discuss why Dirac naturalness does not capture the central dogma of AoS naturalness because it is too restrictive.
\subsubsection{Failures of absolute naturalness}
The SM is very unnatural when adhering to Dirac naturalness, while it is not according to AoS naturalness. 
The broad range of dimensionless Yukawa couplings ($10^{-6}$ up to $1$) would already
suggest that the SM is incomplete.
New dynamics or symmetries should be identified which
explain the fermion masses in a deeper theory respecting
Dirac naturalness.
Moreover, masses of elementary particles (for instance the 
electron mass $m_e \sim 0.511$ MeV and the
proton mass $m_p \sim 1$ GeV) would be unnaturally low compared to the Planck mass $M_{P} \sim 10^{18}$ GeV and other fundamental scales such as the Fermi scale. Both the proton and electron mass would be natural according to AoS naturalness if the physics at low energies ($E \lesssim 1$ GeV and $E \lesssim 0.5$ MeV, respectively) would be quite insensitive to physics at higher energies, rendering $m_p \sim 1$ GeV a natural scale for the proton. The dimensionless parameter satisfies $m_p/M_P = 10^{-18} \ll \mathcal{O}(1)$ and 
both the electron and proton masses
would be highly unnatural according to Dirac, while they do not exhibit a strong sensitivity on UV physics. In other words, absolute naturalness is too restrictive.
\nn
This raises the question:
why is the natural electron mass allowed to be incredibly small compared to the Planck scale 
and with respect to Fermi scale 
$M_F = 555$ GeV
of weak interactions
(since one may object quantum gravity is not well-understood I will focus on the latter problem). The dimensionless parameter $m_e/M_F \simeq 10^{-6}$ is problematically low value for adherents of absolute naturalness.
An important realization 
was provided by 't Hooft: the electron mass is a lot closer to zero than to the Fermi scale! (Wells 2014) 
Many other parameters of the SM turn out to be a lot closer to zero than to the dimensionful scale of the theory too and
't Hooft (1979) argued that such parameters could potentially be given a natural explanation in terms of a symmetry argument. After all, symmetries can make parameters vanish (such as the photon mass $m_\gamma$) 
and therefore explain why $m_\gamma/M_{P} = 0$ is perfectly natural. Could \emph{near symmetries} perhaps account for \emph{low} values of parameters?
\nn
I will now introduce technical naturalness, explain why this constitutes an improved definition of naturalness over absolute naturalness, 
and then assert
that small dimensionless ratios 
(like $m_e/M_F$ and $m_p/M_P$) 
are  - contrary to what Dirac surmised - not necessarily unnatural.
\subsection{Technical naturalness \label{sucsec:technat}}
If certain dimensionless parameters come out to be small at an energy $\mu_0$ then this cannot be accidental according to technical naturalness. These low values instead indicate the existence of a \emph{near symmetry}
(the symmetry of the Lagrangian would be enhanced when setting a term in the Lagrangian equal to zero). A near symmetry could be responsible for  - otherwise unnatural - low values of dimensionful parameters in the theory.  
This led to a refined definition of Dirac naturalness called \emph{technical naturalness} or \emph{'t Hooft naturalness} (named after its proposer):
\begin{quote} 
	\centering 
	The naturalness criterion states that one such [dimensionless and
	measured in units of the cut-off] parameter is allowed to be much
	smaller than unity only if setting it to zero increases the symmetry of the theory. If this does not happen, the theory is unnatural. 
	('t Hooft 1979, p.135)
\end{quote}
Dimensionless parameters can be much smaller than their Dirac natural value, according to 't Hooft, whenever there is an
enhanced symmetry of the theory when the coefficient is set to zero. The natural size of the coefficient $c_0$ would then be
\begin{equation}\label{key}
\boxed{c_0 = \mathcal{S} \times \mathcal{O}(1)} ,
\end{equation}
where $\mathcal{S}$ is the parameter which violates a symmetry of the system, typically satisfying $\mathcal{S} \ll 1$.
Radiative corrections must then be
proportional to the symmetry violation (Craig 2013).
Technical naturalness is based upon the fact that symmetries endowe Lagrangians with constraints which prohibit
the occurrence of
large quantum corrections. 
Imposing the requirement that theories with small parameters $c_0 \ll 1$ need to
possess technical naturalness is of course less philosophically taxing
than demanding that nature is described entirely by absolutely natural parameters (Wells 2013, p.2).
\subsubsection{Why technical naturalness reproduces the central dogma of autonomy of scales naturalness} 
The origin of the different quantum behavior
for massive and massless fermions
lies in the enhanced symmetry when setting the mass parameter to zero. 
Massive spinors and vectors are ``protected'' by chiral and gauge symmetry (respectively) both of which are \emph{near symmetries}. One can show that perturbative corrections to bare mass parameters of ``protected'' particles are proportional to the bare parameters, implying that (Williams 2015, p.16)
$$
\mathcal{O}(\Delta m)
\leq
\mathcal{O}(m_{\text{bare}}) .$$
Because quantum corrections are of the same order of magnitude as the bare term, the technical naturalness criterion reproduces AoS naturalness. In the presence of a near symmetry, the physical observable in equation \eqref{eq:naturalness} does not depend sensitively on $\Lambda$ and would thus be of the same order of magnitude when the UV cutoff energy changes radically by $\Lambda \rightarrow \Lambda'  = \alpha \Lambda \  (\alpha \in \mathrm{R})$ where $\alpha$ may be large (say $\alpha \gg 1$). This guarantees a prohibition of correlations among widely separated energies - UV physics plays a minute role for the magnitude of the physical observable $m_{\text{phys}}$.  Technical naturalness
explains, 
by means of a symmetry argument,
how hierarchies observed in the IR can be protected against large radiative corrections.
\nn
I will now discuss that many parameters of the SM are technically natural and consequently argue that technical naturalness is an improved formalization of naturalness  in comparison to absolute naturalness.
\subsubsection{Successes of technical naturalness}
Let us now put technical naturalness to the test - technical naturalness is capable of explaining the naturally low electron mass. The absence of strong interscale sensitivities for the electron can be successfully explained by the near chiral symmetry. This near symmetry prevents high quantum corrections to the bare mass term from occurring. The electron mass would be highly unnatural according to Dirac naturalness because the electron mass is significantly smaller than the Planck and Fermi scales.
\nn
The masses of the pions, proton and neutron are well-understood in the context of technical naturalness. We know from asymptotic
freedom of QCD (see \S \ref{sectioneft}) that the
perturbative gauge coupling in the UV flows to
strong value at the low scale and confinement happens
at $\Lambda_{\text{QCD}} \sim 1$ GeV. This gives the characteristic scale of
the hadrons in the theory; the proton and neutron
obtain a mass which is approximately equal to this scale (Wells 2013, p.2). 

The smallness of the proton mass compared to the Planck mass can also be understood as technically natural. 	The proton consists of three quarks which are described by QCD. 
The action of QCD is enhanced with a \emph{conformal symmetry} (scale invariance) in the massless quark limit (Dine 2015, p.4) so technical naturalness would entail that $m_p \ll M_P$. That is indeed the case and is reflected by the QCD renormalization group equations (the RGEs; I recall that ``renormalization'' is the statement that the parameters
of a theory vary with the UV cutoff scale $\Lambda$) for the strong coupling $\alpha_s$, specifically
\begin{equation}\label{eq:protonRGE}
{d\alpha_s \over d\log\left(
M_P/E
\right) }
= -2 b_0 \alpha_s^2
\end{equation}
with $b_0$ a real-valued constant. 
One may now ask at what scale
$E \equiv \Lambda$ the coupling becomes of order unity and straightforwardly calculate:
\begin{equation}\label{key}
\Lambda =
m_p
\exp \left[
{
	-2
	\pi
	\over
	b_0
	\alpha_s\left( M_p \right) }
\right].
\end{equation}
The constant $b_0$ is  roughly  7 for  QCD,  so  if
$\alpha_s(M_P)$
is  approximately  0.5,  the  exponential  is
incredibly small. The scale $\Lambda$ is consequently of the order of the proton mass, the proton mass is hence not sensitively dependent on UV physics.
The logarithmic variation of the coupling constant $\alpha_s$ implies that the proton necessarily acquires a mass $m_p = \mathcal{O}$(GeV) even if the UV cutoff is taken around the Planck mass (see Dine 2015, p.4 for computational details).
\subsubsection*{The problem of scalar particles revisited}
Nearly all elementary particles have technically natural masses
within the SM\footnote{We have seen that this indeed holds true for many elementary particles, the proton mass for instance is low compared to the Planck mass due to a spontaneous breakdown of chiral symmetry ('t Hooft 1979, p.138).}, yet the (scalar) Higgs boson violates technical naturalness severely (Wells 2015).
Scalars have 
- in stark contrast with spinors and vectors -
no enhanced symmetry in the massless limit, indicating that massive and massless scalars exhibit identical quantum properties (Williams 2015). 
Due to the absence of a near symmetry, the relevant operator receives perturbative corrections greatly exceeding the order of magnitude of the bare mass:
$$ \mathcal{O}(\Delta m_\varphi) \gg \mathcal{O}(m_{\varphi, \text{bare}}). $$
This poses a tremendous problem for fundamental scalar particles in the SM, which are allowed to receive huge quantum corrections. 
As was discussed in the previous section, scalar particles receive huge quantum corrections because they are quadratically sensitive on UV physics. Because $\Delta m_\varphi \propto \Lambda^2$  rather than $\Delta m_\varphi \propto m_{\varphi, \text{bare}}^2$ (as would have been the case for spin-1/2 and spin-1 particles) we conclude that scalar particles are the most likely candidates for unnatural parameters.
\subsubsection{Violations of technical naturalness \label{subsec:violtech}}
The following three parameters violate technical naturalness. Physicists aim to find solutions to these problem in the context of models with additional symmetries. If this near symmetry from a more fundamental theory is broken in the IR, the low values of the following parameters can be understood and would be natural if $\mathcal{S} \ll 1$.
\subsubsection*{Gravity}
An important shortcoming of technical naturalness has to be addressed: it cannot explain the unnatural value of the cosmological constant $\Lambda_{CC}$ (to be discussed in more detail in section \ref{sectioncosmologicalconstant}). Setting $\Lambda_{\text{CC}} = 0$ does not increase the symmetry of the Lagrangian (Dine 2015, p.20) so gravitational interactions do not respect technical naturalness.
Technical naturalness cannot explain why the ratio of the gravitational to weak force is unnaturally small, \emph{i.e.} (Yao \emph{et al.}, 2006)
\begin{equation}\label{eq:ratiogravweak}
\frac{G_{\text{N}} c^2}{G_F \hbar^2} = 5.7517(82) \times 10^{-32}.
\end{equation}
Since quantum gravity is not yet well-understood, one may object that this problem is irrelevant (this naturalness problem may be solved by a viable theory of quantum gravity). 
't Hooft optimistically concludes that
\begin{center}
[w]e have nothing to say about this fundamental problem, accept [except] to suggest that \emph{only} gravitational effects violate naturalness. ('t Hooft 1979, p.137)
\end{center}
It is however false that gravitational physics is the sole violation of technical naturalness, two other violations of technical naturalness are known to occur in nature.
\subsubsection*{The Higgs mass}
As has already been discused, the Higgs mass it not protected by a near symmetry. 
Technical and absolute naturalness thus coincide in this case and the technically natural physical value would consequently be $m_{\text{H}} \sim c_0 \Lambda^2$, where $c_0 = \mathcal{O}(1)$.
The current LHC restrictions on impose $c_0 < 10^{-4}$ and could be as small as $c_0 \simeq 10^{-32}$ if the electroweak theory remains accurate all the way up to the Planck mass.\footnote{'t Hooft conjectured 
	(in 1979, long before the experimental discovery of the Higgs boson)
	that technical naturalness ``...is the reason why light, weakly interacting scalar particles are not seen'' ('t Hooft 1979, p.136).}
\subsubsection*{The strong CP problem}
The SM poses a puzzle at the level of 
a technically unnatural  marginal (mass dimension four) operator. The electroweak theory is known to break a symmetry called ``charge conjugation parity symmetry'' (henceforth called CP), which is a discrete symmetry (rather than Lie symmetry). Although the strong interactions also permit a CP-violating term, no occurrences of CP violations have been confirmed in the QCD sector (Dine 2007, p.71).
What this implies is that the CP-breaking term (the theta parameter)
should either vanish or be astonishingly small ($\theta \ll 1$), while it can range from $0$ tot $2\pi$ in principle (Donoghue 2007, pp. 5-6). 
\nn
The theta parameter for strong interactions occurs in the following Lagrangian term
\bse
\begin{equation}\label{eq:FQCD}
\mathcal{L}_{\text{QCD,}\theta} = {\theta  \over 16\pi^2 } F_{\mu \nu}^{a}  \tilde{F}^{\mu \nu a}, 
\end{equation}
where  $F_{\mu \nu}^{a}$ is the QCD field strength and its dual is
\begin{equation}\label{key}
\tilde{F}^{\mu \nu a} = {1 \over 2} \epsilon_{\mu \nu \rho \sigma} F^{\mu \nu a},
\end{equation}
\ese
(where $\epsilon_{\mu \nu \rho \sigma}$ is the totally antisymmetric Levi-Civita tensor).
Since this Lagrangian can be written as a total divergence (Dine 2000, p.2), one might expect that it is irrelevant for physics (analogous to how an additional Gauss-Bonnet term in the Lagrangian does not modify the gravitational field equations (Lovelock 1971)).
The equations of motion for QCD are left unmodified as well, but a non-zero theta parameter would in fact have physical consequences.
Despite being a total derivative, $\mathcal{L}_{\text{QCD,}\theta}$ violates CP symmetry. One can show this term consequently contributes to the electric dipole moment (EDM) of the neutron as a function of $\theta$:
$$ d_n = 5.2 \times 10^{-16} \times \theta \ \text{cm}.$$
Our current experimental data concerning the neutron EDM ($d_n < 3 \times 10^{-26} \text{e cm}$) thus introduces a strong limit on theta, namely $\theta \leq 10^{-10}$ (Crewther \et 1979). 
Had nature respected
CP-symmetry in the absence of theta $(\theta \rightarrow 0)$, the small value $\theta \ll \mathcal{O}(1)$ would have been technical natural.
But nature violates CP. Indeed, the phase appearing in the CKM
matrix is of order one, implying that CP is violated by weak interactions.
Technical naturalness thus provides an insufficient criterion to explain the unnaturalness of the QCD theta parameter.
\nn
The strong CP problem has motivated
the theory of axions, in which an extra symmetry removes the strong CP
violation, but requires a very light pseudoscalar boson - the axion - which
has not yet been found (Donoghue 2007, p.7).\footnote{Hossenfelder (2018a) has argued that this is another failure of naturalness. This is not true. Axions would interact very weakly with other matter and would therefore be incredibly hard to detect, to my knowledge we cannot rule out the possibility that these axions constitute (a part of) the dark matter content of the universe.}
Arguably the most popular explanation of the smallness of the $\theta$ parameter involves a hypothetical \emph{axion} particle.\footnote{Another solution is \emph{spontaneous CP violation}, where CP is broken by the expectation of a complex field $\Phi$ in (omitting the indices of the QCD field strength, these can be found in equation \ref{eq:FQCD})
	$$\frac{1}{16\pi^2} \Phi F \bar{F}.$$
This possibility is discussed more elaborately in Dine (2007, \S5.5.2) and Vecchi (2014).
} The original idea has been put forward by Peccei and Quinn (1979) and is referred to as the PQ-solution. I will come back to their idea in chapter \ref{sec:withoutnaturalness}, where I will argue that their idea solves this problem provided that one is willing to buy the ``price'' of the multiverse. This is ensued by other examples where the multiverse approach may prove more fruitful than conventional natural solutions.
\subsubsection{Technical naturalness is too restrictive}
Although technical naturalness is an improvement over absolute naturalness,  I will assert that it is still too restrictive to capture the AoS dogma. 
The latter claim will be backed up by elucidating the inability 
of technical naturalness
to account for the naturalness of the proton mass. Small dimensionless parameters can indeed be AoS natural when protected by a symmetry, however, this is merely a \emph{sufficient criterion}. The QCD scale remains low with respect to $M_P$ due to \emph{dimensional transmutation} and technical naturalness cannot account for \emph{dynamical solutions} for unnatural parameters - it only accounts for symmetry solutions.
\subsubsection*{The QCD scale}
One can come up with compelling reasons as to why technical naturalness would be an impoverished notion of naturalness in comparison AoS naturalness. Although the latter seems less well-defined, it more accurately captures the central features of naturalness as will now be argued. An important objection against technical naturalness is that the near symmetry requirement is \emph{merely a sufficient} (rather than \emph{necessary}) criterion for natural parameters. This will be exemplified in the context of quantum chromodynamics, where the smallness of a parameter has nothing essential to do with a symmetry principle . 
\nn
The QCD scale ($\Lambda_{\text{QCD}}$) is significantly lower than the Planck scale  whereas it is not protected by any symmetries. We would conclude that the QCD scale is technically unnatural. 
The low value of the QCD scale can however successfully
be explained by \emph{dimensional transmutation}, which is a generic feature of QCD dynamics (Banks 2008, \S9.12). Dimensional transmutation renders the QCD scale only mildly sensitive on UV physics.
According to the autonomy of scales definition of naturalness,
the low value of the dimensionless parameter $\Lambda_{\text{QCD}}/M_P$
 would thus be natural, whereas the smallness of this parameter
is inexplicable when adhering to technical naturalness. Technical naturalness is therefore too restrictive to capture the central dogma of naturalness.
\nn
Notwithstanding, technical naturalness remains
incredibly useful. Technically natural parameters 
imply small quantum corrections so these parameters
are AoS natural. However, technically unnatural parameters may also only \emph{mildly dependent on UV physics} and therefore be AoS natural.
\subsection{Concluding remarks} 
Both absolute and technical naturalness aimed to formalize our notion of naturalness as defined in the previous section - I claim that naturalness is better understood in the AoS notion which was defined as in \S \ref{sectionnaturalprohibitionscales}. Technical naturalness, which is more permissive and successful than absolute naturalness, suggests that problems of naturalness can be solved by incorporating additional symmetries into a more fundamental theoretical framework. 
I have discussed that also technical naturalness is too restrictive - it only provide a \emph{sufficient} criterion to determine whether theories are natural. 
Indeed, naturalness problems \emph{may} be solved by incorporating the theory in a more fundamental theory with near symmetries, however,
this ought not necessarily be the case. This claim has been exemplified by the QCD scale: one cannot exclude the possibility that the gauge hierarchy problem may be solved without incorporating additional symmetries into the more fundamental field theory.
\nn
As
a consequence of aforementioned failures of technical naturalness,
scientists' notion of naturalness has gradually evolved away
from aforementioned connection with enhanced symmetries since the 1980s (Grinbaum 2009, p.1). 
This departure from technical naturalness is probably due to the incapability of this notion to account for several low dimensionless parameters, including equation \ref{eq:ratiogravweak}.
The most well-entrenched definition of naturalness was subsequently provided by Barbieri and Giudice, both of whom formulated a fundamentally different definition of naturalness which reproduces equation \ref{eq:naturalness} in the mid-1980s. These scientists implemented
Wilson's notion
of naturalness that “[o]bservable properties of a system should be stable against
minute variations of the fundamental parameters.” 
An example of an observable property of a system is $M_Z^2$.
This parameter is equivalent, up to constants of order unity,
to the Higgs mass and the Fermi constant $G_F
^{-1/2}$ and would be of pivotal importance in discussions of naturalness
(Anderson and Casta\~{n}o 1994, p.13). 
\section{Naturalness as a prohibition of fine-tuning}
\subsection{$\mathbf{c_n \neq \mathcal{O}(1)}$ implies fine-tuning}
In almost every discussion of naturalness, one encounters claims relating unnatural theories to a requisite form of fine-tuning (Williams 2015, p.17). 
Naturalness problems are intimately related to problems of fine-tuning.
The dimensionless coupling $c_0$ of the Higgs field at the Planck scale should be incredibly small, $c_0 = \frac{m_\text{H}^2}{\Lambda^2} \sim 10^{-34}$. 
The dimensionless coupling constant has to be chosen to be unnaturally low for no other reason than being compatible with experimental data. 
Setting the parameters $c_0$ in such a way to solve our problem at hand
is irreconcilable with dimensional arguments which are ``almost always successful in effective field theories'' (Williams 2015, p.17) because the contribution of the operator is not determined by the physical scales which are involved in the problem. 

What this implies is that the Higgs mass has to be severely \emph{fine-tuned}. The bare Higgs mass has to intricately cancel the gigantic quantum corrections in such a way that $m_{\text{H,bare}}^{2} \approx - \Delta m_{\text{H}}^{2}$ such that $m_{\text{H,bare}}^{2} + \Delta m_{\text{H}}^{2}$ gives a small, positive physical mass.
The intimate relation of naturalness to fine-tuning has led to many misconceptions in the literature, most notably the identification of naturalness problem with fine-tuned parameters.
\nn
Several physicists 
(among others, Anderson and Casta\~{n}o (1994), Casas \et (2005),  Donoghue (2007))
conceive of naturalness problem as nothing more than fine-tuning problems. 
Although it is undeniable that fine-tuning takes place in unnatural theories (certain values should take very specific values in order to achieve a delicate cancellation between bare terms and the sum of quantum corrections for no other reason than to match experimental values), I will assert that the identification of naturalness with fine-tuning is too weak and makes a weak argument for taking naturalness seriously. The notion of naturalness as a prohibition of fine-tuning ignores pivotal and qualitatively different aspects of naturalness problems and the latter are moreover likely to be degraded to pseudoproblems. 
\subsection{Fine-tuning problems can be devalued to pseudo-problems \label{subsecfinetuning} 
}
Naturalness as a prohibition of fine-tuning bears resemblance to fine-tuning puzzles in other areas of physics, including the horizon and flatness problems in cosmology and the requisite low entropy of the early universe, which may be degraded to pseudo-problems. I will focus on the latter and discuss why these fine-tuning problems are fundamentally different compared to naturalness problems in the context of EFTs. The problematic feature of naturalness problem is \emph{not its fine-tuned character}, it lies elsewhere.\footnote{The problematic aspect is the strong dependence of low-energy DOFs on high-energy physics.}

The second law of thermodynamics entails that the low entropy of the current universe would exceed its current value if the initial entropy would not have been extraordinarily low, hence one would expect there to have been an astoundingly low entropy within the early universe - called the Past Hypothesis (Albert 2000).
To my knowledge, no compelling reasons have been put forward as to why this low entropy would be required other than to be consistent with the omnipresent low entropy in the universe approximately 17 billion years after the Big Bang. According to several philosophers of physics (\emph{e.g.} Earman 2006) this is not truly a conundrum. It does not cry out for an explanation, since our universe simply happens to be endowed with this feature.\footnote{Others scientists oppose this view. Penrose (2012) argues that this criticism only holds water in non-cyclic universes and constructed a Conformal Cyclic Cosmology in which the low entropy at the Big Bang is entailed by the conformal phase of the universe before the Big Bang in our aeon took place. These topics are beyond the scope of this thesis, what is important is that many scientists believe that the early universe should have been in a low-entropic state while no fundamental laws of nature strictly require or entail this.}
\nn
Likewise, fine-tuning problems in the context of naturalness
have been criticized by Wetterich (1984), Hossenfelder (2017) and Dine (2015), each of whom has argued that naturalness fine-tuning problems  constitute pseudo-problems. Dine argues that problems of fine-tuning merely have metaphysical import: 
\begin{center}
	The  universe  is
described by a single theory, with a single set of degrees of freedom and a single lagrangian
with fixed parameters. Things are the way they
are, and it is not clear why we should be troubled the value of some parameter or other. (Dine 2015, p.28) 
\end{center}
If the parameters are whatever they are they must indeed be considered immutable or ``god-given" numbers. 
It should be kept in mind that many parameters of the SM cannot be calculated in this theoretical framework 
(such as the physical Higgs mass)
and nothing prohibits these parameters from being fine-tuned. Likewise, Wetterich contends that problems of fine-tuning are irrelevant because
\begin{center}
	[w]e do not need to know the exact formal relation between physical and bare parameters (which furthermore depends on the regularization scheme), and \emph{it is not important if some particular expansion method needs fine-tuning in the bare parameters or not}. The relevant parameters are the physical parameters, since any predictions of a model must finally be expressed in terms of these. (Wetterich 1984, p.217, my italics)
\end{center}
I will now put forward arguments as to why these claims are wrong.
\nn
If we come to grips with this fine-tuning problem and set $c_0 \sim m_{\text{H}}^2/\Lambda^2$, this would not entail that the fine-tuning problem is solved. The bare terms remain sensitively dependent on UV physics and therefore receive quadratic corrections which drive the \emph{effective mass} greatly beyond the low physical mass. One would then again have to fine-tune the bare term and sum of quantum corrections in order to keep the effective mass of the same order of magnitude as the physical mass (Williams 2015, p.18).

Several physicists have argued that this is not problematic, since these quantities were not renormalized quantities. Barbieri (2013, p.3) objects that ``we [are] supposed to talk of physical renormalized quantities, with all divergences suitably reabsorbed'' and Burgress (2013, \S1.4) contends put forward a similar objection ``why can't one simply absorb these large correction into a large (but finite) renormalization of the mass $m$?'' Indeed, it seems plausible that the fine-tuning problem would be solved in that case.

However, it would only be solved at a \emph{specific energy scale}. Assume that our EFT is valid up to an energy $\Xi$, in others words, all high energy DOFs between $\Xi$ and $\Lambda$ have been integrated out of the theory. Now we can again construct a low-energy effective theory of this EFT and integrate out all DOFs between a new energy scale $\Xi'$ and $\Xi$. The ``heavy fields'' which have now been integrated out of the theory give corrections to the relevant couplings, the effective Higgs mass in this case, and the fine-tuning problem reappears!
\begin{center}
One can't simply renormalize the mass with every move to a new $S_W'$ [Wilsonian action] appropriate for a new cutoff scale [$\Xi'$]. (Williams 2015, p.18)
\end{center}
The problem is that one should always \emph{choose a single energy scale} where the bare mass is renormalized.
This discussion highlights that fine-tuning is not the central problem of naturalness problems. The truly problematic feature of naturalness problems is that \emph{the RG flow of physical, renormalized parameters remain sensitive to UV physics}.

\subsection{The Barbieri-Giudice measure
\label{sectiongiudicebarbieri}
}
A quantitative analogue of Wilson's formulation that
``observable properties
of a system should not be unstable against minute variations of
the fundamental parameters''
 has been popularized by Barbieri and Giudice. This was accomplished in their seminal (1988) paper, in which the degree of fine-tuning ($\Delta$) has been defined as\footnote{The derivatives are rescaled by $a/M_Z^2$
in order to	remove the dependence of the sensitivity $\Delta$ on the
	overall scale of $a$ and $M_Z$.}
\begin{equation}\label{eq:BG}
\Delta_{\text{BG}}(\mathcal{O}, a_i) \equiv \max \abs{ 
\frac{a_i}{\mathcal{O}}
	\frac{  \partial \mathcal{O} }{ \partial a_i} }.
\end{equation}	
Here, $a_i$ are \emph{all the parameters of the theory} (not necessarily dimensionless) and the degree of fine-tuning is thus defined as the maximal variation of the observable $\mathcal{O}$ (for instance, the mass of the $Z$ boson) with respect to any of these parameters.\footnote{A theory with $\Delta_{\text{BG}} = 10$ exhibits a parameter tuning of 10\%, while $\Delta_{\text{BG}} = 100$ corresponds to a fine-tuning of 1\%, \emph{et cetera}.}
The prescribed methodology to determine $\Delta_{\text{BG}}$ is something like the following procedure (see Feng 2013 for a more extensive discussion):
	\begin{enumerate}[(1):, nolistsep]
		\item{} Pick a model (Standard Model, GUT model, \emph{et cetera}),
		\item{} Select low-energy parameter(s) (\emph{e.g.} the  mass of a $W$ or $Z$ boson),
		\item{} Select high-energy parameters (\emph{e.g.} the SUSY or GUT scale),
		\item{} Take derivatives of low-energy parameters with respect to high-energy
		parameters,
		\item{} Define the amount of ``tuning'' in the model as the maximum value of
		the derivatives.
	\end{enumerate} 
\vspace{3mm}
\noindent
The Barbieri-Giudice prescription became a widely used methodology for quantifying fine-tuning, primarily in the 1990s (Anderson, Casta\~{n}o 1994, p.2).\footnote{Afterwards the \AC measure of naturalness has become more prominent - this notion adheres to a discordant definition of naturalness called landscape naturalness.  
Many papers on theory choice after 2000 use landscape naturalness to compare models (Giudice 2008, p.10).}
\nn
Already at this point, one may argue that an unsatisfactory amount of arbitrariness has slipped into the \BG measure. For instance, it is left to our subjectivity
whether one should select all low-energy parameters or whether one can be selective and 
\emph{how much fine-tuning should be tolerated and still be considered natural}. What counts as ``an acceptable upper bound'' has changed considerably over the years; Barbieri and Giuduce themselved deemed $\Delta \leq 10$ a natural upper bound (Anderson and Casta\~{n}o 1994), while the accepted upper limit gradually shifted towards 20 (Chan \et 1998) many scientists ``would now call 1000 a reasonable value'' (Craig 2013, pp.16-17), probably due to the progressively more stringent constraints on the Minimal Supersymmetric Standard Model (MSSM) and  failures of other BSM models to live up to the constraints imposed by \BG measure.
The initial motivation for $\Delta \leq 10$ was 
based on the belief
that theoretical and experimental values should not differ more than one order of magnitude.\footnote{This is actually closely related to AoS naturalness; significantly smaller or higher values values of fine-tuning would result in a high discrepancy of theoretical and experimental values and do not necessarily entail a decoupling of energy scales.}
Although somewhat arbitrary choices have to be made in this process,  the \BG measure can be justified  insofar
as the measure specifies the sense in which elementary scalar masses are
``unduly sensitive'' to high-energy physics and ascribes a \emph{measure of unnaturalness} to these particles.
\nn 
This \BG measure allowed
Barbieri and Giudice to place upper limits on parameters which are relevant for SUSY breaking, such as upper bounds for supersymmetric particle masses. But is the \BG measure an accurate quantification of naturalness?
Let us for the time being be reluctant in assigning much significance to these upper limits of supersymmetric particle masses and evaluate the philosophy underlying these new
definitions of naturalness. Does the \BG measure capture the central dogma of AoS naturalness?
\subsubsection{Why \BG naturalness does not formalize naturalness as autonomy of scales}
Firstly, naturalness and fine-tuning are put on equal footing in the \BG measure. I have discussed in \S \ref{subsecfinetuning} which this identification ignores important features of naturalness.

Secondly, Barbieri and Giudice have strayed away from the
well-established connection between naturalness and (near) symmetries in the \BG notion of naturalness. Naturalness has turned into a measure of how sensitive observables are to variations of their underlying parameters. A weak connection with ``naturalness as a prohibition of correlations among widely separated scales'' can nonetheless be established, since the UV cutoff scale $\Lambda$ of the EFT is included in the set of parameters $\{a_i\}$. 
Unnatural theories in the sense of section \ref{sectionnaturalprohibitionscales} are highly sensitive on the value of $\Lambda$ and this unnaturalness would then be reflected in high values of $
\partial \mathcal{O} / \partial \Lambda$ (and thus in $\Delta_{\text{BG}}$).
 Naturalness is however no longer necessarily a measure of how sensitive IR parameters depend on UV physics, but a measure of sensitivity on generic perturbations of its underlying parameters. One would obtain a high degree of naturalness if $\partial \mathcal{O} / \partial \Lambda$ is small provided that another derivative (say $\partial \mathcal{O} / \partial M_Z$) is large.
 The \BG measure of naturalness thus provides a discordant notion of naturalness. Additional backup to this claim is provided by the following (and more vigorous) argument.
\nn
I will now discuss a failure of the \BG measure, which lends further support to my claim that the Barbieri's and Giudice's quantification of naturalness does not truly amount to an AoS notion (according to which the proton mass is paradigmatically natural (Dine 2015)).
\subsubsection{Failure of the \BG measure: the proton mass}
Anderson and Casta\~{n}o have assessed whether a simple application
of this formula will always give a reliable measure of fine tuning. They concluded that instances of sensitivity (in the context of the renormalization group flow) cannot be separated  from instances of fine-tuning in the Barbieri-Giudice measure of naturalness (Anderson and Casta\~{n}o 1994).
The lightness of the proton compared to the Planck scale can be understood by examining the renormalization group flow of the QCD coupling constant.
Anderson and Casta\~{n}o (1994, p.3) have shown that the logarithmic running of the 	QCD coupling constant $\alpha_{\text{QCD}}$ yields the following naturalness measure for the proton mass
\begin{equation}\label{barbierigiudicedef}
\Delta_{\text{BG}} \left(m_{p}, g_s(M_p) \right) = \left(
\frac{4\pi}{b_3}
\right) \frac{1}{\alpha_{\text{QCD}}(M_{P}) } \geq 100.
\end{equation}
This result implies that the proton mass is highly fine-tuned. The high value of $\Delta_{\text{BG}}$ is ascribed to the strong sensitivity of the proton mass on physics near the Planck scale by Anderson and Casta\~{n}o. They argued that
\begin{center}
	[t]he proton mass would have exhibited this
	strong sensitivity no matter what its value was, so it makes no sense to say
	that a value near 1 GeV is fine tuned. (Anderson and Casta\~{n}o 1994, p.3)
\end{center} 
Interestingly, their conclusion is correct
in spite of the fact that their explanation is wrong (see \S\ref{sucsec:technat}, p.\pageref{eq:protonRGE}). 
Although the proton mass is paradigmatically natural (Dine 2015, p.4)), the \BG measure obtains a highly unnatural value for this parameter. This indicates that equation \ref{eq:BG} systematically overestimates the required degree of fine-tuning for the physical parameters and does not accurate capture AoS naturalness.
In other words, the \BG measure is an imprecise measure of fine-tuning and fine-tuning is not even an accurate measure of naturalness! 
This entails that one should be reluctant in ascribing any significance to the numerical naturalness values which are predicted by this naturalness measure.\footnote{I will soon assert that one should guard against ascribing significance to numerical values of naturalness of \emph{any measure of naturalness}.}

\section{A Bayesian notion of naturalness: landscape naturalness \label{sec:bayesian} }
 I will argue that physicists' notion of naturalness has departed from the original notion of naturalness since the advent of landscape naturalness in the 1990s (starting with the \AC notion). 
This is seldomly mentioned explicitly in the literature (although it has been stated explicitly in Wells 2018).
Newer definitions of naturalness are based on a \emph{discordant} notion of naturalness  called \emph{landscape naturalness} (of which the prominent one - the \AC measure - will be discussed in \S \ref{sec:anderson}). AoS and landscape naturalness being discordant, I think it is actually misleading to speak of \emph{new measures of naturalness}.

Consequently, it is of fundamental importance to clearly distinguish the original meaning of naturalness and the ``new-age'' landscape naturalness. 
Failing to acknowledge these fundamental differences would be inauspicious for our understanding of naturalness, since
``what the two criteria count as
`natural' \ti{can and will come apart}.'' (Williams 2018, p.57, my italics)

\subsection{The Anderson-Casta\~{n}o landscape
\label{sec:anderson}
}
Anderson and Casta\~{n}o (1994) modified the Barbieri-Giudice measure of naturalness in such a way that situations in which a sensitivity is not due to fine-tuning have been excluded. In order to achieve this, the Barbieri-Giudice measure was \emph{divided by its average value 
$\bar{\Delta}$
over a ``sensible range of parameters"} $a_i$
\begin{equation}\label{eq:AC}
\Delta_{\text{AC}}(\mathcal{O}, a_i) =  \Delta_{\text{BG}}(\mathcal{O}, a_i)/ \bar{\Delta}_{\text{BG}},
\end{equation}
where $\bar{\Delta}_{\text{BG}}$ is an average value of $\Delta_{\text{BG}}(\mathcal{O}, a_i)$.
Anderson and Casta\~{n}o argued that this provides a reliable measure of fine-tuning, because it gives a large value when a quantity
is fine-tuned, while reducing to a value whose order of magnitude is unity when it encounters a ``typical sensitivity" which is not due to fine-tuning (Anderson and Casta\~{n}o 1994, p.4). The latter feature is absent in the Barbieri-Giudice measure; it now arises because our results are divided by an \emph{average sensitivity}.\footnote{Athron and Miller (2007) adjusted the AC measure of naturalness in such a way that models with multiple fine-tuned observables and finite variations of parameters can be considered, this has become a fairly prominent measure of naturalness. That many different naturalness measures have been fashionable highlights that naturalness is hardly quantifiable. I will defend the initial AoS notion further in chapter \ref{chapterviabledefinition}.} 
The resulting ratio 
$\Delta_{\text{AC}}(\mathcal{O}, a_i)$
consequently remains large
for unusually sensitive solutions and will be of order one when “typical” sensitivities (such as the sensitivity of the proton mass on ultraviolet physics) are encountered. 
The Anderson-Casta\~{n}o measure yields $
\Delta_{\text{AC}}\left(m_{p}, g_s(M_P) \right) \simeq 1$ and the proton is thus perfectly natural according to this measure, in accordance with the AoS notion.
\nn
One may wonder how the ``sensible range'' of parameters $a_i$ is chosen. For instance, one could choose those values for parameters 
	for which the experimentally verified predictions of the systems remain unperturbed (not ``unusually unstable'' in Wilson's sense) but one could also choose the parameter range \emph{by fiat}. Since the results 
	are
	dependent on the parameter range, it is imperative to reach consensus as to how a sensible range of parameters is to be understood.
	
	The range over which the parameters are allowed to vary mathematically represents our assumptions (Anderson and Casta\~{n}o 1994, p.5). This is typically taken to be a uniform distribution
	because there usually are no compelling reasons to prefer one value over another for these parameters other than to fit experimental data.
	\begin{center}
		The ``theoretical license'' at one’s discretion when making this choice
		necessarily introduces an element of arbitrariness to the construction. (\emph{ibid}, p.5)
\end{center}
This ``representation of our assumptions'' is exactly what makes landscape naturalness a \emph{Bayesian notion of naturalness}. If one has good reason to deviate from a uniform distribution function, one is free to do so.
\nn
Let us note that Wilson's definition of naturalness has now be modified to: ``Observable properties
of a system should not be \emph{unusually unstable} against minute variations of
the fundamental parameters'' (Anderson and Casta\~{n}o 1994, p.13, my italics). This seemingly inconsequental modification actually has profound implications for fundamental physics. The SUSY scale for instance is now allowed to be significantly higher than is allowed by any of
aforementioned notions of naturalness (Giudice 2008). Again, I advice the reader to guard against ascribing much significance to measures of naturalness. I will now discuss why the \BG measure incarnates a fundamentally novel idea of naturalness which is irreconcilable with the AoS notion. 
\nn
The introduction of $\bar{\Delta}_{\text{BG}}$ into the \AC definition of naturalness ties in naturalness with a Bayesian notion. This \emph{range of parameters} embeds this
 definition of naturalness into the multiverse (also known as the many-worlds ontology), because these parameters are no longer considered to be ``fundamental" - they could have been different. Since Anderson and Cast\~{n}o's intention to infer the plausiblity of numerical values of parameters in our universe from probability distributions (assuming that the parameters could have been different)
 incarnates the idea of the multiverse. 
\subsection{The multiverse}
According to the multiverse ontology, fundamental parameters should not be interpreted as
``god-given numbers'' but as \emph{dynamical variables} taking different values in a landscape of vacuum states (Susskind 2005, Carr 2007).
Our physical model is now conceived of as a one instantation in a broader class of systems with other possible parameters. 
Much of the popularity of the multiverse was brought about by the discovery of the
enormous landscape of possible vacua in string theory (Giudice 2017, p.9).
In the context of naturalness, we conceive of all \emph{possible} values of the parameters as being realized in a parallel universe and 
want to infer the probability distribution function to be located in this particular world. The intuition of a multiverse is in fact shared by many physicists, most notably cosmologists 
(eternal inflation requires a multiverse (Guth 2007))
and string theorists (Schellekens 2013). 

In string theory with some compactified dimensions there are many
types of quantized fluxes (analogous to the magnetic flux in QED) which can take non-equivalent values. These values
determine the potential for vast numbers of possible states (vacua). In each of these vacua, the DOFs and the parameters of the Lagrangian will take different values.\footnote{If there are
enough such states, the parameters will be densely distributed. (Dine 2015, p.21)} Of course, the existence of such a
\emph{landscape} or \emph{discretuum of vacua} remains conjectural (Dine 2015, p.21).
\nn
Anderson and Casta\~{n}o therefore defined naturalness in terms of the \emph{likeliness of a set Lagrangian parameters}, where the likeliness distribution of the fundamental parameters $a_i$ (such as the proton mass) should be deduced from statistical arguments in the vacuum landscape. 
Anderson's and Casta\~{n}o's notion of naturalness marked the advent of
a different foundation of naturalness measures, which are now to be interpreted as measures of \emph{probability}. 
Models are now considered ``natural'' if
and only if they are ``likely'' in the landscape of possible universes. 
\nn
The tricky part is to
understand what kind of \emph{mechanism} singles out our universe.
The general methodology underlying landscape naturalness is as follows (Wells 2018, pp. 48-50);
\begin{enumerate}[1), nolistsep]
	\item {} Firstly, one starts with a large set of vacua across which
	parameters are allowed to vary.
	\item {} One then offers some selection criteria that justifies restricting
	attention to a subset of these vacua. Examples of such selection criteria are:
	\begin{enumerate}[a), nolistsep]
		\item {} a \emph{phenomenologically acceptable criterion} requires that physical results in four-dimensional spacetime are consistent with our experimental data. This implies that one restricts attention to EFTs with
		$D = 4$, no unbroken SUSY, at least 3 quark generations, \emph{et cetera}. 
		\item{}
		an \emph{anthropically acceptable criterion} requires that one restricts attention to EFTs whose values for
		parameters do not rule out observers.\footnote{Although this is the most cogent selection criterion for several naturalness problems (to be discussed in \S \ref{sec:withoutnaturalness}), it is also the most hated one (Giudice 2018, p.8). Barness (2013, p.18) criticizes the anthropic principle because it cannot be falsified.} 
		\item {} A \emph{statistical selection criterion} can be imposed whenever
		the vast majority of the possible vacua have features similar to our universe. Unnatural parameters can then be justified in a statistical fashion.
	\end{enumerate}
\item {} Lastly, one places some \emph{measure of probability} over the selected
vacua and determines most \emph{likely EFTs} to observed in this landscape. 
\end{enumerate} 
Both the phenomenological and anthropical selectrion criterion are \emph{censorship criteria}, which exclude vacua with wrong properties.\footnote{Both the Higgs vev and the value of the CC are plausible in the multiverse when
	adhering to the AP (Weinberg 1987, Agrawal \emph{et al.} 1998), at least in a context in which
	only a limited number of parameters scan (Giudice 2018, p.8).
	Although these solutions are plausible solutions to the hierarchy and CC naturalness problems (as I will argue in \S \ref{sec:withoutnaturalness}), these solutions do not restore naturalness. Instead, these solutions aim to explicate why certain parameters are unnatural. Since the original notion of naturalness has been abandoned in landscape naturalness, a discussion of these multiverse solutions to naturalness problems is deferred to \S \ref{sec:withoutnaturalness}, where ``the post-natural solutions'' will be discussed.}
We should note that the most natural EFT is the most likely EFT in the landscape, a notion which not necessarily coincides with the EFT exhibiting the lowest sensitivity on UV physics. 
``Natural theories'' have therefore been bequeathed a new meaning since the advent of landscape naturalness in the 1990s.
\nn 
Physicists have attempted to tackle problems of naturalness by means of this stringy vacua methodology.
The following quotes from papers in which landscape naturalness has been employed elucidate why this notion of naturalness exhibits fundamentally distinctive features compared to AoS naturalness;
\nn
``An effective field theory (or specific coupling, or observable) $T_1$
is more
natural \emph{in string theory} than $T_2$, if the number of phenomenologically
acceptable vacua leading to $T_1$
is larger than the number leading to $T_2$.'' (Douglas 2004, my italics)
\nn
``If the property in question is common among these \emph{`anthropically
	acceptable' vacua} then the property is natural. By common I mean that
some \emph{non-negligible fraction of the vacua} have the required property. If
however, the property is very rare, even among this restricted class, then it
should be deemed unnatural.'' (Susskind 2004, my italics)
\nn
My claim is that solving these multiverse problems does not amount to solving the naturalness problem that scientists began with in the 1970s.
A simple way to see this is that weak-scale SUSY (which remains
the most prominent natural
solution to the Higgs vev problem) need not be ``stringy natural'' (Susskind (2004) and Douglas (2013)\footnote{This is an extremely strong constraint which very much disfavors
	the natural solutions to the hierarchy problem...perhaps we should
	not be too bothered by this, but we should ask for some more
	fundamental reason why $\Lambda_{\text{SUSY}} \sim 30 - 100$ TeV. Later we are going
	to argue this from \emph{stringy naturalness}.
	(Douglas 2012, my italics)} have shown this) and I will assert their claims can be generalized. Susskind argued that
\begin{center}
``...the most numerous “acceptable
vacua” do not have low energy supersymmetry. \emph{Phenomenological
	supersymmetry appears to be unnatural}'' (Susskind 2004, my italics).
\end{center}
In a more general sense, what AoS naturalness and landscape naturalness count as ``natural'' is inherently different. 
This is because AoS naturalness determines the naturalness of a model as an \emph{intrinsic property of the model} - one can calculate how strongly IR physics depends on UV physics by means of the RGEs underlying the EFT in question. Instead, in landscape naturalness, the ``stringy naturalness'' depends entirely on the \emph{distribution of
vacua}. The ``stringy naturalness" of a model is therefore \emph{extrinsic to the model} itself.
\nn
Attention is increasingly directed toward the possibility that we inhabit a
multiverse (with examples in string theory and eternal inflation) and I find it understandable that scientists aim to embed the naturalness principle into the multiverse and reformulate it appropriately. We should however be aware that these attempts to recast the traditional naturalness principle in a statistical setting are \emph{not conservative
embeddings}.
Both the physical motivation for, meaning of and possible solutions to naturalness problems have altered significantly from our AoS starting point so it is important to avoid conflating these disparate notions of naturalness when assessing the naturalness literature.
\subsection*{Concluding remarks}
I have introduced the most prominent definitions of naturalness in the physics literature of the previous decades up to now and argued that the majority of these fail to capture the central dogma of the AoS notion of naturalness.  I endorse William's (2015, p.23) claim that ``quantative measures involve large amounts of arbitrariness'' (with exceptions for absolute and technical naturalness) and ``render incorrect judgements of paradigmatically natural scenarios.'' Technical naturalness is the exception, but this merely provides a sufficient criterion for the principle of insulation. 
\nn
What I claim is that naturalness does not have to be quantified - the  AoS notion suffices. Although this definition is seemingly ill-defined (why else would scientists have constructed quantitative measures of naturalness?), this definition captures all important aspects of naturalness whereas ensuing measures of naturalness are actually unsuccessful in capturing this dogma.
Interestingly, this AoS understanding of naturalness was explicitly stated in the earliest discussions of naturalness in the 1970s (see Ovrut, Schnitzer 1980 and Susskind 1979) it has been superseded by impoverished notions of naturalness. I aim to restore physicists' confidence in the predictive power of the AoS notion. I will show in chapter \ref{chapterviolations} that both the successes and violations of naturalness can be recognized without adhering to a formal measure of naturalness. First, I will defend the AoS notion of naturalness further by invalidating criticisms regarding its alleged aesthetic and ill-defined character in chapter \ref{chapterviabledefinition}.

\chapter{Defending autonomy of scales naturalness
\label{chapterviabledefinition}
}

Firstly, I will discuss the role of naturalness and aesthetics in \S \ref{sec:aesthetics}. Once this relation has been established, I will examine whether the one should indeed conceive of the natural criterion as an aesthetic criterion rather than acceptable scientific argument. I will conclude that 
one can indeed conceive of naturalness as an aesthetic principle (since an ``aesthetic theory'' is not well-defined)
but such talk of aesthetics is irrelevant in our evaluation of the guiding principle.
I will put forward arguments as to why naturalness is a reasonable criterion to impose on EFTs 
and argue that its foundation does not lie
in arguments of aesthetics. One consequently cannot relegate naturalness problems to pseudoproblems.
\nn
Secondly, I will evaluate Hossenfelder's (2018a) arguments as to why naturalness would be ill-defined. It should be kept in mind that many formalizations of naturalness have been put forward in the academic literature and some of these are discordant (autonomy of scales versus landscape naturalness). Most criticisms of Hossenfelder are directly aimed at landscape naturalness, as will be shown in \S \ref{sec:notwelldefined}, and consequently these criticisms do not pose a threat for the viability of the AoS naturalness notion which I defend in this thesis.
At the end of this chapter I have thus contended that naturalness is \emph{not an aesthetic principle} and \emph{the mathematical meaning of AoS naturalness is perfectly transparent}.

\section{Naturalness and aesthetics \label{sec:aesthetics}}
Giudice (2008, p.1) contends that ``[t]he role of naturalness in the sense of \emph{``aesthetic beauty"} is a
powerful guiding principle for physicists as they try to construct new theories'' (my italics). 
Aesthetic motivations for naturalness are fairly common
in the scientific literature.\footnote{Although many scientists, including those who belittle the physical significance of the principle, do not conceive of naturalness as an aesthetic principle. For instance Hossenfelder merely claims that the principle is mathematically ill-defined, but it does not serve to make physical laws more aesthetical.}
Among others, Anderson, Casta\~{n}o and Riotto (2017, p.3) claim that the principle is an ``aesthetic principle'', Donoghue (2007) contends that naturalness is primarily an ``aesthetic choice'' and, in the same vein, naturalness
 is 
according to
 Athron and Miller (2007, p.2)
 intimately related to a ``question of aesthetics.'' Taking their claims at face value (for the time being), what would be the physical significance and implications of a purely aesthetic foundation for naturalness? In order to answer this question we ought to look at the broader picture and examine whether aesthetics have proven to be a successful heuristic in other disciplines of physics. I 
 then
 will come back to naturalness in \S \ref{sec:naturalnessnotaesthetic} and assert that the principle would be utterly useless if it would amount to a purely aesthetic criterion.
\nn
According to Grinbaum (2009, p.2) ``one must not belittle the place of beauty in the scientist's thinking.'' 
Several scientists have advocated the importance of beauty in the laws of nature (Einstein (1917), Dirac (1933)) while others claim that 
questions of
aesthetics belong to the realm of the arts, not science.
Einstein vividly supported the utility of aesthetics in physics early in
his life, claiming that aesthetics ``may be valuable when
an \textit{already found} truth needs to be formulated in a final form."
Dirac has probably been the most outspoken adherent of mathematical beauty in physics (Kragh 1990, \S14) - the following quotes vividly illustrate his 
aesthetics-inspired
philosophy of physics. 

There is a tradition
at the University of Moscow that distinguished visiting physicists are requested to write 
a self-chosen inscription
on a blackboard, which remains preserved for posterity.
When Dirac visited Moscow in 1956, he wrote that
\begin{center}
	A physical law must possess mathematical beauty.
	
	(Kragh 1990, p.275)
\end{center} 
This short inscription 
accurately
summarizes the philosophy of science which has strongly influenced Dirac's thinking from the mid-1930s on.\footnote{References to `beauty' `ugly', `beautiful' and `ugliness' is  ubiquitous in Dirac's publications (Kragh 1990, p.275).} 
Dirac gradually developed a more extreme (and highly unsettling) philosophy according to which \emph{aesthetics should take precedence
	over Ockham's razor}:
\begin{center}
	The research worker, in his efforts to express the laws of Nature in
	mathematical form, should strive mainly for mathematical beauty.
	He should still take simplicity into consideration in a subordinate
	way to beauty... It often happens that the requirements of simplicity
	and beauty are the same, but when they clash the latter must take
	precedence. (Dirac 1937)
\end{center}
Dirac however leaves it to our subjectivity what is meant exactly with \emph{requirements of beauty} (except that simplicity is typically considered beautiful). 
I will argue in the next subsections that this subjective character of aesthetics is responsible for the 
fruitlessness of aesthetic arguments as a guide for the natural sciences.\footnote{I do endorse Einstein's view that aesthetics may be very valuable in order to \emph{reformulate} ``already found truths" (Einstein 1917), because it is desirable to write down the physical content as succinctly as possible. 
An example will be provided in equation \eqref{eq:SM+GR}. Aesthetics should however not be elevated to genuine research imperatives, as will be argued in \S \ref{subsecaesthetics}.}
	
Nonetheless, Dirac's strong devotion to aesthetics led him to develop a rather extreme philosophy of science:
\begin{center}
it is
more important to have beauty in one's equations than to have them fit experiment. (Dirac 1963)
\end{center}
Of course this statement is wrong. Although the laws of physics may be argued to be beautiful (as will be argued in \S \ref{subsubdefendaesthetics})
it is due to high precision experiments (\eg of the fine-structure constant) and
experimental verification in general that our paradigmatic theories of the universe 
(\eg GR, quantum mechanics, QFT) 
came to be accepted. It is very reasonably to assume that LQG, quantum gravity, string theory, SUSY and other high-energy theories will not be accepted 
by large physics communities
due to their 
potential
inherent beautiful structure either; instead these theories should pass future experimental tests with flying colors.

Nonetheless, aesthetics could still serve as a fruitful guide in physics even if its role is subordinate to that of experimental verification, so it is worth examining whether arguments of aesthetics can help fundamental physics advance. If so, naturalness could retain a scientific character even if it is a purely ``aesthetic principle'' (Donoghue 2007), however I will put forward arguments as to why there is no logical relation between aesthetics and the laws of nature.
\subsection{Aesthetics in physics
\label{subsecaesthetics}
}
I will now evaluate the role of aesthetics in our laws of physics; by putting forward arguments \emph{in favor} of the utility of aesthetics in physics, ensued by arguments \emph{against} this claim. I will argue that the latter criticisms should be taken seriously and that these imply that scientists should not pursue aesthetical laws.
\subsubsection{Defending aesthetics
\label{subsubdefendaesthetics}
}
Einstein's aforementioned claim that aesthetically inspired arguments ``may be valuable when an \emph{already found} truth needs to be formulated in a final form'' (1917) can be rationalized, especially given the unification dream of physicists. Theories should describe as many phenomena as possible, whilst  hinging on as few assumptions - and utilizing as few parameters - as possible. 
Indubitably many scientists consider theories which satisfy these principles to be elegant, since it allows a prodigious phenomenology to be
accurately described by a compact set of parameters. 

Quantum mechanics and special relativity have successfully been amalgamated in quantum field theory. 
The predictive power of QFT, one may argue, exceeds that of quantum mechanics since 
the non-relativistic limit in QFT recovers quantum mechanics - but one can now also describe relativistic processes -
moreover QFT is  the appropriate theoretical framework to describe gravitational interactions too (Wald 1984).
In fact, the entire content of both the Standard Model and general relativity can be expressed in a remarkably compact notion (Donoghue 2007, p.3):
\begin{equation}\label{eq:SM+GR}
\mathcal{L} = -{1 \over 4} F^2 + \bar{\psi} i D \psi +{1 \over 2} D_\mu \varphi D^\mu \varphi + \bar{\psi} \Gamma \psi \varphi + \mu^2 \varphi^2 -\lambda \varphi^4 - {1 \over 16 \pi G_{\text{N}} } \mathcal{R} - \Lambda
\end{equation}
A vast amount of particle physics phenomenology
is thus beautifully organized through a simple set of gauge symmetries. The magic about gauge theory ultimately lies in the richness of its structure and its ability to
produce a great variety of different manifestations out of a simple conceptual principle.
Short-range forces, long-range forces, dynamical symmetry breaking, confinement are all examples of
phenomena which are successfully described by the very same principle. The vacuum structure of gauge theories is incredibly
rich, with $\theta$-vacua, instantons, chiral and gluon condensates, all being expressions
of the same theory. Stated differently, gauge theory is an exquisite tool to create
complexity out of simplicity (Giudice 2017).

Likewise, physicists involved in the high-energy frontier of theoretical physics aim to combine quantum mechanics and gravity into quantum gravity. Among other things, they study Grand Unified Theories (GUTs) which predict that all but the gravitational force are manifestations of one and the same force at the GUT scale (in the deep UV: $E_{\text{GUT}} \simeq 10^{16}$ GeV).
Arguably, one may see beauty in these (potential) unification schemes of nature. Now lend me your ears, I will assert that aesthetics should not be elevated to a leitmotiv to study any kind of model. 
\subsubsection{Criticizing aeshetics}
The reason is simple - one could also put forward several reasons as to why the laws of nature are inelegant. 
\begin{enumerate}[i), nolistsep]
	\item {}
	The simple looking Lagrangian in equation \eqref{eq:SM+GR}
	and its symmetry based origin actually conceals a far less beautiful
	fact. 	The SM contains many incalculable parameters (18 in total), which need to be plugged in by hand (Cahn 1996). Examples include the masses of all the leptons and quarks, the weak
	mixing angles (describing the charge current interactions of quarks and of leptons), the strength of the three gauge interactions,
	the overall scale of the weak interaction, the cosmological constant $\Lambda$ and
	Newton's gravitational constant $G_{\text{N}}$ (Donoghue 2007, p.3). None of these parameters are predicted by
	the SM but have to be determined by experiments.
	\item{} 
	Einstein gravity, while astonishingly successful at the classical level, does not yield a straightfoward quantization since it is nonrenormalizable. The infinitude of diverging intergrals in the quantum formulation of GR is problematic and considered ugly by most scientists. 
	\item {} A naive extrapolation of the weak, strong and  electromagnetic beta functions shows that their magnitudes are only \emph{approximately identical} at the GUT scale. The unification of coupling constant depends on the masses of sparticles (Hossenfelder 2018a, p.10), but with experimental bounds pushing lower mass bounds upwards, the unification can only be obtained by postulating cancellation effects between several quantum corrections (Ellis and Wells 2017). That the gauge unification is not robust under small variations of parameters (so it can only be maintained through a delicate fine-tuning mechanism) may be considered to be an unaesthetic feature of SUSY.\footnote{This is reminiscent of 
	the CC and $\nu$
	fine-tuning problems in the  context of naturalness - I will soon argue that all these parameters may be fine-tuned.}
\end{enumerate}
Moreover, it is simply false that particle physicists are only willing to tolerate ``beautiful'' or ``elegant'' theories.
An insightful example can be given in the context of naturalness:
\begin{itemize}
	\item{}
	Low-energy SUSY remains the most popular solution to the hierarchy problem and would be an incredibly elegant theory \emph{if unbroken}, however
	``many particle physicists openly acknowledge that when broken at low energies (as it must be in order to describe our world), SUSY is a rather unattractive theory''
	(Williams 2015, p.22).
	Although the meaning of ``inelegant'' is equally subjective as that of ``beauty'', Richter's (2006)
	succinct criticism of SUSY shows that one should be very lenient not to conceive of SUSY as inelegant: ``[t]he price of this invention [the SUSY solution to the hierarchy problem] is \emph{124 new constants}... in this case a conceptual nicety was accompanied by an explosion in arbitrary parameters.'' (Richter 2006, p.2, my emphasis) See the following footnote for additional quotes supporting the claim that broken SUSY is an inelegant theory.\footnote{
		``[W]eak-scale supersymmetry is neither ravishingly beautiful (and hasn't been for decades) nor excluded'' as contended by Feng (2013, p.353), while Shifman (2012, p.4) contends that
		``[a]lthough theoretically supersymmetry is a beautiful concept, the corresponding phenomenology was and still is less than elegant.''
	} 
\end{itemize}
Quite generally, natural BSM theories
decrease in simplicity with respect to the SM. Simplicity in this context has a technical meaning: the simplest theory
is the one with the smallest number of degrees of freedom consistent with known facts (Dine 2015, p.28). One may therefore wonder: what is more aesthetical, a simple theory or a natural theory?
\nn
I aimed to convey that there is no logical connection between aesthetics and the paths ``chosen by nature.'' The reason is simple - the meaning of aesthetics hinges primarily on \emph{personal preference} rather than universal conventions.\footnote{Physical laws may \emph{happen to be beautiful}, but also ``ugly laws'' may be true.} 
Whether or not the well-entrenched laws of nature 
in contemporary physics
are deemed aesthetical is in the eye of the beholder, as my previous examples have shown. we cannot expect more fundamental laws to exhibit ``a kind of beauty'' that scientists would agree upon. 
Of course, the most fundamental laws of nature may turn out to be more aesthetical than those from contemporary physics, nonetheless this can only be determined \emph{a posteriori}. 

Since the meaning of aesthetic laws is ill-defined, one should guard oneself against ascribing much significance to aesthetics in physics. Arguments of beauty should certainly not be
elevate to a guide for theory choice (contrary to symmetries, renormalization, \emph{etc.}). 
This deflationary view on the normative role of aesthetics in physics was uttered early on by
Einstein, who correctly surmised that:
\begin{center}
	[aesthetically motivated arguments] fail \emph{always} as heuristic aids.

	(Einstein 1917, my italics)
\end{center}
Grinbaum argued in the same vein as Einstein that ``there is no necessary link between beauty and empirically verified truth'' and argues that the role of aesthetics should not be elevated to that of a ``normal scientific argument'' (Grinbaum 2009, p.18).\footnote{This poses a stark contrast with a fairly common belief among physicists and philosophers that mathematical beauty should be pursued while constructing theories of the universe (see Chandrasekhar's great disussion thereof in \emph{Truth and Beauty} (1987)).}

I will defend Einstein's and Grinbaum's position that arguments of aesthetics are irrelevant for the viability of the corresponding model. 
That aesthetics 
shouldn't play a role in the (possibly non-empirical) assessment of scientific hypotheses is succinctly exemplified by the history of the Dirac's Large Number Hypothesis (LNH). This theory has gone astray because Dirac ascribed excessively much significance to the role of beauty in physics and I will discuss afterwards that arguments of aesthetics are \emph{generically misleading}. 
\subsubsection{Dirac's Large Number Hypothesis
\label{subsec:LNH}
}
Dirac aimed to provide explanations for \ti{very large numbers} in nature (not necessarily in a field-theoretic context)
and contended that large numbers should be related to a single fundamental constant of nature.
Dirac was heavily inspired by Eddington,
 a physicist who had a strong belief that large numbers cannot be accidental peculiarities (Giudice 2008, p.4).
Dirac took this Eddington's line of thought step further
in his 
Large Number Hypothesis (LNH)
and contended that \emph{all large numbers in nature} should be related to a single large number\footnote{In the same fashion, physicists have tried to explain smaller numerical values of parameters which are used in our description of the universe. The (inverse) fine-structure constant $\alpha^{-1} = {4\pi \epsilon_0 \hbar c \over e^2} = 137.035 991 1(46)$ has been subject to many intended derivations of this value. Among many attempts, one finds $\alpha^{-1} = 108\left( 8/1843 \right)^{1/6}$ (Aspden and Eagles 1971) and $\alpha^{-1} = 2^{-19/4} 3^{10/3} 5^{17/4} \pi^{-2}$ (Robertson 1971). 
These numerical coincidences may be considered beautiful (although I personally disagree), but these coincidences are
not particularly illuminating.}, which he chose to be the age of the universe $\tilde{t} \simeq 10^{40}$, expressed in atomic units, known as the \emph{Hubble age}. 
Two numbers that figured prominently
in Dirac's work are measures of mass and force:
\begin{enumerate}[i), nolistsep]
	\item {} The number of massive particles
	(such as protons and neutrons) in
	the visible region of the universe; the
	number has been estimated by Eddington to be about $N_{edd} \simeq 10^{80}$ (Gale 1981, p.157). 
		\item {} A dimensionless form of the gravitational
	coupling constant - which is a
	measure of the strength of the gravitational
	force - and is given by ${G_{\text{N}}m_e m_p \over e^2} \simeq 10^{-40}$ (Giudice 2008, p.5).
\end{enumerate} 
These numerical coincidences fueled Dirac's strong adherence to aesthetics and he claimed that these numbers are related (Gale 1981, p.157). But of course the age of the universe is not a constant so 
 the relations
of the numbers should be continuously
changing. Dirac forestalled
this criticism by arguing that both the
number of massive particles 
and
the gravitational
coupling constant
are \emph{not truly constant} - they change
with time in such a way that the following
relations remain valid
throughout the entire history of the universe. 
$$ \boxed{N_{edd} \sim \tilde{t}^2, \ \ \ \ \ \ \ G_{\te{N}} \sim 1/\tilde{t}}.$$ 
These numerical relations are
too striking to be dismissed as coincidences 
according to Dirac, he therefore proposed that these numerical agreements result
from some unknown causal connection. 
The most striking consequence of the LNH is that the ``fundamental constant'' $G_{\text{N}}$ is, in fact, a time-dependent variable (Ray \et 2007, p.1).
Dirac's arguments had too much of a kabbalistic flavor to many of his contemporaries (Giudice 2008), yet Dicke would be persuaded by Dirac's conlusion if he could come up with compelling causal relations (Gale 1981, p.157). Dicke claimed that the numerical relation between $N_{edd}$ and $G_{\text{N}}$ can arguably be backed up by Mach's principle\footnote{See 
Dicke (1961) for an extensive discussion of Mach's principle in tjis context or
Gale (1981, p.157) for a brief summary.}, but
it is not apparent why the
gravitational coupling constant and the number of
massive particles should possibly be related to $\tilde{t}$. 
\nn
Dicke (1961) then found a better explanation for Dirac's astronomically large numbers, which does not require any time-varying Newton constant.\footnote{A time-varying Newton constant would actually be blatantly irreconcilable with our universe. Life would have not been able to develop on Earth in such a universe; the oceans would boil on Earth after life could have developed (Teller 1948).} The value of the Hubble age is probably
strongly constrained by \emph{anthropic conditions}, Dicke's argumentation was roughly as follows. One
essential condition is that the universe
should have aged enough to allow time
for the creation of elements heavier than
hydrogen, because ``it is well known that
carbon is required to make physicists.''
Since heavy elements are  synthesized in main-sequence
stellar evolution and then dispersed throughout space by supernovae, the Hubble age of an inhabited universe
cannot be shorter than the age of the
shortest-lived star. An estimation of the time required by these processes shows that the three numbers considered by Dirac should indeed be at
least as large as we observe them in an expanding univere (Gale 1981, p.157). They could not be much larger either.\footnote{For instance, if
the Hubble age would have been much greater than
the age of a typical star, most planets which could support life would
have ceased to exist by now (Giudice 2008). } What Dicke's solution shows is that numerical coincidences is nature may be entirely \emph{due to the cosmological history of our universe} and does not necessarily indicate hitherto unknown relations between parameters.
\subsubsection{Implications for naturalness \label{subsec:implicationnaturalness} }
Although Dirac's LNH has no quantum field theoretic origin, three important lessons can be learned from our example which \emph{do} bear consequences for naturalness:
\begin{itemize}
	\item {} ``Aesthetics'' is a subjective term whose meaning may differ depending on the topic and personal preferences.  ``Aesthetic'' meant to Dirac that large numbers 
	(not necessarily in a field-theoretic context)
	can be explained in terms of dimensionless parameters, while 
``aesthetic'' means to me that theories are not
constructed by means of unbridled speculation hinging on arguments of beauty, but by means of elegant, logical deduction.
In the context of naturalness, aesthetic arguments typically imply that models with little fine-tuning 
are to be preferred
over strongly fine-tuned models. But what if nature has chosen to be delicately fine-tuned? Ugliness seems to be the main criticism (supposedly ruling out this possibility) and I assert that this is not a sound criticism. I will elaborate on these statements and explicate why naturalness does not have an aesthetic foundation in \S
\ref{sec:naturalnessnotaesthetic}.
\item {}
The powerful engine behind theoretical physics
has always been (and should always be) logical deduction. Aesthetic speculation should not be pursued, since there is no logical relation between aeshetics and the paths chosen by nature. Karl Darrow surmised that nature would be more ``elegant'' if the atomic nucleus would only contain two subatomic particles (Darrow 1933) but that, of course, does not make it true. 

The same holds true for the arguably appealing thought that the Earth is the center of the universe.  Vilenkin  persuasively argues that:
\begin{center}
	We humans have a well-documented tendency toward hubris, arrogantly imagining ourselves at center stage, with everything revolving around us. We've gradually learned that it's instead we who are revolving around the sun, which is itself revolving around one galaxy among countless others. Thanks to breakthroughs in physics, we may be gaining still deeper insights into the very nature of reality. (Vilenkin 2011)
\end{center}
The ``deeper insights" which Vilenkin alludes to refers to the possibility that our universe could be one out of many universes - \emph{the multiverse theorem}. One should not rule out this possibility simply because one might not like the idea of it.
\item {} Dicke's plausible explanation for the small numbers lies in the cosmological history of our universe. Could it be that also the $G_F/G_{\textbf{N}}$ and \lcc \ are small because of the history of the  universe, that it not truly indicates the existence of BSM physics contrary to what your local dedicated naturalist would assert? I will come back to this thought-provoking question in \S \ref{subsec:selection}.
\end{itemize}
Naturalness problems have recurringly been identified as problems of aesthetics in the scientific literature (Williams 2015, p.20). A plausible explanation 
for this identification
is that naturalness nihilates (or at least ``severely reduces'') the degree of fine-tuning in our models.
Grinbaum (2009, p.17) contends that we ``use it [naturalness] to please the senses by setting off the models in a beauty contest'' and
 Shifman (2012, p.4) argues that the naturalness criterion is ``aesthetic, or, if you wish, philosophic.''  
Interestingly, even those who strongly advocate the utility of naturalness 
often consider their
preference for natural theories to be an aesthetic preference. 
This even includes renowned physicists who \emph{introduced prominent definitions of naturalness} such as
Anderson, Casta\~{n}o and Riotto (1997) and Athron and Miller (2007).

These adherents of naturalness argue that physicists' ``aesthetic preference" for natural theories should nonetheless be taken seriously.\footnote{Of course, that explains why these scientists developed a measure of naturalness in the first place.} Others physicists have taken an opposing stance - they argued that the aesthetic character of naturalness has profound implications which devaluate its utility.
Donoghue (2007, p.1) argues that problems of naturalness ``have a more aesthetic
character'' than other problems in physics and ``\emph{it is [consequently] not as clear that a resolution is required}, yet
the problems motivate a search for certain classes of theories (\emph{ibid}, my italics).''
Donoghue's claim boils down to the following: although we may find natural solutions to several problems occurring in EFTs, there is no fundamental reason why this should be the case.
Naturalness problems would then not truly be physical problems. Taking this position, one may conclude that
\begin{center}
[i]f you do not like it [naturalness as a guiding principle]
you can
ignore it (Shifman 2012, p.4).
\end{center}
This quote provides a strong incentive for us to investigate whether or not the naturalness principle merely has an aesthetic foundation. Naturalness would be an obsolete guide if the principle can only be backed up by arguments of aesthetics. Let me therefore evaluate Grinbaum's criticism of naturalness and put forward arguments to support my claim that the naturalness is neither an aesthetic 
nor a sociological
principle.
\nn
Consequently, if naturalness solely has an aesthetic foundation (Grinbaum's and Shifman's position) problems of naturalness are prone to being identified as pseudoproblems. In this case, violations of naturalness solely imply that nature turns out to be less aesthetical than previously thought. Would this does not run counter to deeply embedded features of theories of physics? It seems that it does not. 
Naturalness would indeed look more like a theorists' prejudice
rather than a ``normal scientific argument'' if the criterion could only be backed up by arguments of aesthetics.
I will however argue that, although natural theories are arguably more aesthetical than unnatural theories\footnote{The former would explain why the Higgs mass and the cosmological constant are low without requiring a delicate cancellation mechanism between $q_0$ and $\Delta q$.}, the foundation of naturalness is in fact much more profound than solely aesthetical. Naturalness is an expectation well-motivated by the EFT framework, because it ought to be imposed to procure the separation of energy scales which is \emph{prima facie} underwritten by the Decoupling Theorem.\footnote{There are however instances where this decoupling of scales might break down (when relevant operators are concerned) so we should take into account the possibility that the most fundamental laws of particle physics are not describable by EFTs. The possibility of naturalness becoming an otiose principle at smaller distance 
will be discussed in \S \ref{sec:withoutnaturalness}	
and has \emph{nothing to do with aesthetics}.} 
\subsection{Naturalness problems are not aesthetic pseudoproblems \label{sec:naturalnessnotaesthetic}}
Shifman's position is closely related to Grinbaum's, although they disagree as to \emph{why} many scientists like naturalness. 
According to Grinbaum, scientists' preference 
for natural theories
admits a sociohistorical interpretation in which \emph{irrational
	sociological reasons} and \emph{contingencies} have elevated the principle to the important role it has acquired in contemporary physics. 
This includes ``down-to-earth sociological factors'' (Grinbaum 2009, p.2)
including ``the choice at the leading universities of professors with a particular taste in physics'' (\emph{ibid}, p.18) and ``the abrupt reversals between fashionable and worn-out research,'' both of which indicate that the physical significance customarily ascribed to naturalness 
is more a matter of a partly circumstancial history than subject to a rigorous epistemology. 
According to Grinbaum, the (supposedly contingent) initial definitions of naturalness have heavily influenced the ensuing development of particle physics: ``[t]hose who are the first to fix the arbitrary convention of what is natural
and what is not, exercise significant influence over those who will follow later'' (\emph{ibid}, p.18).
The ``vague feeling of aesthetical unease'' (\emph{ibid}, p.18) regarding elementary scalars
has initially been expressed by Weiskopf (1939) 
(``Even the Coulombian part of the self-energy diverges'')
and subsequently by Wilson (1971) 
(``It is interesting to note that there are no weakly coupled scalar particles in
nature; scalar particles are the only kind of free particles whose mass
term does not break either an internal or a gauge symmetry'')
and naturalness criteria have subsequently turned into a powerful ``sociological instrument"  (Grinbaum 2007, p.18).\footnote{This has been achieved by Dirac, 't Hooft and later on by Barbieri and Giudice,
whose influental work 
 have provided novel mathematical definitions of the principle (Grinbaum 2007, p.18). }

As a result of this circumstancial history,
Grinbaum claims that naturalness consequently does not have the force of ``normal scientific arguments'' (Grinbaum 2007, pp. 2,18). In other words,
naturalness problems do not truly encompass physical problems (Williams 2015, p.20). 
Now lend me your ears, I come to bury this inaccurate view.
I will assert that a \emph{conflation of the central dogma of naturalness and  qualitive measures of fine-tuning} has led to numerous misconceptions of the physical significance of naturalness and plausibly led scientists to (falsely) conclude that naturalness is either a sociological or aesthetic restriction. It should be noted that these measures of naturalness are actually measures of \emph{fine-tuning} and as a consequence of that have been devalued to aesthetic principles.
\nn
When formal measures of fine-tuning are conflated with naturalness, it becomes apparent why naturalness has been downgraded to sociologically influenced pseudoproblems. 
Fine-tuning problems are, unlike naturalness problems in the AoS notion, easily downgraded to pseudo-problems (Williams 2015, p.21). Whenever models cannot account for the fine-tuning of the theory (\eg the fine-tuning is not explained by a near symmetry), one may object that the theory happens to be fine-tuned and does not cry out for an explanation. One might allege that the problem of fine-tuned lies elsewhere, the theory simply fails to account for the finely-tuned feature. 
One may then deny that this is problematic by adopting
Wetterich's pragmatic position (introduced in \S \ref{subsecfinetuning}) according to which ``it is not important if some particular expansion method needs fine-tuning in the bare parameters or not'' (Wetterich 1984, p.217).
\nn
It is important to realize that criticisms of aesthetics 
only apply to 
\emph{measures of naturalness}, most apparently to definitions where naturalness has been conflated with fine-tuning (such as the \BG measure from \S \ref{sectiongiudicebarbieri} and landscape naturalness from \S \ref{sec:bayesian}). 
\begin{itemize}[nolistsep]
	\item {} First of all, the scientist has to choose a specific measure of fine-tuning, for instance the Barbieri-Giudice, \AC or \AM measure (or a less prominent measure such as $\Delta = \sqrt{\sum_i \Delta_{\text{BG}}(a_i  )}$ which has been employed in Casas, Espinoza  and  Hidalgo (2005)).
	Different measures assign different degrees of fine-tuning (Williams 2015, p.21) and yet different measures are used in the scientific literature to (allegedly) compare models. 
	\item {} One is then free to select parameters from the UV theory and parameters from the IR theory, which are  functions of the high-energy parameters. 
	The degree of fine-tuning is then defined as the ``maximum sensitivity'' of the IR parameters to changes in UV parameters.
	\emph{Common choices} for the IR parameters are the masses of $W$ and $Z$ bosons ($\sim \text{vev} \sim m_{\text{H}}$), but one could equally well choose other IR parameters. This is unsettling because the choice of parameters ``at low and high energy can drastically change the amount of fine-tuning assigned to a model by the chosen measure'' (Williams 2015, p.21).
\end{itemize}
There is an even larger amount of sociologically influenced and somewhat arbitrary choices in the quantification of naturalness measures, namely the acceptable value of fine-tuning. 
\begin{itemize}
	\item{}
This ``...has drifted up over the years, starting around $\sim$ 10 and floating towards 100. Many would now call 1000 a reasonable value.'' (Craig 2013, pp. 6-7)
\end{itemize}
 What this implies is that models which were once deemed ``highly unnatural'' have come to be considered acceptable. This makes Grinbaum's claim that ``naturalness can at best be understood as a sociologically heuristic'' (Grinbaum 2009, p.19) understable.
Also ``[t]his makes naturalness look more like a prejudice of theorists" (Hossenfelder 2017) than a normal scientific argument or ``an expectation well-motivated by the effective field theory framework'' (Williams 2015, p.22).  
I think that this plausibly
explains why many physicists and philosophers of quantum field theory have disparaged the scientific character of naturalness. 
\nn
A reluctant approach where little significance is ascribed to \emph{quantifying the degree of (un)naturalness} - as advocated in the fruitful AoS naturalness notion - allows one to inoculate 
naturalness from aforementioned
sociological and aesthetic criticisms.

I will put forward argument to support my claim that the central dogma of naturalness - that conspiracies between widely separated scales are weak - is not affected by these criticisms. AoS naturalness is required to guarantee the predictive power of EFTs (as was discussed in \S \ref{sectionnaturalprohibitionscales}) and thus serves a well-motivated goal rather than ``to please the senses.'' 
I will now discuss why this reluctant approach can be justified and is, in fact, the best approach to understanding naturalness. An additional reason is that Hossenfelder's criticism that ``naturalness is ill-defined'' (Hossenfelder 2018a) does, in fact, \emph{not apply to the AoS notion of naturalness}. 
\section{Why autonomy of scales naturalness is not ill-defined
\label{sec:notwelldefined}
}
Hossenfelder is one of the biggest sceptics of the naturalness principle. She argues that naturalness is an \emph{aesthetic criterion} which has primarily been constructed to spread the belief that nature is inherently beautiful. The most problematic aspect of naturalness, according to Hossenfelder, is that the principle is not even mathematically well-defined. 
Hosseldelder's position is tantamount to the claim that 
physicists have embraced the principle for too long - the time has come to discard naturalness from our list of guiding principles which are used for our non-empirical assessment of more fundamental laws of nature. I leave the punchline to Hossenfelder: 
\begin{center}
	The focus on ill-motivated theories [theories whose sole motivation is naturalness]
creates a vicious cycle in which we attempt to find evidence for unpromising theories
by experiments which deliver little guidance on the development of better theories, resulting
in more fruitless theories and further experimental null-results. (2018a, p.15)
\end{center}
The AoS notion of naturalness may be said to be 
a rather loose physical
heuristic, but it is certainly not ill-defined.
Chapters \ref{chaptermotivation} 
has shown that a plausible diagnosis for the difficulty associated with the construction of a
satisfactory measure of naturalness 
(which allows us to ascribe a \emph{degree of naturalness})
is that it runs afoul of a well-known Aristotelian
dictum: ``It is the mark of an educated man to look for precision in each class
	of things just so far as the nature of the subject admits''
(Aristotle $\sim$340 BCE).
\nn
Along the same line, Craig has argued that quantative measures of naturalness are otiose:
\begin{center}
One
	frequently comes across models that are constructed using a legalistic
	interpretation of naturalness that fails an intuitive sniff test. I would prefer
	we exercise our physical judgment when weighing naturalness
(...) Perhaps a
reasonable criterion in the context of model-building is to ask if the IR
theory is a generic function of the UV parameters, but [one should] not commit overly
much to specific measures (Craig 2014)
\end{center}
Craig's underlying assumption is that the meaning of natural theories is crystal clear to scientists, while formal measures of naturalness fail to accurately capture the central features of naturalness. My discussion of naturalness is an endorsement of this view. The loose heuristic AoS notion of naturalness which is advocated in this thesis is transparent, but one may argue that it could be cricized for being ``mathematically ill-defined.''  
Craig's recommendation is that ``we exercise our physical judgment when weighing naturalness''(Craig 2014) 
rather than aiming to ascribe an actual degree of (un)naturalness to theories and I endorse this view.
\nn 
The renormalization group equations (RGEs)  accurately capture how strongly IR physics depends on UV physics and should consequently be important for our physical judgment.
Additionally, near symmetries may explain why certain dimensionless parameters are significantly smaller than order unity so technical naturalness should be used as well for our physical judgment.
In the next chapter I will introduce both successes
(\ref{sectionsuccessesnaturalness})
and violations 
(\ref{sec:violations})
of ``naturalness" and put forward arguments as to why these are best understood in the AoS naturalness notion. This notion allows us to (i)
recognize natural parameters in field theories,
(ii) recognize the problematic cases where IR physics depends sensitively on UV physics, and moreover (iii) give us an idea of how strongly naturalness is violated.
 \nn
In order to provide some support to these statements already, I will show that
the extraordinarily strong sensitivity of the Higgs on the UV cutoff scale $\Lambda$ can be shown in the context of the renormalization group equations (RGEs) (see Schwartz (2014, \S 23.6.1) for computational details). The general idea behind these equations is that one can examine the dependency of coupling constant with respect to the cutoff scale $\Lambda$. I will examine two couplings $a_2$ and $a_4$ (where the subscript denotes the dimension of the corresponding operator - the former is a relevant operator, the latter a marginal operator)\footnote{Since $\dim \varphi = \left[\varphi \right] = 1$ and the action has mass dimension 4, the $a_2$ constant corresponds to a constant with mass dimension $4-2=2$ and $\left[
 	a_4
 	\right] = 0$.}, whose RGEs read
 \bse
 \begin{equation}\label{key}
 \Lambda \frac{d}{d\Lambda} a_2 = \Lambda^2 \beta_2( \Lambda^{-2} a_2,  a_4) 
 , \ \ \
 \Lambda \frac{d}{d\Lambda} a_4 = \beta_4( \Lambda^{-2} a_2,  a_4).
 \end{equation}
After redefining the coupling constant in order to make them dimensionless, \ie $a_2(\Lambda) = \Lambda^2 c_2$ and $a_4(\Lambda) = c_4$, these equations can straightforwardly be shown to be equivalent to
 \begin{equation}\label{key}
 \Lambda \frac{d}{d\Lambda}  c_2 + 2c_2 =  \beta_2( c_2,  c_4) 
 , \ \ \
 \Lambda \frac{d}{d\Lambda} c_4 = \beta_4( c_2,  c_4).
 \end{equation}
 The beta functions can be expanded perturbatively in $c_2$ and $c_4$. After expanding up to first-order in the coupling constants, these RGEs read
 \begin{equation}\label{key}
 \Lambda \frac{d}{d\Lambda}  c_2 = A  c_4 + (B-2) c_2 
 , \ \ \
 \Lambda \frac{d}{d\Lambda} c_4 = C  c_4 + D c_2,
 \end{equation}
 where $A,B,C,D$ are real constants ($\ll 1$; otherwise perturbation theory becomes unreliable).
The value of the marginal operator $c_2$ at the UV cutoff is thus given by
 \begin{equation}\label{key}
 c_2(\Lambda) = c_4(\Lambda) \frac{A}{2} \left[
 \left(
 \frac{\Lambda}{\Lambda_\text{H}}
 \right)^2 
 -1
 \right]
 +
c_2(\Lambda_\text{H})  \left[
\left[1 + \frac{AD}{2} \right]
 \left(
 \frac{\Lambda}{\Lambda_\text{H}}
 \right)^{-2} 
 -1
 \right].
 \end{equation}
 \ese
We can immediately see that the dimensionless relevant coupling parameter $c_2(\Lambda)$ exhibits a strong sensitivity on the cutoff scale. Setting $\Lambda = 10^5$ GeV and $\Lambda_{\text{H}} = M_P = 10^{19}$ GeV, we obtain an unsettingly tiny dimensionless parameter $\left(
\Lambda/M_P \right) = 10^{-28}$. 
The severeness of the naturalness violation can immediately be recognized; an infinitesimal change in $c_{2}(\Lambda_{\text{H}})$, for instance $c_{2}(\Lambda_{\text{H}}) \rightarrow c_{2}(\Lambda_{\text{H}} + 10^{-20})$ causes the IR coupling $c_2(\Lambda)$ to jump \emph{eight orders of magnitude}! The AoS notion of naturalness, when supplemented with RGEs and physical intuition, are in fact sufficient to distinguish natural parameters from unnatural ones.

\chapter{Assessing the utility of naturalness
\label{chapterviolations}
}
In order to asses the utility of naturalness, I will first discuss successes (\S \ref{sectionsuccessesnaturalness}) and violations (\S \ref{sec:violations}) of the AoS naturalness guide. Then I will introduce the most prominent natural solutions to these naturalness problems in \S \ref{sec:solvehierarchy} and argue that current LHC data puts the viability of these models under much stress. I will then try to answer several interesting philosophical questions - \emph{could parameters be unnatural} and, if yes, \emph{which consequences would this bear for the ontology of QFTs}? I will put forward arguments as to why \emph{selection criteria on the multiverse} can explain why both the Higgs vev and \lcc \ have to be unnaturally small in \S \ref{subsec:selection}. This forces us to reflect upon the Decoupling Theorem which \emph{prima facie} underwrote an ontology of quantum field theories according to which these can be described in EFTs. This ontology is underwritten by the DT because it entails that widely separated energy scales decouple. I will however conclude in
\S \ref{subsecdecnotsat}
that generic EFTs do not meet the conditions of the theorem and moreover, \emph{even if these conditions are met, the DT is too weak to underwrite decoupling of scales}. This counterintuitive result entails that unnatural parameters may not be accurately described by EFTs and I will speculate on what kind of field theories would be appropriate to describe such parameters. I will shed light on the possibility of theories with UV/IR interplay on \S \ref{subsec:UV/IR}

\section{Successes of the naturalness criterion
	\label{sectionsuccessesnaturalness}}
\setlength{\epigraphwidth}{0.75\textwidth}
\epigraph{[N]aturalness seems to be one of the best-kept secrets of	physicists from the public, a secret weapon for evaluating and motivating theories of the world on its deepest	levels.}{\textit{Philip Nelson (1985)}}
\noindent
Arguably, naturalness provides a \emph{warning sign} that new physics is about to surface at a certain scale in order to circumvent a fortituous cancellation mechanism. 
Indeed; there have been successful applications of the naturalness criterion in the history of particle physics. The predicted  \emph{existence of the charm quark} has been the biggest achievement of naturalness, since the criterion has been used by Gaillard and Lee (1974) to predict the mass of this charm before its experimental discovery. This is discussed below and will be ensued by a discussion of two other successes of naturalness which have only been realized \emph{in hindsight}.
\subsection{The charm quark}
 Fortitious cancellations are circumvented by a natural solution in the context of neutral kaon mixing (Giudice 2008, pp.13-14). I will now discuss how naturalness provided a warning sign that new physics (the charm quark) would emerge at a specific energy scale.
 
 Two neutral kaons exist ($K^0 =  d\bar{s}$ and $\bar{K}^0 = \bar{d}s$) and these composite particles interact via the following diagram ($u,c,t$ means that the quark may either be an up, charm or top quark):
 \begin{figure}[h!]
 	\centering
 	\includegraphics[scale=0.2]{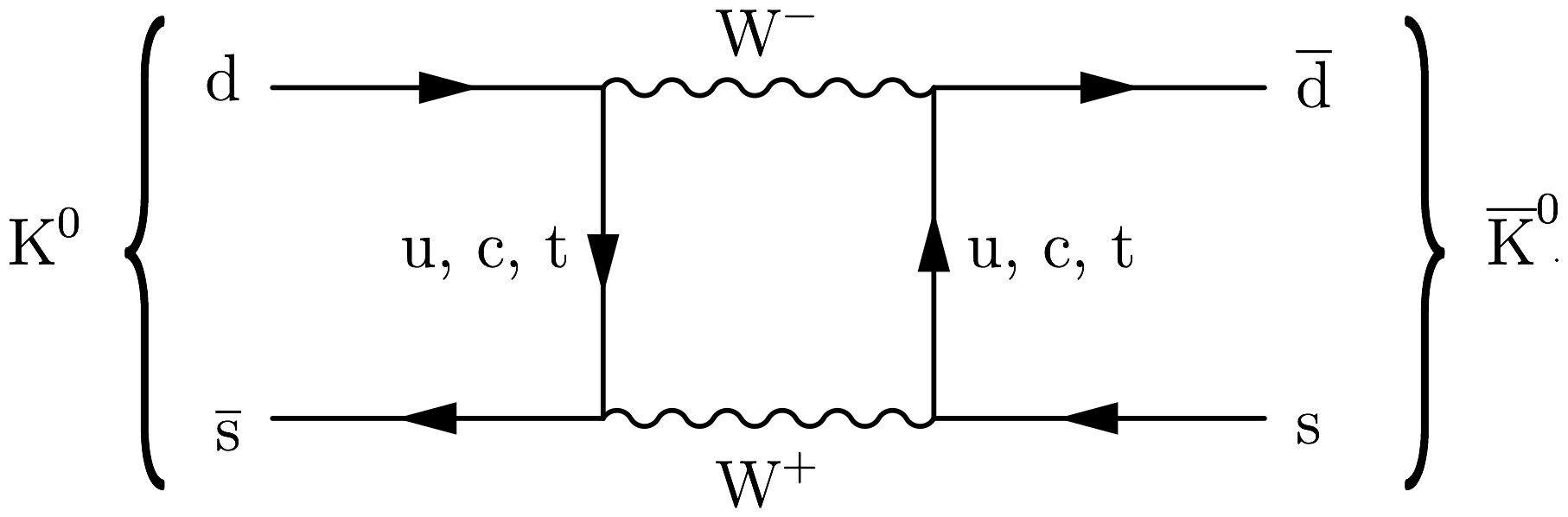}
 \end{figure}

\noindent 
When taking superpositions of these kaons, one can construct the ``long-lived'' and ``short-lived'' kaons ($K_L^0 = (K^0 + \bar{K}^0)/\sqrt 2$ and $K_S^0 = (K^0 - \bar{K}^0)/\sqrt 2$, respectively).
The mass difference between $K_L^0$ and $K_S^0$
has been computed in an EFT which is valid at energies of 
order of the kaon mass (Craig 2017, p.14);\footnote{In equation \eqref{eq:kaon}, $f_K = 114$ MeV is the kaon decay constant and $\sin \theta_C = 0.22$ is the Cabibbo mixing angle.}
\begin{equation}\label{eq:kaon}
{ m_{K_{\text{L}}^{0}} - m_{K_{\text{S} 	 }^{0}} \over m_{K_{\text{L}}^{0}} }
=
{G_F^2 f_K^2 \over 6\pi^2} \sin^2 \theta_C \Lambda^2.
\end{equation}
Note that the quadratic sensitivity on UV physics arises because the kaons are scalars (albeit composite rather than fundamental scalars). After plugging in the constants in equation \eqref{eq:kaon} and imposing that the correction (the RHS) is smaller than the measured value $ (m_{K_{\text{L}}^{0}} - m_{K_{\text{S} 	 }^{0}} )/ m_{K_{\text{L}}^{0}} = 7 \times 10^{-15}$,
one can straightforwardly infer that the cutoff satisfies $\Lambda < 2$ GeV. 
Glashow, Iliopoulos and Maiani (1970) realized that
the quadratic divergence in the kaon mass difference
vanishes in the presence of a 4th type of quark (coined the charm quark).
Indeed, before reaching this energy scale a new particle modifies the $\Lambda^2$ UV-behavior of the field theory. The
charm quark\footnote{Its mass $m_c \approx 1.3$ GeV has been deduced from the discovery of the $J/\Psi$ meson.}
implements the so-called GIM mechanism according to which
flavor-changing neutral currents (FCNCs) are suppressed in loop diagrams
(Glashow, Iliopoulos, Maiani 1970). 
The charm quark modifies the field theory and ameliorates the $\Lambda^2$ behavior above 1.3 GeV in such a way that the quantum corrections remain sufficiently small for the neutral kaon mass difference to be perfectly natural.

This discussion of  $K^0 - \bar{K}^0$ mixing is not merely rationalization - it is historically accurate.
Gaillard and Lee (1974) have actually used this naturalness argument to compute the mass of the charm quark before its experimental discovery (Craig 2017, p.15).
\subsection{A posteriori successes}
In hindsight, we can recognize several other cases in the history of physics where arguments of naturalness - had physicists firmly devoted themselved to the principle - would have predicted the existence of additional particles in other EFTs. I will now discuss two \emph{a posteriori} analyses. The electromagnetic energy of a classical electron poses a naturalness problem 
- the electron
mass is small in comparison to the contribution which stems from the self-energy of its electric field
-
is cured by the coming of the positron and the unnatural pion mass difference is cured by the rho meson (Giudice 2013, p.3).
\subsubsection{Electron mass}
	Lorentz modeled the electron as a spherical (smooth) charge distribution with radius $r$ (Dine 2015, p.2). 
One would expect that the mass
of the electron would, at least, be of the order the self-energy of the electron arising from
its Coulomb field: 
\begin{equation}\label{eq:m_e}
m_e \simeq {e^2 \over 4\pi r}.
\end{equation}	
From this equation we may infer that $r \approx 10^{-10}$ cm, in other words, we obtain the non-sensical conclusions that electrons would be larger than atomic nuclei (Giudice 2008, p.13). Experiments have shown that its radius is many orders of magnitude smaller than radii of nuclei: $r \approx 	10^{-17}$ cm. The self-energy of equation \eqref{eq:m_e} would however greatly exceed the physical mass of the electron by a factor of $5 \times 10^{4}$ 
for such a small radius
(Dine 2015, p.3).
	\nn
Murayama was the first scientist to point out that the electromagnetic energy of an electron requires delicate fine-tuning in order for it to have its experimentally verified value, because it is sensitivitely dependent on UV physics (Murayama 2000). 
This conundrum can efficiently be resolved in two mutually exclusive ways.
\begin{itemize}[nolistsep]
	\item {}
		Different contributions to the electron energy may cancel with an extraordinarily high precision whose occurrence, though not logically excluded, is not enforced by any obvious reasons.
	\item {}	
 Alternatively, ``new physics'' may become relevant below the energy scale $r^{-1} = 4\pi m_e/e^2$, modifying quantum corrections in the UV. 
\end{itemize}
The latter option would abide by naturalness (since the quantum corrections to the electromagnetic energy remain sufficiently small) and experiments have shown that the electron mass is indeed natural. 
The positron\footnote{This is the anti-particle of the electron. We should acknowledge in retrospect that antiparticles were already predicted by Dirac, who worked on a consistent relativistic theory of fermions. His model entailed that each fermion is accompanied by an antifermion.} introduces an additional degree of freedom and severely reduces the dependence of the physical electron mass on high energy physics. The inverse proportionality of mass and radius $m_e \sim 1/r$ is now replaced by a mild logarithmic dependence $m_e \sim \ln\left(r\right)$ (see Dine (2015, p.3) for the full expression).
The radius of the electron can now be taken to be as small as the Planck lenght $l_p \simeq 10^{-35}$ cm whilst satisfying $\mathcal{O}(\Delta m) \leq \mathcal{O}(m_{\text{bare}})$, complying with the naturalness principle.
\subsubsection{Pion mass difference } 
	The difference in mass between the postively charged and the neutral pion is significantly smaller than the mass of the individual pions. From an effective field theoretic point of view one obtains
	\begin{equation}\label{pionmass}
	m_{\pi^{+}}^{2} - m_{\pi^{0}}^{2} = \frac{3 \alpha}{4 \pi} \Lambda^2
	\end{equation}
	which exhibits a quadratic divergence in the UV (the pions are composite \emph{scalars}) and the 
	natural value of the
	mass difference would therefore be large (keep in mind that $\Lambda$ may be all the way up to the Planck mass).
	The experimental value for $m_{\pi^{+}}^{2} - m_{\pi^{0}}^{2}$ is ($35.5$ MeV)$^2$ and, in a reminiscent fashion to the electron mass, we require that equation \eqref{pionmass} does not exceed this value, resulting in $\Lambda = 850$ MeV. Since the EFT breaks down at higher energy scales, new physics should be discovered at energies below, or at, $\Lambda$ if naturalness is obeyed. Again, this turned out to be the case: the existence of a $\rho$ meson (whose mass is $M_\rho = 770$ MeV) was discovered and, more importantly, this particle indeed softens the electromagnetic contributions. There is thus no strong sensitivity on the high energy physics above 770 MeV.
\subsection*{Naturalness does not necessarily reign supreme}
Naturalness has also been successful for the determination of several cutoff scales, including (i) the electron mass beyond which QED emerges  and (ii) the GeV range beyond which hadronic resonances emerge (Giudice 2017, p.3). It would however be an unjustified to point at historical precedent and conclude that all laws of nature are natural. In the previous examples naturalness served as a useful guide, enabling us to infer at which
energy scale a specific EFT would break down and should be replaced by a new physical description. I will now put forward arguments that this may not be the case for the Higgs vev and the cosmological constant - this will raise the important question: \emph{does AoS naturalness sometimes provide bad counsel}? 
\section{Violations of naturalness
\label{sec:violations}
}
Two dire violations of naturalness have been observed in nature 
(the cosmological constant problem in cosmology and 
the Higgs naturalness problem in
particle physics),
both of which could potentially undermine the validity of naturalness. 
This chapter aims to nuance whether
naturalness can ultimately be used as a fruitful guiding principle in our journey of unraveling the fundamental laws of nature and I will conclude that prominent models which attempt to solve naturalness problems are not likely to restore naturalness in the ensuing section. 
\subsection{The Higgs mass
	\label{sechierarchy}
}
CERN's Large Hadron Collider (LHC) delivered in 2012 what looked like (and was later confirmed to be by Atlas (2012)) a Higgs particle, the cornerstone of a decades-long search in order to complete the SM and understand the origin of the electroweak symmetry breaking mechanism.
Now that the Higgs boson has been discovered, the most pressing goal has become to elucidate the origin of this particle. The potential of the Higgs field $\varphi$ in the SM is described by
	\begin{equation}\label{eq:higgspotential}
	V(\varphi) = -{\mu^2 \over 2} \abs{\varphi}^2 + \frac{\lambda}{4} \abs{\varphi}^4,
	\end{equation}
	where the LHC data implies that $\mu \approx 89$ GeV and $\lambda \approx 0.13$.\footnote{The Higgs field in the SM takes a constant value everywhere
		is spacetime. This is called its vacuum expectation value (vev)	 $\upsilon$ and its magnitude has been measured to be $\upsilon = $ 246 GeV (Donoghue 2007, p.4). This is the only dimensionful
		constant in the electroweak interactions and hence sets the scale for
		all dimensionful parameters of the electroweak theory.}
	That the Higgs mass is ``unnaturally small'' is explained as follows.
	Dimensional analyis entails that $\mu^2 \approx M_P^2$ which is nearly 34 orders of magnitude higher than what has been measured (Aad \emph{et al.} 2012; Chatrchyan \emph{et al.} 2012).\footnote{I recall that a small value of $\mu$ would not cry out for explanation if the Lagrangian for the Higgs field would have an enhanced symmetry in the limit $\mu^2 \rightarrow 0$, but this is not the case (Dune 2015, p.6).}
\nn
The SM consequently exhibits a severe violation of naturalness which is closely associated with the spontaneous symmetry breaking mechanism induced by the Higgs boson.
\begin{equation}\label{key}
m_{\text{H,phys}}^{2} = m_{\text{H,bare}}^2 + \Delta m_{\text{H}}^2.
\end{equation}
\begin{figure}[h!]
	\centering
\includegraphics[scale=0.4]{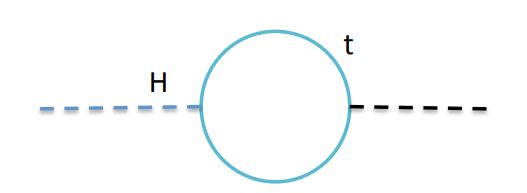}
\caption{The dominant one-loop correction to the Higgs mass involving a loop correction from top quarks. The leading quantum correction in provided by equation \eqref{eq:unnaturalm*}}
\label{fighiggscorrection}
\end{figure}
Scalar masses receive quadratic corrections from every heavy particle to
which they couple (as was shown explicitly in section \ref{sectionnaturalprohibitionscales});
the leading quantum correction to the Higgs mass ($\Delta m_{\text{H, leading}}$) comes from the coupling to the heaviest fermion - the top quark.\footnote{The
coupling is described by $\mathcal{L}_{\bar{t}tH} = \lambda_t H Q_3 \bar{t}$ where $Q_3$ denotes the third quark doublet consisting of the $t$ (top) and $b$ (bottom) quarks.}
This Feynman diagram is displayed in figure \ref{fighiggscorrection}. 
LHC data has unambiguously shown that the physical mass of the Higgs boson corresponds to $m_{\text{H}} \approx 125$ GeV, while quantum corrections exceed this values for cutoff scales of $\Lambda \geq 500$ GeV (Williams 2015). 
If the SM remains valid up to GUT energy scale where a grand unified symmetry is believed to be restored, loop corrections induce a mass at least 13 orders of magnitude larger
than the actual mass of the Higgs boson and the situation becomes even worse if the SM remains valid up to the Planck scale with a corresponding discrepancy of 32 orders of magnitude. 
\nn
Some physicists have argued that the Higgs naturalness problem is not truly a physical problem, claiming that whether or not the naturalness problem 
shows up
would 
be dependent on your choice of regulator. I will now invalidate this claim and argue that the Higgs naturalness problem is truly a physical problem that we should care about.
\subsubsection*{The alleged dependence of the hierarchy problem on one's regularization scheme}
Several physicists\footnote{This deflationary position on this problem is advocated, among others, by Alvaraz-Gaum\'{e} and V\'{a}zquez-Mozo (2012, p.236)} have argued that the Higgs vev problem is not a genuinely alarming problem and I want to clarify why they are wrong. 
These physicists claim (correctly) that the quadratic sensitivity on UV physics does not appear in
other regularization schemes including a dimensional regularization scheme\footnote{In this regularization scheme, one retains field modes of arbitrarily high momenta (no cutoff parameter is introduced), however, the calculations are done for a non-integer number of dimensions ($D - \epsilon$) where 
$D$ is the number of spacetime dimensions one is interested in and 		$\epsilon$ is infinitesimally small. The soon-to-be divergent integrals now have exponents of $\left(1/\epsilon\right)$ and diverge in the $\epsilon \rightarrow 0$ limit. These divergences can now be removed by adding appropriate \emph{counterterms} in the Lagrangian (Williams 2015, p.14).}, where the terms quadratic in $\Lambda$ are absent. Let me repeat for the reader's convenience that the dominant contribution to $\Delta m^2$ without a regularization scheme in \S2.2 was $\frac{g^2}{16\pi^2}\Lambda^2$ and is thus quadratically divergent in the cutoff scale.
Their conclusion that scalar field thus do not exhibit a strong sensitivity on UV physics is however wrong.
Alvaraz-Gaum\'{e} and V\'{a}zquez-Mozo (2012, p.236) falsely alert that 
\begin{center}
as [dimensional regularization] regularizes the quadratic divergences to zero it seems that the whole hierarchy problem results from using a clumsy regulator, and that  by using [dimension regularization] we could shield the Higgs mass from the scale of new physics.
\end{center}
This (alleged) elimination of quadratic divergences has led to many misconceptions in the literature;
\begin{itemize}
		\item{} Wells (2012, p.56) is sceptic as to whether the gauge hierarchy problem constitutes a worrisome naturalness problem: ``Since the quadratic sensitivity of $\Lambda$ can be removed, the hierarchy problem is nothing but a ``knowing-just-enough-to-be-dangerous naive way to look at the Standard Model.'' According to Wells, the hierarchy problem does not cry out for an explanation.
	\item{} Wells concludes in the subsequent year (2013, p.1) that naturalness in general is not to be taken seriously: ``In dimensional regularization, the most common
	technique to book-keep the infinities of the quantum
	field theory, there is no quadratic divergence explicitly
	manifested in the effective theory. Thus, as the argument
	goes, Naturalness is only a fuzzy philosophical notion,
	and should not be taken seriously.''
\end{itemize}
I will now argue why these arguments are blatantly false. 

Firstly,
the hierarchy problem exists regardless of the choice of regulator.\footnote{Although it would not cry out for an explanation if naturalness turns out not to be a fruitful guiding principle; this will be discussed later (in \S \ref{sec:withoutnaturalness}.}.  Using the dimensional regularization scheme one now obtains 
\begin{equation}\label{key}
\Delta m^2 = \frac{g^2}{16\pi^2}M^2,
\end{equation} 
where $M$ is the heavy fermion mass which has been integrated out of the EFT because $M \gg \Lambda$ (Williams 2015, p.15). 
We therefore run into the same AoS naturalness problem; the scalar obtains a physical mass whose order of magnitude is well beyond the cutoff scale of the EFT because $m_{\text{phys}}^2 \sim M^2$.

Secondly, Alvaraz-Gaum\'{e}, V\'{a}zquez-Mozo and Wells have equated naturalness problems with ``problems of divergences'' (quadratic for the Higgs mass, quartic for the cosmological problem - to be  discussed in \S \ref{sectioncosmologicalconstant}). Although  problems of divergences are intimately related to problems where IR theory exhibits an unduly sensitivity on UV physics,
naturalness problems should be thought of as
the former rather than the latter. The reason is obvious; different regularization schemes may remove the quadratic divergences of integral but the physical Higgs mass retains its unduly 
sensitivity on UV physics.
Although the exact dependence of parameters on UV physics is generically dependent on the regularization scheme, performing calculations with a different regulator does not eliminate the hierarchy problem at all. This is easily recognized when adhering to the AoS naturalness notion. This is less philosophically taxing than requiring that the Higgs should divergence in terms of positive powers of the \emph{cutoff} - what matters according the AoS notion is that \emph{the physical mass is expected to exceed the UV cutoff}.
\subsubsection*{Why the hierarchy problem matters}
I have argued that the electroweak hierarchy problem arises because an elementary scalar particle couples to a \emph{heavier particle}. The heaviest hitherto observed fermion is the top quark, but heavier fermions may exist in nature (a plethora of them would exist if our world is described by SUSY).
We are therefore led to concluse that additional hierarchy problems will emerge
whenever exotic massive
states, whose mass terms are invariant under the EW gauge group, are coupled
to the Higgs field (Giudice 2013, p.2).
These exotic particles would render the Higgs naturalness problem more severe.

Moreover, this hierarchy problem need not be unique - the Higgs
boson is indeed the only scalar particle we must reckon with in contemporary particle physics but additional fundamental
scalar particles\footnote{Although the currently observed Higgs may still be a composite particle.}
may be required by various patterns of high-energy symmetry breaking. 
It has been argued that additional fundamental scalars may be found in both the dark energy and axion sectors, and scalar `inflatons'  may have been responsible for cosmological inflation in the early universe (Giudice 2018, p.10). Each of these particles would initiate their its naturalness problem whenever it couples to heavier non-scalars.
\subsection{The cosmological constant
	\label{sectioncosmologicalconstant}
}
Both astronomers and astrophysicists have presented compelling evidence
that the energy density of the universe is largely in some hitherto unfamiliar form;
dark matter constitutes about 30\% of the energy supply through some non-baryonic pressureless matter and about 60\% of the total energy is in some form with negative pressure known as dark energy (Dine 2007, p.72).
The nature of dark energy remains
mysterious, but many scientists surmise that it is equivalent to the 
the cosmological constant (CC) or equivalently the energy of the vacuum.\footnote{Observations in themselves do not directly entail that the dark energy must be	a cosmological constant. Many models have been built that relax the lorentz invariance assumption of dark matter, for instance by considering slowly varying scalar fields $\phi^i$ (Burgess 2013).}
\nn
What is the vacuum's energy, and how does it gravitate? This deceptively simple question
is relevant because a vacuum energy would affect the way the universe expands in
an observable way (Burgess 2012, p.2).
If the CC is indeed nothing but the vacuum energy, it introduces a naturalness problem where IR physics depends extemely delicately on UV physics. This problem resides in the fact that the cosmological constant is a relevant operator with even lower mass dimension than the Higgs mass operator in the effective Lagrangian underlying General Relativity
\begin{equation}\label{key}
\mathcal{L} = \int d^4x \sqrt{-g} \left[ \mathcal{R} + \Lambda_{\text{CC}} \right],
\end{equation}
where $\mathcal{R}$ is the Ricci scalar (which is irrelevant for our discussion) and the the important part for my discussion of naturalness
is the cosmological constant.
\nn
The instability of relevant operators is due to large radiative corrections. The instability, of course, becomes worse the lower the dimension of the relevant operator is. The Higgs mass is determined by $\phi^2$ (mass dimension 2) but one
can consider an even lower dimensional operator 
in EFTs: the operator $\mathcal{O}_2 \equiv \phi^0 = 1$ has mass dimension 0. This corresponds to an overall additive constant in the Lagrangian (equivalently: the zero-point
energy in the Hamiltonian) which corresponds to a physically irrelevant additive shift in the energy in flat Minkowski
space,\footnote{This shift is not always irrelevant - the Casimir effect hinges on differences in vacuum energies.} 
yet this term becomes physically
relevant when the theory is coupled to gravity: it corresponds to the cosmological constant term $\Lambda_c$ in the Einstein field equations (Duncan 2012, p.578). 
The corresponding coefficient $a_{-2} = c_{-2} \Lambda^4$ 
(see equation \ref{eq:couplingsa_n})
receives quartic contributions from integrating out field modes from the UV scale $\Lambda$ down to the (much lower) scale of astronomical phenomena:
$$\Lambda_{\text{CC}} = \Lambda_{\text{CC,bare}} + c\Lambda^4,$$
where $c$ is of order unity (Donoghue 2007, p.5).
Taking the cutoff scale at the Planck mass
one would expect 
(because of dimensional analysis)
that $\Lambda_{\text{CC}}
\simeq M_P^4$ which is roughly 120 orders of magnitude
beyond the experimentally verified $10^{-47} \ \text{GeV}^{4}$ (Giudice 2013, p.3, Dine 2015, p.19).\footnote{\label{footnote:	CC}
Why the theoretical value of $\Lambda$ should be huge is well-understood from a theoretical point of view. Both the  zero-point
energy for bosons and the energy of the filled Dirac sea for fermions contribute to the vacuum energy and contribute like (Dine 2015, p.72)
$$\Lambda_{\text{CC}} = {1 \over 2} \sum_{\omega} (-1)^{F} \int_0^\Lambda {d^3k \over (2\pi)^3 } \sqrt{k^2 + m_\omega^2},$$
where all terms are \emph{quartically divergent}.
The sum is over all particle species $\omega$ (including spins), which is just the sum of
the zero-point energies of the oscillators (of each momentum).
Imposing a cutoff scale around the Planck scale, one obtains $\Lambda_c \simeq 10^{54} \text{ GeV}^{4}$.

Experimentally, the vacuum energy density would have to have this small but nonzero value in order to be consistent with i)
	tests of the distance-redshift relation (using distant supernovae) 
	and ii)
	a spatially flat universe (Planck Collaboration 2013). Einstein’s equations imply the total energy density must be \emph{critical}. One can consequently infer what the contribution from the vacuum energy should be.}  
This discrepancy remains one of the biggest mysteries in fundamental physics. 
\nn
When imposing naturalness on the CC, an energy threshold should exist around $2.4 \times 10^{-3}$ eV
beyond which new physics should emerge.\footnote{$\Lambda_{\text{CC}}^{1/4} $ $= \sqrt[\leftroot{-2}\uproot{2} 4]{
		3.3 \times 10^{-11} \ \text{eV}^{4} }
	$ $=
	2.4 \times 10^{-3}$ eV.	
	The value of the CC is small in particle physics
	units, but it is substantial in units relevant to the present cosmological epoch in cosmology. Indeed, the CC
	has only ``recently'' 
	(in the past few billion years)
	become important, and it will keep dominating
	the energy density henceforth.} The lack of any
empirical evidence whatsoever for the existence of new physics around this energy scale lends support to the conjecture that the cosmological constant does not respect the naturalness criterion, as was already noted by 't Hooft (1979); IR physics is unduly sensitive on UV physics.
\nn
A daunting attempt at such a theory is being made by Sundrum (1999),
the possibility of new dynamics at this extremely low scale
which modifies gravity but not any other interactions is however highly contrived (Donoghue 2007).
\nn
Since a viable theory of quantum gravity is still beyond our reach, it may be infeasible to tackle the CC problem in contemporary physics. 
The energy scale of the cosmological constant lies in the deep IR and several attempts to tackle the problem from an IR perspective have been putward in the literature where modifications of gravity at large
distances
are studied.\footnote{A well-studied IR modification is one where inverse
	d'Alembertian corrections ($1/\square$) are introduced in the gravitational action. An example of such a theory is massive gravity, where an $m_{\text{gr}}^2 \mathcal{R}/ \square^2$ term is added to the Einstein-Hilbert action (see Maggiore 2014, Jaccard \et 2013 and Modesto, Tsujikawa 2013). These theories typically exhibit a reduction in the gravitational field at large
	distances (Conroy 2017, p.17). Modifications of gravity in the deep IR are unlikely to solve the CC problem, because the zero-point energy is UV divergent (it thus already receives small contributions from the IR).}
The conceptual problem associated with the CC however come from quantum fluctuations in the deep
UV, see footnote \ref{footnote:	CC}. Attempts to solve this UV problem with the traditional methods used for the Higgs are
hopeless - the UV cutoff is significantly lower than any energy scale entering gravitational theories (and GR is known to be accurate up to energies greatly exceeding $\Lambda_{\text{CC}}$, so the quadratic sensitivity of the CC on UV physics cannot be eliminated).

Usually, large disagreements with data immediately
spawn many candidate theories, designed to remove the discrepancy without
ruining the success of other experimental tests. Yet, for the vacuum energy there is not even one candidate modification of the EFT which provides a potentially viable way to reconcile observations with predictions (Weinberg 1989, Carroll 2001,
Burgess 2013, p.3). This is fundamentally different from the strong-CP violation angle $\theta$ and the higgs vev $\nu$, for which dynamical solutions have been proposed to obain natural values for these parameters.
\nn
Multiverse solutions have the most promising prospects of solving the CC puzzle. A discussion of these solutions is however deferred to \S \ref{sec:withoutnaturalness}, since these explanations do not aim to restore naturalness in the EFT. Adherents of the multiverse instead aim to explain why the CC could only have very particular values in our universe, \eg because of anthropic arguments.
\section{The current role of naturalness in theory choice \label{sec:solvehierarchy} }
\setlength{\epigraphwidth}{0.75\textwidth}
\epigraph{While the discovery of the Higgs boson has completed a chapter in the book of science, it
	has also crystallized new conceptual problems that cry out for solutions.}{\textit{Gian Francesco Giudice 2017, p.7}}
\noindent
Although the CC problem poses a bigger naturalness problem than the Higgs vev, 
the difference in order of magnitude is relatively unimportant. What is important is that both problems cry out for an explanation, however, no viable dynamical solutions have been proposed for the CC problem as was already discussed in \S \ref{sectioncosmologicalconstant}. I will therefore explain natural solution to the electroweak hierarchy problem in this section.
\nn
The unnaturally low Higgs mass requires delicate cancellations between the sum of quantum corrections and bare mass term. Many solutions to this hierarchy problem have been put forward in the literature during the last decades and  I will briefly discuss the most prominent hypotheses. 
The collective effort of physicists to understand the role of naturalness in electroweak physics was rewarded
with extraordinary and fundamentally new ideas in theoretical physics. Attempts to resolve the Higgs naturalness problem have led to theoretical models in which both the fundamental structure of spacetime and the pattern of fundamental forces were rigorously reformulated (Giudice 2017, p.3). 
Examples of such theoretical models are
technicolor, supersymmetry, stringy models with extra dimensions, composite Higgs models, all of which enforce
a fundamentally new interpretation of physical reality. These theories may entail a huge particle zoo at smaller distances, imply that all particles are ultimately composed of single strings (with different excitations) or that our 3+1-dimensional world is merely an effective field theoretic approach of a higher-dimensional world at smaller distances. All these naturalness-inspired models have opened new avenues in the exploration of the
particle world and many models predict new physics around the TeV scale. 
These avenues have been enthusiastically pursued by high-energy experimental research, where
the LHC (due to its advanced technology: $\sqrt s = 13$ TeV during the last run)
will be the critical step in this journey.
No evidence of dynamics which is able to cure the ``sickness'' (w.r.t. naturalness) of the Higgs has been
found up to date, while particle colliders are approaching the energy threshold where new physics is expected to be hidden.
\nn
The requirement that BSM models be
natural has heavily influenced model-building in high-energy physics for nearly 40 years.
The measured value $m_\text{H} = 125.15 \pm
0.24$ GeV is however a bit high for SUSY models and a bit low for composite models (like little Higgs models), making theoretical interpretations rather uncomfortable because these options are neither unequivocally favored,
nor excluded (Buttazzo \et 2014).
\subsection{Little Higgs models}
Among the approaches in which the idea of dynamical electroweak symmetry breaking is purportedly reconciled with the existence of a light Higgs particle (compared to the scale of interactions)
are models in which the Higgs is considered to be an approximate Goldstone boson (see Arkani-Hamed \et 2001, Georgi and Kaplan 1984 and Kaplan \et 1984). The pivotal
idea of these \emph{little Higgs models} is that new strong interactions arise at an energy
scale $\mathcal{M}$ (around the TeV scale), where these strong interactions possess an approximate global symmetry which is moreover
spontaneously broken, giving rise to Goldstone bosons through Goldstone's theorem (this is of course reminiscent of what happens in the electroweak theory). One of the bosons corresponding to this broken symmetry acts as the SM Higgs
boson. The gauge interactions of the SM necessarily break these symmetries and
give rise to a Higgs potential (Dine 2013, p.10). 
The induced Higgs mass term is way too large, although this can be circumvented by introducing
additional features in such a way that the approximate symmetries are violated only
by \emph{combinations of additional gauge symmetries}.
This may prove challenging, but does not rule out the possibility of little Higgs models in itself.
Accounting for the SM fermion masses (electrons, muons, \emph{et cetera}) is however incredibly challenging as well (Arkani-Hamed \et 2001), implying that these models most likely do not comply with the successes of the SM.
\subsection{Supersymmetry}
There are three
phenomenological considerations which have traditionally been taken as the primary leitmotivs for weak-scale
supersymmetry (SUSY): the Higgs hierarchy problem, grand unification of gauge couplings at high energies (Dimopoulos and Georgi 1981 \& Dimopoulos \et 1981) and weakly interacting massive particles (WIMPs) as dark matter candidates (Giudice 2008). I will mainly focus on the relation of SUSY to the Higgs hierarchy problem, although the `WIMP miracle' will be discussed (albeit briefly) too because it bears important implications for the viability of low-energy SUSY.
\nn
It would be desirable to construct theories in which the masses of elementary
scalar fields are protected by symmetries in order to explain the low mass of the Higgs boson.
A possible solution to the hierarchy problem in the SM imposes a novel spacetime symmetry which predicts the existence of a new
boson for every known fermion, and a new fermion for every known boson, where the new particles are called supersymmetric partners (spartners).

In globally supersymmetric models, there are two basic
types of multiplets (Dine 2015, p.13): 
\begin{enumerate}[i), nolistsep]
	\item {}
chiral multiplets, consisting of a complex scalar and a spin-1/2 fermion, 
\item {} vector multiplets which consist of a chiral fermion and a gauge boson.
\end{enumerate}
The RG flow
of superpartners would eliminate the quadratic divergences of the Higgs mass, because fermionic and bosonic loops have \emph{opposite signs and identical magnitudes} (for unbroken SUSY) - 
the quadratic divergence of the Higgs would therefore no longer be problematic because the contribution from the supersymmetric top quark counteracts this divergence (subdominant divergences are cancelled in an identical fashion by SUSY particles). Supersymmetry
is however at best a broken symmetry, because no superpartners that have been observed so far, indicating that 
(chiral/vector) multiplets exhibit no mass degeneracy  (Anderson and Casta\~{n}o 2008, p.1). 
\subsubsection*{Broken SUSY} 
Superpartners
could have gauge invariant mass terms if supersymmetry is softly broken, and
the masses of supersymmetric partners (\emph{spartners}) can be made arbitrarily heavy provided that the scale of
supersymmetry breaking is sufficiently high. Since the masses of particles and sparticles differ, the total vacuum contribution to the Higgs mass does not cancel exactly; one can however show that the naturalness problem is strongly ameliorated compared to the SM hierarchy problem. 
The quadratic divergence (SM) is replaced by a logarithmic divergence in broken SUSY (Feng 2004, p.6);
\bse
\begin{align}
m_{\text{H}}^2 &= m_{\text{H},0}^2 - \frac{1}{16 \pi^2} \lambda^2 \Lambda^2 + \frac{1}{16 \pi^2} \lambda^2 \Lambda^2 - \cdots  \label{eq:1a} \\
 &\simeq m_{\text{H},0}^2 
 -
 \frac{1}{16 \pi^2} \left[
 m_{\tilde{f}}^2 - m_{f}^2
 \right] \ln \left(
 {\lambda / m_{\tilde{f}}}
 \right). \label{eq:1b}
\end{align}
\ese
and is based on two assumptions:
\begin{enumerate}[i), nolistsep]
	\item{} The first \emph{assumption} is that the dimensionless couplings $\lambda$ of SM and spartners are \emph{identical}, otherwise we would not obtain the logaritmic divergence in equation \ref{eq:1b} from equation \ref{eq:1a}.\footnote{Equation \eqref{eq:1b} holds true both for broken and unbroken SUSY. One can easily check that the all quantum corrections vanish if $m_f = m_{\tilde{f}}$ for all particles since $m_{\tilde{f}}^2 - m_f^2 = 0$.}
	
	\item {} A \emph{second} (and progressively more contrived) \emph{assumption} is that the masses of spartners are \emph{not too far above the electroweak scale}; otherwise even the logarithmic divergence gives rise to an unaccountably large discrepancy between theoretical and experimental results for $m_{\text{H}}^2$.
\end{enumerate}
Feng argues that ``the quantum corrections are reasonable even for very large
$\Lambda$'' (Feng 2004, p.6),
but this quote is outdated.
LHC data has shown that quantum corrections exceed the physical value of the Higgs in the most viable broken SUSY ($\simeq$ TeV) models (Giudice 2018).
\nn
As for the CC problem; in SUSY we get a result proportional to the fourth power of the supersymmetry-breaking
scale, which is an ameliorated result in comparison to the fourth power of $M_P$. Nonetheless, even the lowest conceivable SUSY breaking scale is many orders
of magnitude larger than the observed dark energy (Dine 2015, p.20). The mismatch remains 60 orders of magnitude in the ``best case scenario'' and 90 orders of magnitude for high-scale SUSY models (Feng 2004, p.15). Only deranged scientists would claim that SUSY could provide a suitable solution to the CC problem.
\nn
Although the prospects of a natural SUSY Higgs and/or CC appear dim,
naturalness as a guiding principle has led to two distinctive, impressive and concrete successes
within the minimal supersymmetric standard model (MSSM): the prediction of gauge
coupling unification (Marciano and Senjanovic 1982 and Dimopoulos and Georgi 1981) at an energy scale $E_{\text{GUT}} \sim 2 \times 10^{16}$ GeV, tantalizingly close to the
Planck scale (Dimopoulos, Raby and Wilczek 1981). 
This is a great success because unification is pursued by physicists and the three forces of the SM can now be united in a single theoretical framework.
The second success is that many SUSY theories contain a well-suited
\emph{dark matter candidate} in the lightest neutralino, which is customarily called the Weakly Interacting Massive Particle (WIMP) miracle. Its cosmic density is remarkably close to what it should be (Arkani-Hameda and Dimopoulos 2005, p.1) but the particle has not been observed yet.\footnote{Hossenfelder (2017a) considers this to be another failures of naturalness. This cannot be said, because the weakly interacting character of dark matter renders it incredibly hard to detect. To my knowledge, the WIMP miracles cannot be ruled out by current data.}
\nn
The primary leitmotiv to consider Little Higgs models and SUSY models was to ameliorate the Higgs hierarchy problem, however, these solutions suffer from a lack of expected experimental results. 
Now that the utility of naturalness is becoming more and more contrived (and many physicists (\eg Hossenfelder 2018) argue that we should abandon this heuristic) 
we may ask ourselves whether it would be worth studying non-natural BSM models (which could potentially solve other problems of the SM - \eg the massive neutrinos problem). Interestingly, 
several BSM models have been put forward in which naturalness is entirely abandoned, including 
split supersymmetry 
and high-energy SUSY.
\subsection{Concluding remarks}
We were led to conclude in this chapter that the AoS notion of naturalness enables scientists to recognize both the successes and failures of naturalness successfully. In the latter case, it also allows us to get a grasp of how badly naturalness was broken - no quantative measure of naturalness is needed for that purpose. I afterwards discussed two dire violations of naturalness, both of which are known 
to occur in our well-entrenched EFTs
for several decades already. Physicists have long taken it for granted that one or the other natural solution to the Higgs vev problem would be experimentally verified at CERN. 

No experimental clues for SUSY or other dynamical mechanisms which could stabilize the electroweak scale have been found thus far.
The Standard Model also reigned supreme at the most recent run of the LHC,
posing a problem for most natural BSM models according to which ``exotic" BSM particles should show up around the TeV scale. This has led physicists and philosophers of quantum field to critically reflected upon the principle and has led to a surge of distrust in validity of the principle (\eg Dine (2015), Williams (2015), Giudice (2017), Hossenfelder (2018)). 

The absence of novel particles around the TeV scale
was already predicted by several physicists and has, among other things, led to the proposal of Split Supersymmetry in the literature
(Giudice and Romanino 2004, Arkani-Hamed and Dimopoulos 2005, Arkani-Hamed,  Dimopoulos,  Giudice
and Romanino 2005). In this supersymmetric model, \emph{gauge coupling
unification and dark matter are taken as leitmotiv} and the (un)naturalness of the model is deemed entirely unimportant: 
\begin{center}
	Why should the new states appear at the weak scale, if the hierarchy
	problem is not the guiding principle of the analysis? (Giudice and Romanino 2004, p.3)
\end{center}
This theory has several interesting features and
quite distinctive signatures at collider experiments. If Split Symmetry would be confirmed by the LHC, it would thereby
provide tangible experimental evidence against the naturalness principle (Giudice 2008, p.18).
The same line of thought has been pursued in models of SUSY with heavy superpartners around the PeV scale (Wells 2003, Wells 2004) whose spectrums, similar to Split Supersymmetry, are fundamentally different from those of ``standard BSM models'' whose greatest leitmotiv has been naturalness.
\nn
The potential violation of naturalness induced by the unnaturally low Higgs mass
and 
dim success rates of
most natural BSM models
force us to reflect upon the 
fruitfulness of the
naturalness guiding principle in our pursuit to find the laws governing nature at the smallest distances. We have reached a phase of confusion in which the physics community is divided into two camps.
The utility of
naturalness is disputed by scientists and philosophers in the first camp, whereas adherents of the naturalness principle in the second camp contend that the laws of nature are natural and construct novel natural BSM models. In this chapter, I will put forward several reasons as to why naturalness, when conceived of as a prohibition of correlations among widely separated scales, should no longer heavily influence model-building in high-energy particle physics. 
\nn
Both Hossenfelder, Williams and Giudice have argued that naturalness may become otiose in the near future of particle physics. Their comparable conclusions (differing in whether naturalness \emph{may cease to} or \emph{will necessarily cease to} remain a powerful guiding principle) conceal that the underlying arguments are fundamentally different. These are in fact mutually incompatible and deserve further attention. A shared belief of these physicists is that there are no fundamental reasons as to why nature should respect naturalness at even smaller distances, but they disagree about whether naturalness problems may nonetheless still have natural solutions despite the fact that nature does not necessarily respect naturalness.
\begin{itemize}
	\item {}	Hossenfelder contends that naturalness serves to spread the dogma that nature is inherently beautiful.\footnote{Ironically, Hossenfelder does not consider naturalness to be an aesthetic criterion.} The problem of naturalness according to Hossenfelder is that ``ill-defined criteria like naturalness that don't even make good working hypotheses'' (Hossenfelder 2017).
	Naturalness has no solid foundation according to Hossenfelder  
	and suffers from problems of vagueness due to its alleged ill-defined character.
	\item {}
	Giudice 
	and Williams
	contend that it has been perfectly reasonable for physicists to have embraced the naturalness principle in the past. 
	They consider the naturalness principle to be well-defined and argue that ``naturalness is a physically transparent and well-motivated expectation within the effective field theory framework" (Williams 2015, p.26).
	The principle may however have outlasted its use. Williams and Giudice argue that the most fundamental theories of nature may not be describable by EFTs and that this would entail that the most fundamental laws may be unnatural (Giudice 2018). Additionally, they allow for the possibility that the unnatural Higgs mass is not ``cured" by a natural BSM theory. Their argumentation requires a delicate discussion, this will be done in \S \ref{sec:withoutnaturalness}.
\end{itemize}
I have already invalidated Hossenfelder's criticism regarding 
the ill-defined definition of
naturalness in \S \ref{sec:naturalnessnotaesthetic}. Giudice and Williams are right not only in that naturalness is well-defined but also in claiming that it is a well-motivated expectation within EFTs - I will put forward arguments in favor of both claims in the next section. The motivation for naturalness however rests on historical precedence (since decoupling of widely separated scales are ubiquitous in nature) and I will argue that dimensionless parameters are not necessarily natural.
\nn
Also Giudice and Williams claim that naturalness could cease to provide fruitful guidance in the future of particle physics. In \S \ref{sec:withoutnaturalness}, I will put forward several reasons as to why this may indeed be the case.  If naturalness would however no longer be a powerful guiding principle we would have to reshape our inventory of useful guiding principles.
I will assert in \S \ref{sec:withoutnaturalness} that the remaining naturalness problems can potentially be solved by selection criteria. The sensitivity of IR parameters on UV physics would be retained in these unnatural explanations, but the plausibility of these parameters can be inferred from a multiverse perspective.
I then show in \S  \ref{subsecdecnotsat} that the Decoupling Theorem does not underwrite an ontology of QFT in which all field theories can be described by EFTs. I then introduce the possibility of unnatural parameters being describable in field theories with UV/IR mixing in \ref{subsec:UV/IR}.\footnote{Of course, this is rampantly speculative. Future LHC runs may prove me wrong.}
\section{Can we do without naturalness?
	\label{sec:withoutnaturalness}
}
If particles colliders (most likely the LHC) rule out dynamical solutions to the Higgs naturalness problem at the weak scale, the naturalness problem is obviously
not eradicated. Instead, it would be imperative to ``reconsider the guiding principles that have been used for decades to address the most
fundamental questions about the physical world'' (Giudice 2017, p.5) and contemplate whether naturalness remains fruitful to describe physics at the smallest distances. 

Giudice, who formerly was a strong advocate of naturalness and published the prominent \AC measure in the 1990s, has already lost his confidence in naturalness as a guide. He claims 
in his novel (2017) paper
that this guiding principle will probably become otiose, because there are no compelling reasons why the most fundamental laws of nature should respect this criterion. Giudice distinguishes the current ``naturalness era'' from the ``post-naturalness era'' (in which naturalness is no longer a guide) which theoretical physics is currently entering according to Giudice.

Other physicists have already started to contemplate 
how our current concept of naturalness could potentially be reshaped
in post-natural times.
The explanation for the Higgs naturalness may not lie in
a to-be-discovered symmetry but instead lie within the cosmological evolution of our universe, similar to Dicke's explanation for Dirac's large numbers (Giudice 2017, p.8). 
Multiverse solutions are capable of providing a fundamentally different (non-natural) resolution than dynamical solutions
to the two greatest naturalness problems: that of the CC and the Higgs vev.
\subsection{Selection criteria in the multiverse \label{subsec:selection}}
Physicist Nima Arkani-Hamed (2012) has claimed that
``If neither supersymmetry nor any other sort of natural
solution...appears in the data...[t]his would...give theorists a strong
incentive to take the ideas of the multiverse more seriously.''
I will provide arguments in favor of his claim in sections \ref{subsec:selection} and subsequently discuss the implications of unnatural parameters in \S \ref{subsec:UV/IR}.
\nn
On the one hand there are the Higgs vev and strong CP-violating angle 
fow which dynamical solutions have been proposed (without experimental verification) and on the oter hand there is the CC problem, for which few good dynamical suggestions have been proposed (Burgess 2013). This has put the principle of naturalness under considerable stress and physicists have started to explore other kinds of solutions to these naturalness problems. One of the most popular approaches is an embedding of our universe into a multiverse (like in landscape naturalness) and impose suitable selection criteria.
\nn
Weinberg, who was inspired by suggestions by Banks (1985) and Linde (1986),
has argued that the observed universe is
part of a larger structure, which has  been coined the ``multiverse.'' 
Parameters in the multiverse are not ``god-given parameters'' but 
take a range of values, which are essentially randomly distributed (Dine 2015, pp. 20-21). 
I have already mentioned that compactification of string theory with fluxes provides
a model in which a landscape emerges. The number of
possible flux types is typically large (a couple of hundreds or more), and these fluxes can range over
many discrete values. There may be many stationary points of the
effective action for each choice of flux (Dine 2013, p.23). One can therefore build up an exponentially large number of states. This is the setting for Weinberg's solution of the CC problem.
One could in fact take an inventory of the multiverse and infer which universes would contain observers (that is to say, humans). 
Weinberg has argued that a low value for \lcc \ is perfectly reasonable when imposing the \emph{anthropic principle} (AP).
\subsubsection*{Anthropic selection}
The first application of AP to explain unnatural parameters was
due to Linde and Weinberg (1987, 1996), applied to the cosmological constant even
before the this parameter was known to be non-zero. 
Weinberg in particular
gave a physical condition noting that, if the CC was much
different from what it is observed to be, galaxies could not have formed.\footnote{I recall that \lcc is one of the ingredients that governs the expansion
	of the universe. Although its value is small when expressed in the natural units employed in particle physics, it has recently become the \emph{dominant energy density in our universe}. In the previous cosmological epochs of the universe (including the ``important'' epochs in which galaxies have formed) the CC has been small compared to the matter and radiation densities. If the CC would have been a lot higher, it would have dominated in earlier cosmological epochs already.}

Although we lack knowledge of the exact requirements for life to develop, many patches of the multiverse \emph{indubitably} exclude the possibility of life inhabiting that universe.\footnote{The anthropic selection is intimately related to the fact that it is not surprising that life in the universe can only develop in extraordinary small areas of our total universe (at the surfaces of planets which contain liquid water, the surface temperature of the planet probably also plays an important role).} 
Weinberg argues, in my opinion persuasively, that a minimum requirement for life to develop should be that the universe contains galaxies (Dine 2015, p.21). This minimum requirement is already sufficient to predict an ``unnaturally low'' \lcc.
\nn
What Weinberg argues is that the low value of the CC has been \emph{essential} for the development of life in our universe; cosmic structures would not have formed had it been much larger. The rapid expansion of the universe would have a destabilizing effect of structure formation: ``[i]f it had been of its natural scale of $(10^3\text{ GeV})^4$
the universe
would have collapsed or blown-apart (depending on the sign) in a small
fraction of a second'' (Donoghue 2007, p.7). 
In order for the universe to expand slowly enough that galaxies
could form, the CC would have to be within roughly an order of magnitude of its observed value. In other words, an anthropic selection criterion would explain why we find ourselves in this particularly rare patch of the multiverse with a small, positive \lcc.
\nn
This antrophic line of thought has also been pursued in the context of the Higgs vev - the electroweak TeV scale may be
anthropically selected as well (see Lawrance \et 2014 for an extensive overview). 
Life would be impossible for all vevs excepts those lying in a fairly narrow window around $\upsilon = 246$ GeV - its experimentally verified value.
Life requires the complexity which comes
from having many different atoms available to build organisms, 
but these would be absent in the majority of universes with $\upsilon \neq 246$ GeV (Donoghue 2007, p.7).
The reason for this is that both quark and lepton masses ($m_i$) increase linearly with the Higgs vev through $m_i = \Gamma_i \upsilon / \sqrt 2$ (where $\Gamma_i$ are the Yukawa couplings).
Several universes would facilitate the emergence of stars (rendering life possible, at least in principle), but their compositions would be fundamentally different from those in our universe. Nucleosynthesis would take place at a way slower rate, preventing a high abundance of heavy elements (Dine 2015, p.22 and Lawrance \et 2014) and for most values of the vev elements higher than hydrogen would not exist at all (Donoghue 2007, p.7). 
\nn
The argument is roughly as follows:	
all SM parameters except for $\upsilon$ are held fixed. As the vev increases in magnitude, all the quark masses grow so the neutron and proton masses increase as well. The neutron-proton mass splitting gets larger and this value is calculable (Hoyle \et 1953, p.9). The most model-independent
constraint on the vev comes from the value when the neutron-proton
mass splitting becomes larger than the $10$ MeV per nucleon that binds
these nucleons into nuclei (Hogan 1999, p.9). This happens when the vev is about 5 times the observed value (Donoghue 2007, p.7). When this happens, all bound neutrons will decay to protons (Agraval \et 1998). Nuclei consisting solely of protons are known to be unstable so nuclei will fall apart into hydrogen. The complex nuclei will no longer exist (Agrawal \et 1998).

This discussion of AP is of course far from exhaustive. The important message that I aimed to convey to the reader is that naturalness problems may be solved \textbf{while not retaining naturalness}
by applying AP to an elaborate landscape of string vacua.\footnote{Tighter constaints takes into account the calculation of the binding energy, which decreases
	as the vev increases. Another constraint (of comparable
	strength) stems from the need to have stable deuterium. Deuterium
	has been involved in the formation of heavier elements in primordial nucleosynthesis and subsequently in nucleosynthesis in stars. The interested reader is referred to Agraval \et (1998) and Donogue (2007) for great discussions of these topics.} 
\nn
An intriguing feature of the Higgs
naturalness problem is that it can be rephrased from an overt problem of UV sensitivity to a 
less trivial problem of \emph{criticality}.
It is worth exploring this line of thought to further assess the fertility of the multiverse framework.
\subsubsection*{Self-organized criticality} 
\begin{figure}
	\centering
	\includegraphics[scale=0.3]{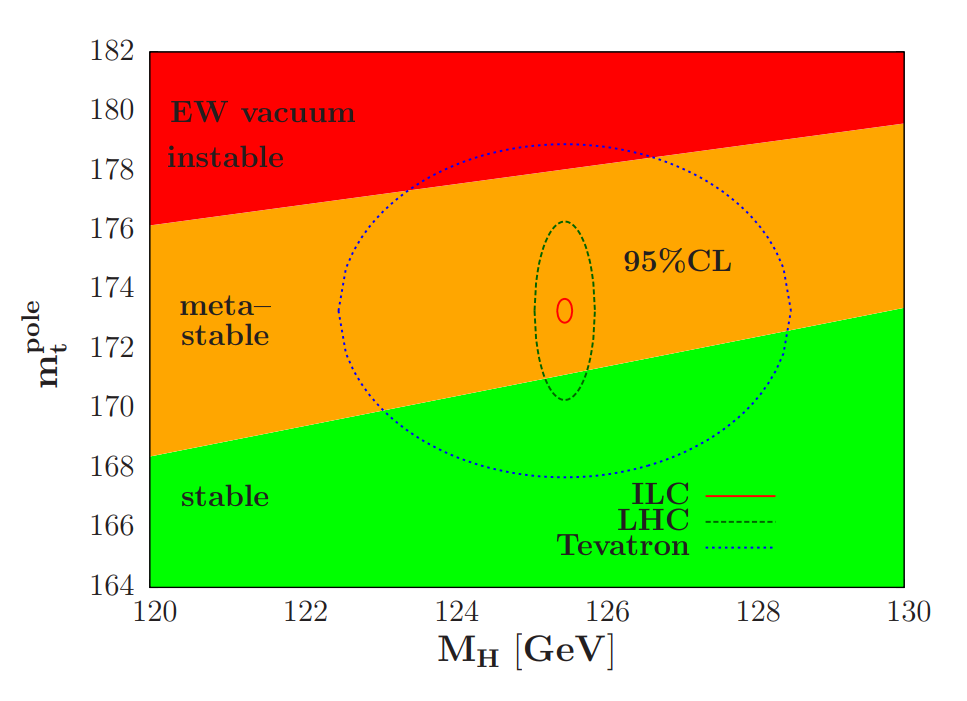}
	\caption{\ti{The current top quark
			and Higgs mass measurements (at the Tevatron and LHC) show that the electroweak vacuum is most likely metastable.} Figure taken from Alekhin \et (2013).}
	\label{fig:higgsvacuum}
\end{figure}

Let us recall that the Higgs potential
contains a wealth of information\footnote{
	In fact, one can write	
	$$V(\varphi) =  (\text{const}) - { m_{\text{H, bare}}^2 \over 2} \varphi^2 + {\lambda \over 4} \varphi^4,$$ where $(\text{const}) = \varphi^0$ is a constant which introduces a physically irrelevant shift in the vacuum energy. When coupled to gravity, it becomes the cosmological constant so one can even relate the cosmological constant problem to the Higgs potential.}; the constant term introduces the cosmological constant problem, the quadratic term the naturalness problem and the quartic term the instability of the SM electroweak vacuum.
It has been pointed out by Giudice and Rattazzo (2006) that the vacuum expectation value of the Higgs vev $\upsilon$
(which is a combination of the parameters of the Higgs potential: $\upsilon = \sqrt{-\mu^2/\lambda}$)
is on the verge of a phase transition. The \emph{order parameter} (see Huang (1987, \S 16.1) for a clear discussion of this concept) of this phase transition can be expressed in terms of the coefficient $\mu^2$ which occurs in the Higgs potential. If $\mu^2$ is negative the symmetry is spontaneously broken,  if $\mu^2$ is positive the symmetry is restored and $\mu^2 = 0$ defines the \emph{critical point} (Giudice 2008, p.18).\footnote{One may recognize this from the Ginzburg-Landay description of ferromagnetism, indeed both cases are completely analogous. In that case, the dipoles are oriented randomly for $T>T_C$ where $T_C$ is the (critical) Curie temperature. The dipoles however align and induce a spontaneous magnetization for temperatures below the Curie temperature (Giudice 2008, p.18), breaking the rotational symmetry of the Lagrangian.}  Due to dimensional analysis, we would expect $\abs{\mu^2}$ to be of the order of magnitude $\Lambda^2$, however, its experimentally verified value is remarkably close to the critical point $\mu^2 = 0$.
\begin{center}
	Let
	us consider the phase diagram for electroweak symmetry breaking, in the space of these
	parameters [the few free parameters of the fundamental theory at the Planck scale which correspond to
	the discrete set of vacuum expectation values of the moduli fields in string theory]. Over the bulk of the parameter space, $\abs{m_{\text{H}}^2}$ is expected to be of order $M_P^2$, and
	therefore either $\expval{H} \sim M_P$ or $\expval{H} = 0$ depending on the sign of $m_{\text{H}}^2$. The hierarchy problem
	is now simply stated as: if the critical line separating the two phases is not special from the
	point of view of the fundamental theory, why are the parameters in the real world so chosen
	as to lie practically atop the critical line?
	(Giudice and Rattazzi 2006, p.1)
\end{center}
The measured Higgs vev is thus quite special in the landscape, so it is statistically unlikely when a uniform distribution  function would be employed. 
The vev is however \emph{near-critical}: the Higgs vacuum does not reside in the
configuration of minimal energy, but instead in a \emph{metastable state} close to a phase transition. Not only the Higgs vev, but also the (physical) Higgs mass is on the verge of a phase transition (see figure \ref{fig:higgsvacuum}). This has been pointed out by Buttazzo \emph{et al.} (2014, p.39) who thoroughly studied this condition of near-criticality in terms of the SM
parameters at $M_P$.
The Higgs potential might undergo a self-tuning process, where critical boundaries 
(the critical values occurring in the Higgs potential)
act as
attractors, just as in the mechanism of \emph{self-organized criticality} (Bak \emph{et al.} 1987). 
\nn
This description was rather technical and an example would probably help the reader grasp the idea of self-organized criticality. Several systems have been discovered in classical physics which have the tendency to evolve into critical states, even if outside agents do not force them to evolve in that direction. That is roughly what self-organized criticality amounts to (see Bak \emph{et al.} 1987 for a more extensive discussion of the subject).
The archetypal example to illustrate this idea is a heap of sand to which grains of sand are continually added. As the pile grows
\begin{center}
it reaches a condition where catastrophic sand slides occur
after the addition of just a single grain 

(Giudice 2008, p.19)
\end{center} 
and this implies that the system arranges itself to be (and remain) near-critical as long as more sand grains are added to the heap. Many seemingly unrelated phenomena in classical physics exhibit this kind of self-organized criticality: traffic jams, river bifurcations, the distribution of earthquake intensity and many more applications of self-organized criticality have been discovered (Giudice 2008).

The concept is now reintroduced in the \emph{quantum realm} and the situation becomes much more complicated in this context.
Could a pattern of self-organized criticality with respect to electroweak scale
bring the SM towards the condition of a large hierarchy $G_F/G_{\text{N}} \ll 0$? If such a kind of self-criticality would operate in nature, it would obviously
not respect the naturalness criterion. Much more interestingly, it would \emph{not be captured by an effective-theory
approach} because the microphysical description would
fail to appropriately account for large-scale correlations (Giudice 2008). 
One should keep in mind that, in the case of the sand pile, individual grains
would be used to describe the avalanches in the sand pile. These would be useless because avalanches occur at all scales (between the
size of a single grain and the size of the whole pile). Avalanches of all sizes obey a power-law distribution
and therefore the dynamics of the system can no longer be understood in terms of individual
grains.
\nn
The near-criticality of the Higgs led Buttazzo \emph{et al.} (2014, p.33) to conclude that:
\begin{center}
	an explanation of near-criticality almost necessarily requires the
	existence of an underlying statistical system. This drives us towards the multiverse as the
	most convincing framework in which one can address the issue
\end{center}
and it is indeed true that an ensemble of theories is required to realize an idea of self-organized criticality in the context of the electroweak scale. 
The multiple near-criticalities of the Higgs parameters could potentially be solved in the multiverse, however, the ontology behind this explanation of the Higgs vev would be fundamentally different from that underlying anthropic considerations. The process of selection our universe would be fully determined
\emph{by dynamics} in the case of self-organized criticality.
\nn
Other explanations for the hierarchy problem which do not aim to reduce the unnatural character of the Higgs boson (contrary to SUSY) are \emph{dynamical selection} (\eg Graham \et 2015) and \emph{statistical selection} (\eg Arvanitaki \et 2017) which are beyond the scope of this thesis (although the corresponding methodologies are briefly summarized in this footnote\footnote{\label{footnote:selectrioncriteria}
	\begin{itemize}[nolistsep]
		\item {}	\textbf{Dynamical selection} explains the near-criticality by means of an additional field:
		When introducing a slow-rolling field (the relaxion) which scans the Higgs mass parameter, the electroweak phase transition could have generated a back-reaction which is able to stop the evolution near the critical point.
		\item{} \textbf{Statistical selection} explains the near-criticality by adhering to the Anthropic Principle:
		A plethora of vacua is generated in the vicinity of a special value for electroweak
		breaking. The cosmological constant can \emph{only in that regime} scan finely enough to be within
		the narrow range of anthropically allowed values. Interestingly, this relates the 
		seemingly unrelated
		hierarchy problem and CC problem.
	\end{itemize}	
}
for the interested reader).
Neither of these explanations have anything fundamental to do with 
the original AoS naturalness. 
Whichever ever of these mechanisms (if any) is truly responsible for the near-criticality of the Higgs parameters, it certainly does not reduce the degree of unnaturalness of the scalar Higgs. \emph{The Higgs boson retains its unduly sensitivity on UV physics, although the reason for its fine-tuned parameters can potentially be understood in the multiverse approach.} 
\subsubsection*{Implications and potential dangers of selection criteria}
These explanations for unnatural parameters
enfeeble the philosophy that natural theories are required in order to account for otherwise unnatural parameters of the universe. Instead, miscellaneous selection criteria in the multiverse might account for the unnatural values of dimensionless parameters in our universe. 

If any of aforementioned selection criteria would be operative, \emph{the naturalness principle would not be
	operative}. For instance, the unnaturally low value of the electroweak scale and CC would not predict any new physics around the meV and TeV scales, contrary to dynamical solutions to these problems.
The SM may remain accurate up to significantly higher energies, since AP could select a fairly narrow window of universes in which the Higgs mass is low. Sensitive dependence on UV physics would no longer be problematic because the universe \emph{may be fine-tuned} through a fortuitous cancellation between the bare terms and quantum corrections, because this is required to enable intelligent life.\footnote{I recall that fine-tuning problems do not run counter to deeply embedded features of theories (Williams 2015, p.21).}
\nn
It is fair to note
that not all cosmologists and philosophers
of science assent to the utility of
the anthropic principle, or even to its
legitimacy.
There are (at least) three potential dangers in adopting selection criteria upon a multiverse:
\begin{enumerate}[i), nolistsep]
	\item {} Selection criteria may turn into a dogma which impedes scientists from searching for more traditional, dynamical solutions
	to naturalness problems. Among others, Woit (2017) has raised this concern ``The problem with such things as string-theory multiverse theories is that `the multiverse did it' is not just untestable, but an excuse for failure. Instead of opening up scientific progress in a new direction, such theories are designed to shut down scientific progress by justifying a failed research program.''
	\item {} It remains debatable under which exact circumanstances life could develop. What is life?\footnote{Some 
	scientists claim that life is necessarily carbon-based life, while others take very liberal views on this topic. Physicist Sean Carroll argues that \emph{life may be just information processing}.} (Barnes 2012, p.3) As Carroll (2006) puts it ``We don’t even fully understand life on this planet, nor do we understand it on the other planets in theuniverse that hold life (if any), nor do we understand it within the other possible universes (if any).''
	\item {} It is difficult to falsify multiverse hypotheses along with its selection criteria.
\end{enumerate}
Regarding the first worry:
selection criteria have indeed found many applications. These hypotheses are arguably fruitful to elucidate essential properties of cosmological inflation.\footnote{Inflation suffers from several fine-tuning problems (see Planck Collaboration (2013)) which are actually intimately related to the exlusion of life-hostile environments.}
A calculation by Carter shows that if gravity had been stronger or weaker by only one part in $10^{40}$, life-sustaining stars like the sun could not have formed (Collins 1999, p.49). The AP might therefore plausibly explain the strength of gravitational interactions.
\nn
Lo and behold, the multiverse does not always come to the rescue. There is at least one remaining naturalness problem which does not permit any kind of environmental justification.
While the small cosmological
constant (Weinberg 1987), the small Higgs vev (Agrawal \et 1998), the existence of dark matter and the scarcity
of anti-matter (Tegmark \et 2005) may plausibly be justified by selection criteria, the
strong CP problem\footnote{This problem revolves around the smallness of the neutron electric dipole moment as was discussed in \S \ref{sectiontechnicalnaturalness}.}
certainly cries out for a dynamical explanation. This has been pointed out by Vecchi (2014, p.16), Dine (2013, p.22), Donoghue (2004, 2007) and Banks \et (2004):
\begin{center}
	For any possible value of $\theta$ in the allowed range from $0 \rightarrow 2\pi$, there
	would be little influence on life. The electric dipole moments that would be
	generated could produce small shifts in atomic energy levels, but would not
	destabilize any elements. Even if a mild restriction could be found, there
	would be no logical reason why $\theta$ should be as small as $10^{-10}$. \emph{Therefore the
	idea of a multiverse does nothing to solve this fine-tuning problem}. (Donoghue 2007, p.8, my italics)
\end{center}
The last claim of Donoghue's is not quite true. It is true that the strong CP naturalness problem has to be solved in a more conventional way, for instance by promoting $\theta$ into a dynamical variable (most probably an axion) as is done in the Peccei-Quinn (PQ) solution of the strong CP problem (Peccei and Quinn 1977), but the multiverse may still prove valuable in this context. What I want to stress in my following discussion is that 
it is realized by surprisingly few physicists that the idea of a multiverse is already part of the established toolkit of theoretical physics. And this toolkit can certainly provide solutions to naturalness problems. 
\nn
The unnaturally low value of the parameter could be selected by some to-be-discovered underlying dynamics in the PQ solution (Giudice 2017, p.8).  
According to Peccei and Quinn (1977), the potential for the axion is generated \emph{after the QCD chiral phase transition} which took place in a quite epoch in the history of the universe. Before
this phase transition, the axion (and thereby the effective value of $\theta$) effectively took randomly different values in different
parts of space, which were subsequently blown up by cosmic inflation into different patches of the universe. \emph{This PQ-axion actually incarnates the idea of the multiverse}. 
The
energy which is stored in the axion oscillations around its minimum 
depends on the initial conditions of the axion and thus varies from patch to patch. We are led to conclude that, although $\theta$ will eventually converge to zero everywhere,
the different patches of the ``multiverse" contain non-equivalent physical information.
The PQ-solution is still studied in contemporary physics, \eg by Jeong and Shin. These authors claim that ``[t]he [PQ] relaxation mechanism, which solves the electroweak hierarchy problem
without relying on TeV scale new physics, \emph{crucially depends on how a Higgs-dependent backreaction
potential is generated}'' (Jeong and Shin 2017, p.1, my italics). This Higgs-dependent backreaction actually amounts to invoking the \emph{dynamical selection criterion} which was introduced in footnote \ref{footnote:selectrioncriteria} - this has to be studied in the context of the multiverse as well. 

What I aimed to convey in this section is that, notwithstanding the diffidence many physicists feel
towards the multiverse (Giudice 2017, p.8), it is a constructive and fertile framework which could certainly help scientists to reformulate open questions in fundamental physics, including naturalness problems. This does not entail that physicists would no longer look for more convential solutions to naturalness problems.\footnote{The fact that many physicists feel diffident about the whole multiverse approach already implies that many physicists would actually prefer solutions in which unnaturally low dimensionless parameters are protected by symmetries.}
\nn
I now come back to the potential dangers ii) and iii) which are lurking in the multiverse approach, which will be discussed very briefly. 
\begin{enumerate}[i), nolistsep]
\setcounter{enumi}{1}
	\item{}  \textbf{It remains debatable under which exact circumanstances life could develop.} Indeed, this is true. Physicists are however often very careful when applying AP and only impose \emph{minimal requirements for life to emerge}, including the existence of planets. It is well possible that both the Higgs vev and CC take \emph{maximum values which are still consistent with more refined antrophic requirements}. In this case one might argue that naturalness is still operative, since \emph{these unnatural values would take the values closest to their natural values} (in other words, their \emph{most natural value}) while allowing for intelligent life as well.
	\item{} \textbf{It is difficult to falsify multiverse hypotheses along with its selection criteria.} 
	Hossenfelder criticizes the multiverse theorem on this ground: ``Without making contact to observation, a theory isn’t useful to describe the natural world, not part of the natural sciences, and not physics" (Hossenfelder 2018b).
On the other hand, I think that the fact that the two most severe violations of naturalness can be solved by selection criteria does entail that multiverse solutions should be taken seriously. Some physicists even assert that these solutions constitute ``evidence for the multiverse'' (Hall and Nomura (2007). The position that I advocate is \emph{agnosticism}. Perhaps the multiverse is out there, perhaps it is not, but at the time being we have no means at our disposal to refute any of these statements.\footnote{Perhaps we will at the future. It is not unlikely that a multiverse could leave an observable imprint in the CMB for instance.}
\end{enumerate}
The multiverse entails that several parameters may be unnatural and it is worth studying the implications of such unnatural parameters. Would they entail that widely separated scales not necessarily decouple in QFTs?
The sensitive behavior of relevant operators to UV physics seems to undermine a central dogma of EFTs, namely \emph{that phenomena occurring at widely separated energy scales should decouple}. If the possibility of IR parameters being sensitively dependent on UV physics is truly excluded by the Decoupling Theorem (DT), one may wonder how violations of naturalness are possible. I will now argue that the DT is not generically applicable to EFTs and, even more interestingly, \emph{that even in those cases in which the DT holds true, it is too weak to underwrite a decoupling of energy scales unless the EFT in question is natural}.
\subsection{The Decoupling Theorem is not necessarily satisfied
\label{subsecdecnotsat}
}
As was already outlined in \S \ref{subsecdecoupling}, Appelquist's and Carazzone's Decoupling Theorem 
tells that, given a QFT that satisfies certain conditions (these will be discussed soon), one can construct an appropriate EFT for low-energy physics by integrating out all the fields with masses above a high-energy threshold. The absence of these fields in the EFT results into a modification of the bare terms, since these take into account all effects 
from high-energy physics beyond the cutoff scale. 
\nn
I have already discussed the DT and alluded to the fact that this theorem has been criticized on several grounds in the literature.
The Decoupling Theorem indeed licenses an ontology of QFT
which is characterized by such quasi-autonomous domains, \emph{provided that the following two conditions which are required for the proof of the DT always hold true in EFTs}:
\begin{enumerate}[1), nolistsep]
	\item{} The ``full theory'' (UV theory) is perturbatively renormalizable (Bain 2013).
	\item{} 
	The ``renormalization condition'' which Hartmann (2001, p.283) referred to is a Mass-dependent Subtraction (MS) regularization scheme (Georgi 1992, p.3).\footnote{
		The reason why this regularization scheme is said to be mass-dependent is the following. Each divergent integral $\int_0^\infty d^Dp \ \kappa(p),$
		where $D$ is the spacetime dimension and $\kappa(p)$ is a function of momentum, can be rewritten as \emph{an infinite sum plus a finite sum}:
		$$ 
		\int_0^\infty
		d^Dp \ \kappa(p) 
		= \int_0^\Lambda d^Dp \ \kappa(p)
		+ \int_{\Lambda}^{\infty} d^Dp \ \kappa(p).$$	
		By introducting renormalization constants,
		the infinite term can be absorbed into a redefinition of the parameters. These constants turn out to be dependent on the heavy masses ($m_2$) appearing in the UV theory; hence this regularization scheme is referred to as a \emph{mass-dependent scheme}.} 
	The second assumption is therefore that a MS regularization scheme suffices to prove that UV physics decouples.
\end{enumerate}
I will now elucidate why the assumptions underlying the DT are not met by generic EFTs.
\subsubsection*{Perturbative renormalizability}
The EFTs of the SM are perturbatively renormalizable, implying that this assumption is met for energies $E < \Lambda$.
The EFT approach however dictates that one includes all terms which are consistent with Lorentz invariance, the symmetries of the theory and cluster decomposition (Hartmann 2001). These terms may be either renormalizable or non-renormalizable, the non-renormalizable terms are (due to dimensional analysis) suppressed by powers of the cutoff $\Lambda$ and can therefore be ignored in most cases, except in the deep UV. This however does entail that ``[v]irtually any EFT...will be non-renormalizable'' (Williams 2016, p.26).\footnote{Although the SM is perturbatively renormalizable, nonrenormalizable terms may appear at higher energy scales. An example of such a nonrenormalizable term is the baryon violation term (abbreviated `bvt')
	$$ \mathcal{L}_{bvt} =  M_{\text{BSM}}^{-2} Q \sigma^{u} \bar{u}^* L \sigma_\mu \bar{d}^*, $$
	which allows processes such as $p \rightarrow e\pi$. 
	The meaning of the symbols is unimportant for my discussion, except for $M_{\text{BSM}}$ which denotes a ``beyond the standard model scale''.
	Experiments which take place deep underground have set
	limits of order $10^{33}$ years on this process, the scale $M_{\text{BSM}}$ must therefore be
	larger than $10^{15}$ GeV. Other irrelevant operators which may occur in the SM are terms which give corrections to the muon magnetic moment (called $\mathcal{L}_{g-2}$) and a flavor-changing term $\mathcal{L}_{\text{fc}}$ (both of which are extensively discussed in Dine (2007, \S4.1)), these nonrenormalizable terms are however heavily suppressed at energies below $ 100$ TeV.
} 
It is currently unknown whether the UV-completion of the SM is perturbatively renormalizable, so the first criterion may not be satisfied by more fundamental laws of nature.\footnote{Additionally, the quantum formulation of Einstein gravity is perturbatively nonrenormalizable and would thus not necessarily exhibit a decoupling of energy scales.}
The DT is not  generically applicable to EFTs, violations of AoS naturalness could consequently occur in nature.
\subsubsection*{The MS regularization scheme}
The SM is however perturbatively renormalizable (and perhaps its UV-completion turns out to be renormalizable too) so we ought to assess whether a mass-dependent (MS) regularization scheme is sufficient to prove that widely separated energy scales decouple (I recall that this is the second assumption of the DT).
Whether a MS regularization scheme suffices has been disputed by several scientists (\emph{e.g.} Bain 2013, Williams 2015, Hossenfelder 2018).
Georgi (1993, p.225) and Manohar (1997, p.329) have argued that MS regularization schemes
do not allow one to ignore the potentially infinite number of
irrelevant terms in the effective Wilsonian action to be extended from tree-level
calculations to higher-order loop corrections. 
Not being able to truncate the EFT to a finite list of terms renders MS regularization schemes particularly cumbersome. Many authors consequently recommend adopting a mass-independent
renormalization schemes ($\overline{\text{MS}}$) like dimensional regularization and minimal subtraction (Williams 2015, p.31).\footnote{EFTs which have been constructed by means of mass-independent renormalization schemes are called \emph{continuum EFTs} (Georgi 1993).}
 One however encounters heavy fields $m_\xi \gg \Lambda$ in $\overline{\text{MS}}$ regularization schemes, the implications of which are profound:
\begin{center}
	The presence of heavy field terms in an effective action
	employing a mass-independent renormalization scheme prevents the application of
	the Decoupling Theorem (Bain 2013, p.7)
\end{center}
If the validity of DT depends on a particular choice of renormalization scheme, should we still believe in a decoupling of scales in QFTs?
The answer to this question is not straightforward. 
\subsubsection*{Assessing the validity of the Decoupling Theorem}
Let me first put forward arguments as to why we should still believe in this decoupling of scales.
Our surmise that vastly separated scales do separate in QFTs actually finds ample support - even in the context of mass-independent regularization schemes. Calculations which are done in renormalization schemes wherein a decoupling of scales is not manifest yield remarkable agreements with experiments \emph{when the decoupling of scales is put in by hand} (Williams 2015, p.26). This great empirical success justifies this artificially introduced decoupling of scales in $\overline{\text{MS}}$ renormalization schemes according to Georgi (1993) and Bain (2013). I think that does not amount to a genuine justification, yet it does show that the laws of nature \emph{typically allow to be described in terms of quasi-autonomous regimes}.
The SM is remarkably insensitive to physics at the smallest distances because its predictions have been confirmed with ravishing precision. This has led
Cao and Schweber (1993), 
Castellani (2002) and
Bain (2013) to endorse the ``quasi-autonomous domains'' ontology according to which high energy physics generically decouples from IR physics.
According to these physicists, decouplings of scales are thus ubiquitous not only in classical physics abut also in QFTs. 
\nn
My claim is that their arguments are internally inconsistent.
The violations of AoS naturalness introduced in \S \ref{sec:violations} made it particularly salient that parameters in unnatural low-energy field theories ($\nu$, \lcc) are unduly sensitive to minute variations of UV parameters.\footnote{One may find this statement feeble since ``a variation of UV parameters is not a physical process" (Hossenfelder 2018a) - fortunately one can also claim that the IR parameters are unduly sensitive on the exact value of $\Lambda$.} This vitiates a notion of quasi-autonomous domains according to which the scalar is ``stable and not disturbed by whatever happens at higher energies'' (Williams 2015, p.32).
The Higgs's strong sensitivity on UV physics is ineliminable in every unnatural theory containing a Higgs and the same holds true for the CC. After all, these correlations among widely separated energies is exactly what gives these effective theories the moniker ``unnatural.'' 
\nn
The implications are twofold, the first pertains to inability of the DT guarantee a decoupling of scales. 
Although the SM satisfies the conditions of the DT, the unnatural
field theories of the
SM are not necessarily decomposable into quasi-autonomous domains. The DT holds true in this case, but still allows an interscale sensitivity which the DT allegedly prohibits. This implies that failures of naturalness have ontological significance\footnote{This was already noticed by Williams (2015, p.32)} - \emph{the DT is too weak to underwrite an ontology of QFT which is characterized by quasi-autonomous domains}. This decoupling of scales only occurs when one additionally imposes that the corresponding EFT \emph{ought to be natural}.

The second implication is of grand importance. Since irrelevant parameters retain an extreme sensitivity on UV physics, \emph{unnatural parameters might not be described accurately by EFTs}. The sensitivity of unnatural parameters to UV physics does, of course, not vanish when switching from a mass-depedent regularization scheme tot a mass-independent scheme. This refutes Cao's and Schweber's claim that the DT implies an ontology of 
successive self-contained energy shells whose union describes all layers of nature - in order to describe relevant operators a field theory is required whose allowed energy values range from the IR to the deep UV.\footnote{This also invalidates Bain's (2013, p.241) claim that ``continuum EFTs are, by themselves, capable of supporting an ontology of quasi-autonomous domains."} The extreme interscale sensitivity of relevant operators is blatantly irreconcilable with any reasonable definition of ``quasi-autonomy'' (Williams 2015, p.33).
\subsection{UV/IR mixing
\label{subsec:UV/IR}
}
 Although it is customarily assumed in classical physics that widely separated scale decouple in classical physics, nobody knows why this should be the case. 
 Exception to the rule could consequently occur and,
 in fact, scales do not decouple in \emph{chaos theory}.\footnote{Two identical sets of initial condition measurements - which according to Newtonian physics would yield identical results - in fact lead to vastly different outcomes in chaos theory, because the macroscopic trajectories are highly dependent on microphysics.}
The previous successful applications of the \emph{idea of insulation} in EFTs may be considered circumstancial -
no relevant operators had been encountered before the discoveries of the Higgs boson and cosmological constant. 
\nn
Their unnatural values run counter to the central dogma of EFTs according to which one can 
study
successive `self-contained energy-shells' and make sense of these separately. 
A theory which manifests an UV/IR-interplay would be beyond the grasp of EFTs, since this holistic character would violate its inner logic.\footnote{Novel fundamental scalars may await us at even smaller distances and I think it is important to accept the possibility that several dimensionless parameters in our universe may simply be unnatural and exhibit delicate interscale sensitivities.}
In EFTs, the fundamental theory in the far UV knows nothing about the theory in the IR.
\begin{center}
From this perspective, one might hope to work around the hierarchy problem
by linking the far UV and the far IR. This would represent a sharp departure from
effective field theory, and the challenge is to make the departure well posed. (Craig 2017, p.42)
\end{center}
Fortunately, concrete examples of theories exhibiting UV/IR mixing have been put forward in the literature which will enhance our understanding of this 
mixing. Arguably the most prominent example has been introduced in the context of
quantum gravity and aims to describe microscopic black holes. 
We can imagine accelerating two protons to $E \sim 10^{18}$ GeV (Planckian energies) and
have these composite particles collider in order to create quantum black holes - they would be Planck-length-sized (Giudice 2017, p.11). 
We might hope to probe distances shorter than the Planck length $l_p$ in the convential way - by increasing the energy of the two protons even further (so above the Planck energy). 
More energetic protons mean more massive black holes, which have larger radii.
When increasing the energy of these particles we would, contrary to our intuition, 
create
\emph{larger and
	larger black holes}.
We would then probe larger instead of smaller distances and hence depart from the UV-regime. ``Exciting the theory in the UV really probes the
physics of the IR.'' (Craig 2017, p.43)
\nn
The possibility of quantum gravity exhibiting an UV/IR interplay has recently been confirmed by Lust and Palti (2017). 
Lacking a viable theory of quantum gravity, we cannot understand yet
what bearing this might have for
problems of naturalness such as
the electroweak hierarchy problem. It would therefore be fruitful to look
into QFTs which exhibit a similar UV/IR mixing structure.
UV/IR mixing is a generic feature of non-commutative geometries and has been discussed in great detail by Minwalla, Van Raamsdonk and Seiberg (2000, \S6). 
Craig (2017, pp.42-45) provides a good discussion and interpretation of this field theory and explains why it exhibits UV/IR-mixing. What is important for my discussion is that such a kind of field theory represents a striking breakdown of a Wilsonian EFT, where
Wilsonian renormalization fails terribly (Craig 2017, p.44). This kind of field theory would require physicists to abandon their habit of viewing QFTs as being decomposable into quasi-autonomous energy shells.
\nn
Craig argues that field theories which exhibit an UV/IR-interplay may describe physics at the smallest distances
\begin{center}
	[i]f the hierarchy
	problem is solved by radically new ideas in quantum field theory, I am willing
	to bet that it will proceed somewhere along these lines of UV/IR mixing. (Craig 2017, p.45)
\end{center}
The same may be true for the cosmological constant, whose energy scale is set by $\Lambda_{\text{CC}} = 2.4 \times 10^{-3}$
eV affects physics at large distances, so
in the deep IR.\footnote{I am talking about astronomical and cosmological distances here.}
Modifications of gravity at large
distances have been put forward, which attempt to tackle the CC problem from an IR-perspective. The
\emph{conceptual problem} of the cosmological constant however come from \emph{quantum effects in the deep
UV}, see footnote \ref{footnote:	CC}. Giudice has argued succinctly that
\begin{center}
	This confusion among scales is at the basis of the problem. It is a big source of confusion
	because the systematic approach of effective field theories has taught us how to separate
	energy scales in successive shells and make sense of the theory at each shell separately. The
	cosmological constant seems to resist this approach. Naturalness is an offspring of effective
	field theory and so it is not surprising that the difficulty we are encountering with the effective
	theory description leads to a problem with naturalness (Giudice 2017)
\end{center}
and I have hopefully compelled the reader that we cannot rule out this possibility. AoS naturalness has been incredibly successful in the past, but there are no guarantees that it will reign supreme in the laws of nature which describe physics at the smallest distances.
\subsubsection*{Future LHC runs might restore our confidence in naturalness}
Testing the naturalness principle at and beyond the weak scale at LHC will indubitably have
far-reaching consequences for particle physics, and will probably be decisive in whether or not naturalness will remain a useful guiding principle for the evaluation of field theories at the smallest distances.
The discovery of new physics around the TeV scale could possibly restore our faith in the naturalness criterion. 
So far, the message from the LHC has not been encouraging, since the data collected at the previous run was in complete accordance with the Standard Model. 
The final verdict will however have to wait for higher-energy LHC runs in the near future.

\chapter{Conclusions
\label{chapterconclusions}
}
The most cogent definition of naturalness is provided by an autonomy of scales notion.
Autonomy of scales naturalness 
\begin{enumerate}[i), nolistsep]
	\item {}
provides a uniform notion which undergirds a myriad prominent naturalness conditions, 
\item {} is a reasonable criterion to impose on EFTs, 
\item {} the successes and violations of naturalness are best understood when adhering to this notion of naturalness.
\end{enumerate}
Rather than an aesthetic criterion, AoS is deeply rooted in the logic of EFTs and guarantees that widely separated scales decouple. Although the fruitfulness of EFTs hinges on this assumption, this decoupling of scales is not at all entailed by the effective field theory framework since quantum corrections may be arbitrarily large. AoS naturalness implies small quantum corrections and is therefore a reasonable criterion to impose on EFTs, with justification on both theoretical  and empirical grounds.
The AoS dogma guarantees that effective field theories yield meaningful results (theoretical justification) and is often respected because
effective field theories are ubiquitous and ravishingly successful in physics (empirical justification).
\nn
Naturalness has however been criticized on several grounds, where  criticisms of naturalness as an ``aesthetic criterion'' (Donoghue 2007) and naturalness being ``ill-defined'' (Hossenfelder 2018a) 
most notably
undermine the scientific character and validity of the principle. I have invalidated these criticisms.
I have asserted that naturalness can only be criticized for being an aesthetic/sociological principle when formal measures of naturalness and their use in physics communities are conflated with the central dogma of naturalness - the former may indeed be argued to be sociologically-influenced and somewhat arbitrary - however these formal measures of naturalness are significantly less successful than AoS naturalness. I have argued that AoS naturalness is deeply rooted in the logic of effective field theories and may therefore not be said to be a purely aesthetic principle.
The latter allows physicists to recognize both natural and unnatural parameters when the principle is used along with the renormalization group equations and it was reasonable for physicists to endorse this naturalness principle - on both theoretical and empirical grounds. The principle has been successful in the past and, among other things, enabled Gaillard and Lee (1974) to predict the mass of the charm quark before its experimental discovery.
\nn	
Yet the two most severe violations of naturalness have not been unequivocally solved by natural extensions of the well-entrenched effective field theories of the Standard Model and General Relativity. A myriad of dynamical solutions has been proposed for the Higgs naturalness problem without experimental verification, while very
few viable solutions have been proposed for the cosmological constant problem. 
These failures have put naturalness under considerable stress and physicists have started to explore other kinds of solutions to these naturalness problems. One may argue that particle physics has therefore entered a phase of crisis in which the fruitfulness of both naturalness and effective field theories has become more contested. While naturalness has fueled most of the BSM model during the last few decades, physicists are starting to surmise that parameters ought not necessarily be natural. Split supersymmetry is an early example of a theoretical framework in which the naturalness guide was abandoned, nowadays many physicists surmise that the SM may remain accurate up to energies far above the TeV scale. Selection criteria in the context of the multiverse have actually revealed that both the Higgs vev and \lcc \ could not have taken natural values in this universe.
Unnatural parameters would then have ontological consequences for quantum field theory.
\nn
We have learned that the laws of both classical and quantum physics usually allow to be decomposed into quasi-autonomous energy shells. information of UV physics is in those cases completely irrelevant in order to describe IR physics. 
No compelling reasons have been put forward as to why widely separated scales should generically decouple, although some scientists argue that this decoupling of scales in the quantum realm is entailed by the Decoupling Theorem. 
I have however shown that the Decoupling Theorem \emph{does not underwrite an ontology of quasi-autonomous energy shells even in those cases where EFTs meet the assumptions of this theorem}. 
Chaotic phenomena provide an exception to this rule in classical physics and exceptions may be found in the quantum realm as well.
Physicists are starting to endorse the possibillity that effective field theories may cease to remain accurate in order to describe unnatural parameters.
Field theories exhibiting some kind of UV/IR interplay may solve this problem.
An example of a field theory with UV/IR mixing has been proposed by Minwalla \et (2000) and such field theories may find applications in particle physics and quantum gravity due to aforementioned violations of naturalness.
\nn
My discussion of field theories with UV/IR interplay is, of course, rampantly speculative but it is certainly an explanation worth investigating. We should also keep in mind that future LHC runs may restore our faith in AoS naturalness when deviations from SM predictions are found at smaller distances.

\chapter*{References}
\addcontentsline{toc}{chapter}{References}
\begin{itemize}[nolistsep]
	
	\item{ATLAS Collaboration (2012), \textit{Observation
		of a New Particle in the Search for the Standard
		Model Higgs Boson with the ATLAS Detector at the
		LHC.} Phys. Lett. B \tb{716P}, pp. 1-26 [arXiv:1207.7214].}
	
	\item{}  Agrawal, V., Barr, S.M.,  Donoghue, J.F. and  Seckel, D. (1998), \textit{The Anthropic principle and the mass scale of
	the standard model}, Phys. Rev. D \tb{57} (1998) 5480 [arXiv:hep-ph/9707380].
	
	\item {} Aizenman, M. (1982),	\ti{Geometric Analysis of $\phi^4$ Fields and Ising Models i and ii}. Communications in Mathematical Physics, 86(1): pp. 1-48.	

\item {} Albert, D.Z. (2000), \ti{Time and Chance}, Cambridge, MA: Harvard University Press.

\item {} Alekhin, S., Djouadi, A. and Moch, S. (2013), \ti{The top quark and Higgs boson masses and
	the stability of the electroweak vacuum}.Phys. Lett. B\tb{716} (2012) 214.

	\item{Anderson, G.W. and Casta\~{n}o, D.J. (1994) \textit{Measures of fine tuning}, 	Phys. Lett. B \textbf{347}:300-308,1995.}
	
	\item{Anderson, G.W., Casta\~{n}o, D.J. and Riotto, A. (1997), \textit{Naturalness lowers the upper bound on
			the lightest Higgs boson mass in supersymmetry}. Phys. Rev. D, \tb{55}:2950–2954, 1997. \url{https://arxiv.org/pdf/hep-ph/9609463.pdf}.}

\item {}  Appelquist, T. and Carazzone, J. (1975), \textit{Infrared singularities and massive fields}, Phys. Rev. D \tb{11},
28565 (1975).

\item {Aristotle ($\sim 340$BC), \ti{The Nicomachean Ethics}. University of Chicago Press; Reprint edition (April 23, 2012).  }

\item{Arkani-Hamed, N. (2012), \textit{The future of fundamental physics}. Daedalus 141(3), pp.53–66.}
	
\item{} Arkani-Hamed, N., Cohen, A.G. and Georgi, H. (2001), \ti{Electroweak symmetry
breaking from dimensional deconstruction}. Phys.Lett., B\tb{513}:232–240, 2001.

	\item{Arkani-Hamed, N. and Dimopoulos, S. (2005),
	\textit{Supersymmetric Unification Without Low Energy Supersymmetry And Signatures for Fine-Tuning at the LHC},	
		JHEP 0506 073 (2005).}
	
		\item{
		Arkani-Hamed, N.,  Dimopoulos. S, Giudice, G.F. 
		and Romanino, A. (2005), \textit{Aspects of Split Supersymmetry}. A., Nucl. Phys. B \tb{709} 3 (2005).
}

\item{
Arkani-Hamed, N.,  Motl, L., Nicolis, A. and Vafa, C. (2007), \textit{The String landscape, black holes and gravity as the
weakest force}, JHEP 0706 (2007) 060 [arXiv:hep-th/0601001].
}

\item {} Arvanitaki, A., Baryakhtar, M.,  Huang, X., Van Tilburg, K. and Villadoro, G. (2014). \ti{The last vestiges of
	naturalness}. Journal of High Energy Physics 2014(3), 22.
	
	\item{Arvanitaki, A., Dimopoulos, S.,  Gorbenko, V., Huang, J.  and Tilburg, K. (2017), \textit{A small weak scale from a small
		cosmological constant}, JHEP 1705 (2017) 071 [arXiv:1609.06320].}
	
	\item {Aspden, H. and Eagles, D.M. (1972), \textit{Aether theory and the fine structure constant}.  Phys. Lett. A \tb{41}, 423 (1972).}
	
\item {Atlas Collaboration (2012), \ti{Observation of a new particle in the search for the Standard Model Higgs boson with the ATLAS detector at the LHC}, Phys. Lett. B\tb{716} (2012), pp. 1-29.}	
	
	\item{Athron, P. and Miller, D.J. (2007), \textit{New measure of fine tuning}. Phys. Rev. D \tb{76}:075010, 2007. }	
	
\item {}  Atick, J.J. and Witten, E. (1988), \ti{The Hagedorn Transition and the Number of Degrees of Freedom of
	String Theory}, Nucl. Phys. \textbf{B310} (1988) 291.

\item{Banks, T.C.P.} 
\begin{itemize}[nolistsep]
	\item {} (1985), \ti{Quantum Gravity, the Cosmological Constant and All That...}
	Nucl. Phys., B\tb{249}:332, 1985.
	\item {}
(2008) \textit{Modern Quantum Field Theory}. Camrbidge University Press, Cambridge, 2008.
\end{itemize}

\item{Banks, T. and  Dixon, L.J. (1988), \ti{Constraints  on  String  Vacua  with  Space-Time
	Supersymmetry.}
	Nucl.Phys., B307: pp. 93–108, 1988.}

\item {} Banks, T., Dine, M. and Gorbatov, E. (2004), \ti{Is there a string theory landscape?}
JHEP, 0408:058, 2004.

	\item{Banks, T.  and Seiberg N. (2011), \textit{ Symmetries and Strings in Field Theory
			and Gravity}, Phys. Rev. D \tb{83} (2011) 084019 [arXiv:1011.5120].}

	\item{Barbieri, R. and  Giudice, G.F. (1988) 
	\ti{Upper Bounds on Supersymmetric Particle Masses.}
	Nucl. Phys, \textbf{B306}, 63 (1988).}

\item {Barrow, 	J.D. (2002), \ti{The Constants of Nature: The Numbers That Encode the Deepest Secrets of the Universe}. Vintage; 1st edition.}

	\item{Barrow J.D. and Tipler F.J. (1986), \textit{The Anthropic Cosmological Principle}. Oxford: Clarendon
		Press.}

\item{} 	Batterman, R. (2013), \textit{The tyranny of scales}. In Oxford handbook of philosophy of physics. Oxford: Oxford University Press.

	\item{Biswas, T.,  Koivisto, T. and Mazumdar, A. (2013), \textit{Nonlocal theories of gravity:  the flat space propagator}, 	arXiv:1302.0532 [gr-qc].}

\item{Biswas, T., Gerwick, E., Koivisto, T. and Mazumdar, A. (2012), \textit{Towards singularity and ghost free theories of gravity.} Phys. Rev. Lett. \tb{108}:031101, 2012.}

	\item{Bondi, H. and Gold, T. (1948). \textit{The Steady-state Theory of the Expanding Universe.}		MNRAS, 	\textbf{108}, pp. 252-270.}

	\item {Burgess, C.P.} 
	\begin{itemize}[nolistsep]
		\item {}
	(2003), \textit{Quantum Gravity in Everyday Life:
		General Relativity as an Effective Field
		Theory}. 	arXiv:gr-qc/0311082.
	\item {} (2013), \ti{The Cosmological Constant Problem:
		Why it’s hard to get Dark Energy from Micro-physics}. Lectures given to the Les Houches Summer School "Post-Planck Cosmology," 8 July - 2 August 2013. 52 pages. 	arXiv:1309.4133 [hep-th].
		\end{itemize}
	\item{Buttazzo, D., Degrassi, G., Giardino, P.P., Giudice, G.F., Sala, S. Salvio, A. and Strumia, A. (2014), \textit{Investigating the near-criticality of the Higgs boson}, CERN-PH-TH-2013-166, FTUAM-13-20, IFT-UAM/CSIC-13-081, IFUP-TH. Online access at 	arXiv:1307.3536 [hep-ph].}
	
	\item{Cahn, R.N. (1996), \textit{The eighteen arbitrary parameters of the standard model in your everyday life}. Rev. Mod. Phys 68, 951.}
	
	\item {} Callaway, D.J.E. (1988), \ti{Triviality Pursuit: Can Elementary Scalar Particles Exist?} Phys.Rept. 167 (1988) 241 RU-87-B1-20.
	
	\item {}  
	Cao, T.Y. and Schweber, S.S. (1993),
	\ti{The conceptual foundations and the philosophical aspects of renormalization theory}. Synthese, 97(1):pp.33-108, 1993.

	\item {} Castellani, E. (2002),	\ti{Reductionism, emergence, and effective field theories}  Studies in History and Philosophy of Modern Physics, 33(2):251-267, 2002.
	
	\item {} Castellani, E. and Borrelli, A. (2018), \ti{The Practice of Naturalness:
		An Historical-Philosophical Perspective}. Presentation given at CERN, February 29th. \url{https://indico.cern.ch/event/630393/contributions/2805717/attachments/1610685/2557435/BorrelliCastellani.pdf}, retrieved on 05-09-2018.
	
	\item{Carr, B. (2007), \textit{Universe or Multiverse?} Cambridge University Press, 2007.}
	
	\item{Carr B.J. and Rees M.J. (1979), \textit{The anthropic principle and the structure of the physical world}. Nature, 278, 605.}
	
	\item{Carroll, S. M.} 
	\begin{itemize}
		\item {} (2001), \ti{The Cosmological constant}, Living Rev. Rel. 4 (2001) 1 [astro-ph/0004075];
		\item {}
	(2006), \textit{Is our universe natural?} Nature \textbf{440}(7088), 1132.
		\end{itemize}
	
	\item{Carter B. (1974), \textit{Confrontation of Cosmological Theories with
			Observational Data}. In IAU Symposium, Vol. 63, Longair M. S., ed., D. Reidel, Dordrecht, pp. 291–298.}
	
	\item{ Casas, J.A., Espinoza, J.R. and  Hidalgo, I. (2005), \textit{Implications for new physics from fine-tuning
		arguments. II Little Higgs models.} JHEP, 03:038, 2005.}

	\item {Chan, K.L., Chattopadhyay, U. and  Nath, P (1998), \textit{Naturalness, weak scale supersymmetry,
		and the prospect for the observation of supersymmetry at the Fermilab Tevatron and at
		the CERN LHC}. Phys. Rev. D, \tb{58}:096004, 1998.}
	
	\item {}  Chandrasekhar, S. (1987). \ti{Truth and Beauty}. Chicago University Press, 1987.
	
	\item{Chatrchyan, S. \et. (CMS Collaboration) (2012), \textit{Observation
		of a new boson at a mass of 125 GeV with
		the CMS experiment at the LHC.} Phys. Lett. B \tb{716},
		30-61 [arXiv:1207.7235].}

\item {Conroy, A. (2017), \ti{Infinite Derivative Gravity:
			A Ghost and Singularity-free
			Theory.} 	arXiv:1704.07211 [gr-qc]. 
	}

\item {} Craig, N. (2013), \ti{The state of supersymmetry after run i of the LHC}. 	arXiv:1309.0528 [hep-ph].
	
	\item {Crewther, R.J., Vecchia, P.D., Veneziano, G.  and Witten, E. (1979), \ti{Chiral Estimate of the Electric Dipole Moment of the Neutron in Quantum Chromodynamics
	}. Phys.Lett. \tb{88}B (1979) 123}.
	
	\item {} Darrow, K. (1933), \ti{Contemporary advances in physics}, XXVI. Bell System Technical Journal,
	12:288–230, 1933. Quoted in (Kragh 1990, p. 267).
	
	\item{Dicke, R.H. (1961), \textit{Dirac's Cosmology and Mach's Principle}. Nature 192, 440.}	
	
	\item {}  Dimopoulos, S.  and Georgi, H. (1981), \ti{Softly Broken Supersymmetry and $SU(5)$}. Nucl. Phys. B\tb{193}, 150 (1981)
	
	\item {} Dimopoulos, S., Raby, S.  and Wilczek, F. (1981), \ti{Supersymmetry and the scale of unification}. Phys. Rev. D\tb{24}, 1681 (1981).
	
	\item{Dine, M.} 
\begin{itemize}[nolistsep]
	\item{} (2000), \ti{TASI Lectures on The Strong CP Problem}. SCIPP-00/30. 	arXiv:hep-ph/0011376.
	
	\item{}	(2007), \ti{Supersymmetry and String Theory: Beyond the Standard Model}. Cambridge University Press.
	
	\item{} (2015), \textit{Naturalness Under Stress}	
		Annual Review of Nuclear and Particle Science. 34 pages.
		arXiv:1501.01035 [hep-ph].
\end{itemize}
	
	\item{Dirac, P.A.M.} 
	\begin{itemize}[nolistsep]
		\tiny{
		\item{} 
		(1937),		\textit{The Cosmological Constants}. Nature 139 (1937) 323; Proc. Roy. Soc. London A 165 (1938) 198.
		
		\item {} (1938), \textit{The relation between mathematics and physics}. Proceedings of the Royal
		Society (Edinburgh), 59:122–129, 1939. Quoted in [41, p. 277].
	
	\item {} 
	(1963),
\textit{The Evolution of the Physicist's Picture of Nature}, Scientific American, Vol. 208, No.
	5, May 1963.	
			}
	\end{itemize}

	\item{Donoghue, J.F.} 
	\begin{itemize}[nolistsep]
		\item {} (1994) \textit{General relativity as an effective field theory:
		The leading quantum corrections}, 	Phys. Rev. D \tb{50}:3874-3888,1994.
			\item {} (2004), \it{Dynamics of M theory vacua.} Phys.Rev., D\tb{69}:106012, 2004.
		\item{} 	(2007) \textit{The fine-tuning problems of particle physics and anthropic mechanisms}.
		In Carr, B., chapter 15, p. 231.
	\end{itemize}

	\item{
Douglas, M.R.} 
\begin{itemize}[nolistsep]
\item {} (2004), \textit{Basic results in vacuum statistics}. Comptes Rendus Physique 5(9-10), pp. 965–977.	
	
	\item {} (2013), \ti{The string landscape and low energy supersymmetry.} In Strings, Gauge Fields,
	and the Geometry Behind: The Legacy of Maximilian Kreuzer, pp. 261–288. World Scientific.

\end{itemize}

\item{Duncan, A. (2012), \textit{The conceptual framework of quantum field theory}. Oxford University Press.}
	
	\item{Earman, J. (2006), \textit{The `Past Hypothesis': Not even false.} in Studies in
		History and Philosophy of Modern Physics 37 (3), pp. 399-430.}
	
	\item{Einstein, A. (1917). Letter to F. Klein, 12 December 1917. Quoted in Pais, A. (1982).}
	
	\item{Feng, J.L.} 
	\begin{itemize}[nolistsep]
		\item {} 	(2004), \textit{Supersymmetry and Cosmology}. Lectures given at the 2003 SLAC Summer Institute: Cosmic Connections to Particle Physics, in the Proceedings SLAC-R-702 and to appear in Annals of Physics. 	arXiv:hep-ph/0405215.
		\item {} (2013), \ti{Naturalness and the status of supersymmetry}. Annual Review of Nuclear and Particle Science, 63(1):pp. 351-382, 2013.
	\end{itemize}
	
	\item{Friederich, S. \& Lehmkuhl, D. (2015), \textit{Particle physics after the Higgs discovery: philosophical perspectives}. Studies in 	 History and Philosophy of Modern Physics, 51. pp. 69-70. ISSN 1355-2198.}

\item {Gaillard, M.K.  and Lee, B.W. (1974), \textit{Rare decay modes of the $K$ mesons in gauge theories}. Phys. Rev. D \tb{10}, 897 (1974).
}
	
\item{Gaillard, M.K., Grannis, P.D. and Sciulli, F.J. (1999), \textit{The standard model of particle physics}. Rev. Mod. Phys. \tb{71}, S96.}
	
\item {}  Gale, G. (1981), \ti{The anthropic principle}. Scientific American 245(Dec): 154.	
	
\item{Georgi, H. (1993) ``Effective Field Theory'', Annual Review of Nuclear and Particle
	Science 43: 209-52.}	

\item {} Georgi, H. and Glashow, S. (1973), \ti{Gauge Theories Without Anomalies}. Phys.
Rev.D7 2457 (1973).

\item {} Georgi, H. and Pais, A. (1974), \ti{Calculability and naturalness in gauge theories}. Physical Review D, vol. 10, Issue 2, pp. 539-558.
	
\item{Gildener, E.  (1976), \ti{Gauge-symmetry hierarchies}. Phys. Rev. D \tb{14}, 1667 (1976).}	
	
	\item{Giudice,  G.F.}
	\begin{itemize}[nolistsep]
		\item{} 	 (2008), \textit{Naturally speaking: The naturalness criterion and physics and LHC}. arXiv:0801.2562 [hep-ph].
		
		\item{} (2013), \textit{Naturalness after LHC8}. 2013 Europhysics Conference on High Energy Physics (EPS). 	arXiv:1307.7879 [hep-ph].
		
		\item{} (2017), \textit{
	The Dawn of the Post-Naturalness Era	
	}. contribution to the volume "From My Vast Repertoire" - The Legacy of Guido Altarelli. 	arXiv:1710.07663 [physics.hist-ph].
	\end{itemize}

		\item{Giudice, G.F.  and Romanino, A. (2004)
		\textit{Split Supersymmetry}.
		Nucl. Phys. B \tb{699} 65 (2004).} 
	
	\item {Glashow, S.L., Iliopoulos, J.  and Maiani, L. (1970),
	\textit{Weak Interactions with Lepton-Hadron Symmetry}.	
		Phys. Rev. D 2, 1285 (1970).}
	
	\item{Graham, P.W., Kaplan, D.E.  and Rajendran, S. (2015), \textit{Cosmological Relaxation of the Electroweak Scale}, Phys.
		Rev. Lett. 115 (2015) 221801 [arXiv:1504.07551].}
	
	\item{Grinbaum, A. (2009), \textit{Which fine-tuning arguments are fine?} \url{https://arxiv.org/pdf/0903.4055.pdf}. }

\item{} Guth, A.H. (2007), \ti{Eternal inflation and its implications}. J.Phys.A40:6811-6826, 2007.

	\item{Hartmann, S. (2001). \textit{Effective field theories, reductionism and scientific explanation.} Studies in History and Philosophy of Modern Physics 32: pp. 267–304.}

	\item{
Hogan, C.J. (2000). Why the universe is just so. Reviews of Modern Physics, 72, 1149–1161. }

\item {} Holton, G. (1973),  \ti{Thematic Origins of Scientific Thought}. Harvard Univ.
Press.

	\item{'t Hooft,  G. (1979), \textit{
	Naturalness, Chiral Symmetry, and Spontaneous Symmetry Breaking.}  In
		\textit{Proc. of 1979 Carg\`{e}se Institute on Recent Developments in Gauge
		Theories}, New York, 1980. Plenum Press.}

	\item{Hossenfelder, S.}
	\begin{itemize}[nolistsep]
		\item{} (2017) 
		\textit{Naturalness is dead. Long live naturalness.} \url{http://backreaction.blogspot.com/2017/11/naturalness-is-dead-long-live.html}, retrieved at 03.06.2018.
		
		\item{} (2018a) \ti{Screams for Explanation: Finetuning and Naturalness in the Foundations of Physics.} 	arXiv:1801.02176 [physics.hist-ph].
		
		\item{} (2018b) \ti{A Theory with No Strings Attached: Can Beautiful Physics Be Wrong?} [Excerpt], Scientific American. \url{https://www.scientificamerican.com/article/a-theory-with-no-strings-attached-can-beautiful-physics-be-wrong-excerpt}, retrieved on 16-06-2018.
	\end{itemize}	
	
	\item{Hoyle, F., Dunbar, D.N.F., Wenzel, W.A. and Whaling, W. (1953), \textit{The 7.68-Mev State in C12}. Phys Rev 92, 1095.}

	\item{Iliopoulos, J. (1979), In \ti{1979 Einstein Symposium}, Berlin, Springer-Verlag.}

	\item {Jaccard, M., Maggiore, M. and Mitsou, E. (2013), \ti{Nonlocal theory of
		massive gravity}. Phys.Rev., D\tb{88(4)}:044033}

	\item{Kallosh, R., Linde, A.D., Linde,  D.A.  and Susskind, L. (1995), \textit{Gravity and global symmetries}, Phys. Rev. D \tb{52}
		(1995) 912 [arXiv:hep-th/9502069].
}	

\item {} Kaplan, D.B., Georgi, H. and Dimopoulos, S. (1984), \ti{Composite Higgs Scalars}.
Phys.Lett., B\tb{136}:187, 1984.

\item {} Kaplan, D.B. and Georgi, H. (1984), \ti{Composite Higgs and Custodial SU(2).}
Phys.Lett., B\tb{145}:216, 1984.

\item {Kragh, H. (1990), \textit{Dirac: A Scientific Biography}. Cambridge University Press.}

\item{Kuhn, T.S. (1962) \textit{The Structure of Scientific Revolutions}, University of Chicago Press, Chicago 1962.}

\item {} Lawrence, J.H., Pinner, D. and Ruderman, J.T. (2014), \ti{The Weak Scale from
BBN.}	arXiv:1409.0551 [hep-ph].

	\item{LIGO Scientific and Virgo Collaborations (2016),
	\textit{Observation of Gravitational Waves from a
		Binary Black Hole Merger}, Phys. Rev. Lett. \textbf{116} (2016)
	no.6, 061102.
}

\item {} Linde, A.D. 
\begin{itemize}[nolistsep]
	\item {}
(1986), \ti{Eternally Existing Selfreproducing Chaotic Inflationary Universe.}
Phys.Lett., B175:395–400, 1986
\item {} (1990), \ti{Inflation And Quantum Cosmology}, Prepared for Les
Rencontres de Physique de la Vallee d'Aoste: Results and Perspectives in Particle Physics, La Thuile, Italy, 26 Feb - 4 Mar 1989.
\end{itemize}
	\item {Lovelock, D. (1971), \ti{The Einstein Tensor and its Generalizations.} Journal of
	Mathematical Physics, \textbf{12}(3), 1971.
}

\item{Lust, D.  and  Palti, E. (2017), \textit{Scalar Fields, Hierarchical UV/IR Mixing and The Weak Gravity Conjecture},
arXiv:1709.01790 [hep-th].
}

\item {Maggiore, M. (2014), \ti{Phantom dark energy from non-local infrared modifications
	of General Relativity}. Phys.Rev., D\tb{89}:043008, 2014.
}

\item{Manohar, A. (1997) ``Effective Field Theories", in Perturbative and Nonperturbative
	Aspects of Quantum Field Theory, Lecture Notes in Physics, Vol 479/1997, Springer:
	311-362. arXiv: [hep-ph/9606222].}

\item {} Marciano, W. J.  and Senjanovic, G. (1982), \ti{Unification of Couplings}. Phys. Rev. D\tb{25}, 3092 (1982).

\item{ 
	McCulloch, M. (2016),
\textit{Critique of Verlinde's Gravity}. Retrieved from \url{http://physicsfromtheedge.blogspot.com/2016/11/critique-of-verlindes-gravity-1.html} at 26-06-2018.}

\item {}
Merz, J. (1904), \ti{A History of European Thought in the Nineteenth Century}. London: Blackwood.

\item{Minwalla, S., Van Raamsdonk, M. and Seiberg, N. (2000), \textit{Noncommutative
	perturbative dynamics,} JHEP 02 (2000) 020, arXiv:9912072
	[hep-th].
}

\item {Modesto, L. and Tsujikawa, S. (2013), \ti{Non-local massive gravity.}
	Phys.Lett., B\tb{727}:48–56, 2013.}

\item {} Montvay, I. and Munster, G. (1997), \ti{Quantum fields on a lattice}. Cambridge University Press, Cambridge.

	\item{Murayama, H. (2000) \ti{Supersymmetry Phenomenology.} arXiv:hep-ph/0002232.}

	\item{Ovrut, B. \& Schnitzer, H.J. (1980), \textit{New approach to effective field theories.} Phys. Rev. D, 21:3369-3387, Jun 1980.}

	\item{Nelson, P. (1985), \textit{Naturalness in Theoretical Physics},  American Scientist
		Vol. 73, No. 1 (Jan-Feb 1985), pp.60-67.}

	\item{Nightingale, J.D. and Foster, J.A. (2005), \textit{A Short Course in General Relativity}.  Springer; 3rd edition.
	}

\item {}  Noether, E (1918). \ti{Invariante Variationsprobleme}, Nachr. D. K\"{o}nig. Gesellsch. D. Wiss. Zu G\"{o}ttingen, Math-phys. Klasse. 1918: pp. 235–257.

	\item{Pais, A. (1982), \textit{`Subtle is the Lord...': The Science and the Life of Albert Einstein}. Oxford
		University Press, 1982.}
	
	\item{Peccei, R.D.  and Quinn, H.R. (1977),
		\textit{CP Conservation in the Presence of Pseudoparticles}. H.R. Phys. Rev. Lett. 38 (1977) 1440; Phys. Rev. D16 (1977) 1791.}
	
	\item{Peebles, P. (1971), \textit{Physical Cosmology}. Princeton Univ. Press.}

	\item {} Penzias, A.A. and Wilson, R. W. (1965), \ti{A Measurement of Excess Antenna Temperature at 4080 Mc/s.} The Astrophysical Journal. 142: 419.
	
	\item {} Penrose, R. (2012), \ti{Cycles of Time: An Extraordinary New View of the Universe}. Vintage; Reprint edition (May 1, 2012).
	
	\item {Peskin, M.E. and  Schroeder, D.V. (1995), \ti{An Introduction To Quantum Field Theory}, Avalon Publishing.
	}

	\item {} Planck Collaboration (2013), \ti{Planck 2013 results. XXII. Constraints on inflation.} 	arXiv:1303.5082 [astro-ph.CO]
	
	\item{Price, H. (1997), \textit{More Apt to be Lost Than Got.} In: Time's Arrow and
		Archimedes Point, Oxford University Press, New York, pp. 22-48.}
	
	\item{Quigg, C. (2007), \textit{Spontaneously symmetry breaking as a basic of particle mass}, Reports on Progress in Physics 70:1019-1053.}

\item {} Randall, L. and Sundrum, R. (1999), \ti{Large Mass Hierarchy from a Small Extra Dimension}. Physical Review Letters. 83 (17): 3370–3373. 

\item{Ray, S., Mukhopadhyay, U. and
Ghosh, P.P. (2007), \textit{Large Number Hypothesis: A Review}. 	arXiv:0705.1836 [gr-qc].}

\item{Richter, B. (2006),
\textit{Theory in particle physics: theological speculation versus practical knowledge}
Physics today: 59(10):8-9, 2006.}

\item {Robertson, B. (1971), \ti{Wyler's Expression for the Fine-Structure Constant 
		$\alpha$}.
	Phys. Rev. Lett. \tb{27}, 1545 (1971).
}

\item {} Ross, G.G., Schmidt-Hoberg, K. and Staub, F. (2012),
\ti{The Generalised NMSSM at One Loop: fine Tuning
and Phenomenology}. JHEP 1208, 074 (2012)
[arXiv:1205.1509].
	
	\item{Schellekens, A.N. (2013). \textit{Life at the interface of particle physics and string theory.} Reviews of Modern
		Physics, 85, 1491–1540.}
	
	\item{Schwartz, M. (2014), \textit{Quantum field theory and the standard model}. Cambridge University Press, Cambridge.}

\item{Shifman, M. (2012), \textit{Reflections and impressionistic portrait at the conference ``frontiers beyond the standard model,'' ftpi, oct. 2012. 	arXiv:1211.0004 [physics.pop-ph].}}

\item{Sol\`{a}, J. (2015), \textit{Fundamental Constants in Physics and Their Time Variation}. Preface to the Special Issue on Fundamental Constants in Physics and Their Time Variation. 	arXiv:1507.02229 [hep-ph].}
	
\item {} Stelle, K.S. (1977) \textit{Renormalization Of Higher Derivative Quantum Gravity}, Phys. Rev. \textbf{D16}, pp. 953-969.

\item {} Sundrum, R. (1999), \ti{Towards an effective particle-string resolution of
the cosmological constant problem}, JHEP 9907, 001 (1999)
[arXiv:hep-ph/9708329].

\item{ Susskind, L.} 
\begin{itemize}[nolistsep]
	\item {}
 (1970), \textit{Dynamics of spontaneous symmetry breaking in the Weinberg-Salam theory}. Phys. Rev. D \textbf{20}, 2619. 
\item{(2004), \textit{Naturalness and the landscape}. arXiv preprint hep-ph/0406197}.
\item{(2005), \textit{The Cosmic
	Landscape: String Theory and the Illusion of Intelligent Design}, Little, Brown and Company, New York
	2005.}
\end{itemize}
\item{Swetman, T.P. (1972), \textit{Are fundamental constants really constant?} Physics Education v7 n7 (September 1972): pp. 411-412.}

\item {} Tegmark, M.,  Aguirre, A., Rees, M. and Wilczek, F. (2006) Phys. Rev. D 73, 023505 (2005), \ti{Astrophysics
	Dimensionless constants, cosmology and other dark matters}
[astro-ph/0511774].

\item {} Teller, E. (1948), \ti{On the Change of Physical Constants}. Phys. Rev. 73, 801 (1948).

\item {
Vecchi, L. (2014),	\ti{
	Spontaneous CP violation and the strong CP problem}.	arXiv:1412.3805 [hep-ph].}

\item{Verlinde, E.P. (2016),
	\textit{Emergent Gravity and the Dark Universe}. 	arXiv:1611.02269 [hep-th]}

\item { Wald, R. M. (1984) \textit{General Relativity}, doi:10.7208/chicago/9780226870373.0.
}

\item{ Weinberg, S.} 
\begin{itemize}[nolistsep]
	\item{}
(1976), \textit{Implications of dynamical symmetry breaking}. Phys. Rev. D \textbf{13}, 974.

\item{} (1979), \ti{Gauge Hierarchies}.  Phys. Lett. B \tb{82}, 387 (1979).

\item {} (1972), \ti{Approximate Symmetries and Pseudo-Goldstone Bosons.}  Phys. Rev. Lett. \tb{29}, 1698 – Published 18 December 1972.

\item{} (1987), \textit{Anthropic Bound on the Cosmological Constant}, Phys. Rev. Lett. \tb{59} (1987) 2607.

\item {} (1996), \ti{Theories of the cosmological constant},
	astro-ph/9610044.

\item {} (2000), \ti{A Priori probability distribution of the cosmological constant.}
Phys.Rev., D61:103505, 2000.

\item{} (2007), \textit{Living in the multiverse}. Universe or multiverse, 29–42.
\end{itemize}

\item{Wells, D.W.} 

\begin{itemize}[nolistsep]
	\item {} (2003) \ti{Implications of supersymmetry breaking
	with a little hierarchy between gauginos and
	scalars}. hep-ph/0306127 (SUSY 2003, Tucson).
\item {} (2004), \ti{PeV-Scale Supersymmetry}. Phys.
	Rev. D\tb{71}, 015013 (2005) [hep-ph/0411041].

\item {(2013), \textit{The Utility of Naturalness, and how its Application to
		Quantum Electrodynamics envisages the Standard Model and Higgs Boson}, Studies in History and Philosophy of Modern Physics, 49: 102–108. doi:10.1016/j.shpsb.2015.01.002.}
\end{itemize}	
	
	\item{Wetterich, C. (1984), \ti{Fine Tuning Problem and the Renormalization Group}. Phys. Lett. B \tb{140} (1984) 215-222 CERN-TH-3528.}
	
	\item { Weiskopf, V.F. (1939), \ti{On the Self-Energy and the Electromagnetic Field of the Electron}. Phys. Rev. \tb{56}, 72.}
	
	\item{Williams, P.}
	\begin{itemize}[nolistsep]
		\item {} (2015),		
		\textit{Naturalness, the autonomy of scales, and the 125 GeV Higgs}. Studies in History and Philosophy of Science Part B: Studies in History and Philosophy of Modern Physics,	Volume 51, August 2015, pp. 82-96.

		\item{} (2018), \ti{Two Notions of Naturalness}. Presentation given at CERN at 28-02-2018, lecture notes can be accesses online at \url{https://indico.cern.ch/event/630393/contributions/2819907/}. Retrieved at 18-07-2018.
	\end{itemize}

\item{Wilson, K.G. (1971), \ti{The Renormalization Group and Strong Interactions}. Phys. Rev. 
	D \tb{3:}1818, 1971.}

\item{Witten, E. (2017), \textit{Symmetry and Emergence}. arXiv:1710.01791.
}

\item {} 
Woit, P. (2017), \ti{Why String Theory Is Still Not Even Wrong}. Scientific American. \url{blogs.scientificamerican.com/cross-check/why-string-theory-is-still-not-even-wrong/}.

\item {Yao, W. M. \et [Particle Data Group], \ti{Review of Particle Physics} J. Phys. G \tb{33}, 1 (2006).}

\item {Zee, A. (2010), \ti{Quantum Field Theory in a Nutshell.} Princeton University Press; 2nd edition. } 
\end{itemize}	

\newpage

\appendix

\include{appendixhierarchy}

\end{document}